\documentclass[12pt]{article}
\pdfoutput=1

\usepackage{graphicx}
\usepackage{epsfig,amsmath,amsfonts,amssymb,cite}
\usepackage[usenames]{color}
\usepackage[a4paper,left=2.cm,right=2.cm,top=2.5cm,bottom=2.5cm]{geometry}

\newcommand{\beq}{\begin{equation}}
\newcommand{\eeq}{\end{equation}}
\newcommand{\be}{\begin{equation}}
\newcommand{\ee}{\end{equation}}
\newcommand{\beqa}{\begin{eqnarray}}
\newcommand{\eeqa}{\end{eqnarray}}
\newcommand{\beqar}{\begin{eqnarray*}}
\newcommand{\eeqar}{\end{eqnarray*}}
\newcommand{\bea}{\begin{eqnarray}}
\newcommand{\eea}{\end{eqnarray}}

\newcommand{\Or}{\mathcal{O}}

\def\O{{\cal O}}

\numberwithin{equation}{section}

\begin{document}

\allowdisplaybreaks
\setlength{\unitlength}{1mm}
\thispagestyle{empty}
 \vspace*{2cm}

\begin{center}
{\bf \LARGE Hairy black holes and solitons in global AdS$_5$}\\

\vspace*{1.5cm}

{\bf \'Oscar J.~C.~Dias$^{a}\,$}, {\bf Pau Figueras$^{b}\,$},
{\bf Shiraz Minwalla$^{c}\,$}, \\
\vspace*{0.2cm}
{\bf Prahar Mitra$^{c}\,$}, {\bf Ricardo Monteiro$^{d}\,$}, {\bf
Jorge E.~Santos$^{e}\,$}
\vspace*{0.8cm}

{\it $^a\,$Institute de Physique Th\'{e}orique,\\
CEA Saclay, 91191 Gif sur Yvette, France}\\[.3em]

{\it $^b\,$ DAMTP, Centre for Mathematical Sciences, University
of Cambridge,\\
Wilberforce Road, Cambridge CB3 0WA, United Kingdom}\\[.3em]

{\it $^c\,$Dept. of Theoretical Physics, Tata Institute of
Fundamental Research,\\
Homi Bhabha Rd, Mumbai 400005, India}\\[.3em]

{\it $^d\,$ The Niels Bohr International Academy, The Niels Bohr
Institute, \\
University of Copenhagen, Blegdamsvej 17, DK-2100 Copenhagen, Denmark}\\[.3em]

{\it $^e\,$ Department of Physics, UCSB, Santa Barbara, CA
93106, USA}\\[.3em]

\vspace*{0.5cm} {\tt oscar.dias@cea.fr,
p.figueras@damtp.cam.ac.uk, \\
minwalla@theory.tifr.res.in, pmitra@physics.harvard.edu, \\
monteiro@nbi.dk, jss55@physics.ucsb.edu} \\

\end{center}

\vspace{1cm}

\begin{abstract}

We use a mix of analytic and numerical methods to exhaustively study a class of 
asymptotically global AdS solitons and hairy black hole solutions in negative 
cosmological constant Einstein Maxwell gravity coupled to a charged massless
scalar field. Our results depend sensitively on the charge $e$ of the 
scalar field. The solitonic branch of solutions we study hits the 
Chandrashekhar limit at finite mass at small $e$,  but extends to arbitrarily 
large mass at larger $e$. At low values of $e$ no hairy black holes exist. 
At intermediate values of $e$ hairy black holes exist above a critical charge. 
At large $e$ hairy black holes exist at all values of the charge.  
The lowest mass hairy black hole is a smooth zero entropy 
soliton at small charge, but a (probably) singular nonzero entropy hairy 
black hole at 
larger charge. In a phase diagram of solutions, the hairy black holes merge 
with the familiar  Reissner-Nordstr\"om$-$AdS black holes along a curve that is 
determined by the onset of the superradiant instability in the latter family.
\end{abstract}

\noindent
\vfill \setcounter{page}{0} \setcounter{footnote}{0}
\newpage

\tableofcontents

\section{Introduction}
\label{sec:intro}

The AdS/CFT correspondence maps asymptotically AdS solutions of Einstein's 
equations to states of a dual conformal field theory \cite{Maldacena:1997re,Gubser:1998bc,Witten:1998qj}. Static bulk 
solutions map to `phases' of the CFT.  A thorough investigation of 
all static gravitational solutions consequently permits complete 
understanding of the phase structure of the dual large $N$ field theory, 
and so is of considerable interest. 

In this paper we follow \cite{Basu:2010uz} to study 
AdS Einstein Maxwell gravity interacting with a charged massless scalar field. 
The bulk theory we study is governed by the Lagrangian 
\begin{equation}\label{sysaction}
S = \frac{1}{8\pi G_5}\int
d^5x\sqrt{-g}\left[\frac{1}{2}\left({\cal R}[g]+12\right) -
\frac{1}{4}{\cal F}_{\mu\nu}{\cal F}^{\mu\nu} -
|D_\mu\phi|^2\right]
\end{equation}
where $ D_\mu = \nabla_\mu - ieA_\mu$, $e$, the charge of the scalar 
field, is a free parameter, and we have set the cosmological length $\ell\equiv 1$. The action \eqref{sysaction}, sometimes called the AdS Abelian Higgs model, may be regarded as a simple toy model for the charged scalar dynamics of systems that appear in concrete examples of the AdS/CFT correspondence (see e.g. \cite{Gubser:2009qm,Gauntlett:2009dn,Bhattacharyya:2010yg}). There has been great interest in this model since it was realized that it allows for a phase transition between the familiar Reissner-Nordstr\"om$-$AdS (RN AdS) black holes and hairy black holes, i.e. solutions with a scalar condensate \cite{Gubser:2008px}. 
In the AdS/CFT dual to this system, such condensate spontaneously breaks the 
global $U(1)$ boundary symmetry, and so corresponds to a superfluid phase 
of the boundary CFT.  Bulk solutions that asymptote to $AdS$ with planar 
sections (i.e. Poincare patch $AdS$) are dual to phases of the dual field 
theory on spatial $R^3$, and have been extensively explored within  the context of the 
AdS/CFT correspondence to learn something about condensed matter phenomena \cite{Hartnoll:2008vx,Hartnoll:2008kx}; see \cite{Hartnoll:2009sz,Horowitz:2010gk,Hartnoll:2011fn} for reviews. The solutions we study in this paper are dual to 
superfluid phases of the boundary field theory on a spatial $S^3$. The results of 
this paper must, of course, reduce to the planar model in the limit that 
the radius of the $S^3$ is taken to infinity; we will see below that this 
is indeed the case.

In this paper we perform a thorough analysis of the properties of three 
classes of 
static charged solutions of the action \eqref{sysaction}. The solutions we study 
all asymptote to global $AdS_5$ space (i.e. $AdS_5$ with spherical sections), 
and so map to phases of the `dual large $N$ theory' on $S^3$.  We investigate
the properties of our solutions as a function of their mass, charge, and 
also of the free parameter $e$ in our model.

The first class of solutions we study in this paper is the set of 
`ground state solitons' discovered in \cite{Basu:2010uz}. These `solitons' 
are static lumps of a scalar condensate in global $AdS_5$ space.
\footnote{The solitonic solutions map to `bose condensates' in the dual CFT on $S^3$.}  
At any given value of $e$ these solitonic solutions appear in a one parameter
family labeled by their charge; the mass of solitonic solutions is determined
as a function of their charge. At infinitesimal values of the charge these 
solitonic solutions are extremely simple; they reduce to $AdS_5$ space 
perturbed by the lowest energy linearized mode of the scalar field.\footnote{The 
harmonic time  dependence of the mode is gauged away yielding a 
static solution, at the expense of turning on the value of $A_0$ at the boundary.} In \cite{Basu:2010uz} this linearized solitonic solution 
was used as the starting point of a perturbative construction for 
solitonic solutions in a power series in their charge $Q$. 
In this paper we have used numerical techniques to extend the construction of 
\cite{Basu:2010uz} to arbitrarily large values of the charge $Q$. 
Using the analytic construction of \cite{Basu:2010uz} as the starting 
point of our numerical construction, we slowly iterate to larger values 
of the charge. Our results agree perfectly with those presented in 
\cite{Basu:2010uz} at small charge. However, we find some surprises at large 
charge.

Our numerics, presented in section \ref{sec:solitons}, show that the 
qualitative properties of solitonic solutions differ depending on the value of the parameter $e$.  For $e>e_{solcrit}$, where 
$e_{solcrit}$ is a critical charge, the solitonic branch of solutions continues 
all the way to infinite charge; in other words there exist solitons at 
arbitrarily large charge. On the other hand for $e<e_{solcrit}$, 
the solitonic branch of solutions terminates in a `Chandrashekhar' singularity 
at a finite value of the mass $M_{crit}$ and charge $Q_{crit}$ 
(see \cite{Bhattacharyya:2010yg} for similar behaviour in a closely related
context). The approach of the solitonic solution to the critical value 
is characterized by a spiral behaviour and critical exponents familiar 
from the study of boson stars (see \cite{Bhattacharyya:2010yg}). Numerically
we find that $e_{solcrit}$ is very close to $\sqrt{\frac{32}{3}}$, a value that 
we will encounter again below.

As $e$ is cranked up to $e_{solcrit}$ from below, our numerical results appear 
to indicate that $M_{crit}$ and $Q_{crit}$ diverge in a manner proportional 
to  $ \ln (e-e_{solcrit})$. 
We have drawn this conclusion by noting that a graph of  $Q_{crit}$ versus 
$\ln (e_{solcrit}-e)$ is very nearly a straight line at small $e_{solcrit}-e$. 
Because of the slow growth of the logarithm, however, 
we have not been able to verify this 
divergence in a more 
straightforward manner by simply checking that $Q_{crit}$ grows arbitrarily
large as $e$ is taken arbitrarily close to $e_{solcrit}$. In the absence of analytic understanding here, our results on this point should be treated 
as tentative, subject to further confirmation; in particular we have 
not really ruled out the (unlikely sounding) possibility that $Q_{crit}$ 
remains finite at $e=e_{solcrit}$.

The second type of solutions we consider in this paper is the very well known 
RN AdS charged black hole solutions. We study the stability of these solutions
to the condensation of the scalar field in \eqref{sysaction}, as a function
of the black hole mass, black hole charge, and parameter $e$. Our analysis
of this question (which uses a mix of analytic and numerical methods) 
yields the following results. RN AdS black holes are never unstable to 
scalar condensation when $e^2 \leq 3$.  On the other hand when 
$3 \leq e^2 \leq \frac{32}{3}$, near-extremal black holes display an instability
at large enough charge, i.e. when $Q \geq Q_0(e^2)$. 
In the range $3 \leq e^2 \leq \frac{32}{3}$ the minimum charge for
instability, $Q_0(e^2)$, is a monotonically decreasing function 
of $e^2$, with $Q_0(3)=\infty$ and $Q_0(\frac{32}{3})=0$. Black holes 
with $Q \geq Q_0(e^2)$ -- and for all values of the charge when $e^2 \geq 
\frac{32}{3}$ -- are unstable in a mass band around extremality.
The function $Q_0(e^2)$ itself appears to undergo a phase transition 
at $e^2=\frac{32}{9}$ in this range, associated with a change in the 
qualitative nature of the instability mode at this value of $e^2$. 
We pause momentarily to discuss this in more detail. 

The instabilities  that leads to scalar condensation about RN AdS black holes
are easiest to study in the case of extremal black holes and occur in 
at least two varieties. The first kind of instability
is localized entirely within the near horizon $AdS_2$ geometry of the 
extremal black holes. This instability occurs because the scalar field 
acquires a near horizon effective mass that violates the Breitenl\"ohner-Freedman bound of the near horizon $AdS_2$ \cite{Breitenlohner:1982jf}, and turns 
out to be the dominant instability for extremal black holes whose horizon 
radius is larger than the radius of curvature of the ambient $AdS_5$ \cite{Dias:2010ma}. 
However, for solutions asymptoting to global AdS 
(as opposed to planar solutions) the near horizon region is not always the 
determining factor for instability. Ref. \cite{Basu:2010uz} found that for small black holes (compared to the AdS scale) the instability
mechanism is not dominantly near horizon. The instability is better  
thought of as an AdS superradiant instability of charged scalar fields.$^{2,}$\footnote{Recall that a mode $e^{-i \omega t}$ of a scalar field with charge $e$ increases its amplitude by scattering off a charged black hole with chemical potential $\mu$ if  $\omega < e \mu$
\cite{Denardo:1973df,Bekenstein:1973mi,Damour:1974qv,Gibbons:1975kk}. 
In an asymptotically global AdS spacetime, this leads to an instability 
since the
outgoing wave is reflected back onto the black hole and scatters again, further
increasing its amplitude \cite{Hawking:1999dp,Cardoso:2004hs,Basu:2010uz}. }
As we argue below, for extremal black holes, the switch between 
these two different types of instabilities leads to a non analyticity
in the function $Q_0(e^2)$ at a particular finite value of $e^2$, most likely 
$e^2=\frac{32}{9}$. 
We mention that there is a closely related ongoing programme to find the phase diagram associated to the AdS scalar superradiance for rotating black holes, rather than charged \cite{Press:1972zz,Hawking:1999dp,Cardoso:2004hs,Cardoso:2004nk,Kunduri:2006qa,Dias:2011at,Stotyn:2011ns}.
In the future it would be interesting to combine these  studies and eventually find charged rotating AdS hairy black holes.

Consider an unstable RN AdS black hole of charge $Q$ and mass $M$. Triggering 
the instability leads 
to a decay process that presumably settles down, at infinite time, to a 
stable static configuration of mass $M$ and charge $Q$. By the Hawking area 
increase theorem the resultant configuration has an event horizon and 
so is a black hole. As the instability involves condensation of the 
scalar field, the resultant black hole solution 
is immersed in a sea of scalar hair. Such solutions are sometimes 
called `hairy' black holes, and must exist at every value of $M$ and $Q$ 
for which RN AdS black holes are unstable. Small hairy black holes 
in the Lagrangian \eqref{sysaction} were constructed perturbatively 
(in an expansion in the charge of the solutions) in \cite{Basu:2010uz}. 
The key observation in \cite{Basu:2010uz}  is that infinitesimally small hairy black 
holes may accurately be thought of as a `superposition' of a small 
RN AdS black hole and the scalar soliton. Using this superposition 
configuration as a starting point, hairy black hole solutions may be 
constructed in a perturbative expansion in the charge of the 
black hole solutions.  Working in the small charge limit at any fixed value of 
$e^2 > \frac{32}{3}$, the authors of  \cite{Basu:2010uz} found that the lowest mass black hole solutions are simply the smooth, 
zero entropy solitonic solutions described earlier in this introduction. 
These configurations 
may be thought of as the {\it infinite} temperature limit of hairy black 
holes.

One of the main results of the current paper is that the nature of hairy black 
holes, in the zero temperature limit, changes discontinuously as a function of
the charge (see Section \ref{sec:largee}) when $e^2 \geq \frac{32}{3}$. 
At low values of the charge -- in fact for all $Q \leq Q_{c_2}(e^2)$, 
where $Q_{c_2}(e^2)$ is a monotonically increasing function with 
 $Q_{c_2}(\frac{32}{3})=0$ -- 
the lowest mass hairy black holes indeed reduce to the solitons, in agreement 
with the perturbative results of \cite{Basu:2010uz}. For $Q \geq Q_{c_2}(e^2)$, 
however, the lowest mass hairy black holes do not reduce to the soliton. 
The new extremal solutions at these charges appear to have finite entropy 
and zero temperature. They may well be singular in the strict extremal limit; in fact, regular extremal hairy black holes are not allowed in the system under study \cite{FernandezGracia:2009em}. 
It turns out that 
for $e^2=\frac{32}{3}(1+\theta)$, $Q_{c_2}(e^2) \sim  {\cal O}(\theta)$. 
As a consequence, the phenomenon described in this paragraph -- the
discontinuous change in the nature of extremal black hole solutions as 
a function of their charge -- is visible at small charges in a perturbative 
expansion in $\theta$ (see Section \ref{sec:largee}). 
At larger values of scalar charge $e$, this phenomenon is 
also visible in our numerical construction of these hairy black 
holes (see Section \ref{sec:largee}). 

We now turn to the spectrum of hairy black hole solutions 
for $ 3 \leq e^2 \leq \frac{32}{3}$. As we have explained above, in this 
range hairy black hole solutions exist only for $Q \geq Q_0(e^2)$. 
When $e^2=\frac{32}{3}(1-\theta)$, it turns out that $Q_0(e^2) \sim 
{\cal O}(\theta)$. Consequently hairy black holes of charge 
$\sim {\cal O}(\theta)$ may be constructed in perturbation theory in 
$\theta$ as in the previous paragraph 
(see Section \ref{sec:middlee}). The results of this analysis at small 
charge, combined with our numerical construction of these solutions 
at large charge, allow 
us to conclude that the lowest mass hairy black holes at fixed charge 
are never solitons. As in the previous paragraph, they appear to be 
possibly singular zero temperature finite entropy solutions.

In summary, in this paper we have presented a rather complete picture of 
the solution space of a family of solitons, RN AdS black holes and a family 
of hairy black holes solutions. The solutions we have constructed 
certainly do not exhaust the set of static charged solutions in 
\eqref{sysaction}. For instance it was demonstrated in \cite{Basu:2010uz}
 that there exist an infinite number of `excited' solitonic solutions 
(one based 
on each of the linearized excitations of the scalar field about
 global $AdS_5$ space). Depending on the value of $e^2$, some of these 
solitons may also be obtained as the lowest mass limit of a class of 
`excited' hairy black hole solutions. At least at small charge, however, 
these new excited hairy black hole solutions are all unstable and 
so are unimportant for thermodynamical purposes. We feel that it is 
plausible that the thermodynamics of the system described by 
\eqref{sysaction} is dominated by one of the three solutions studied 
in this paper at every value of mass, charge and $e$, although a verification
of this suggestion would require further work. 

The results of  this paper appear to throw up a numerical 
coincidence for which we do not have an explanation. The value 
$e^2=\frac{32}{3}$ appears to be special for two different reasons. First it
marks the smallest value of the scalar charge for which one can construct 
arbitrarily small hairy black hole solutions. Second it appears, within
numerical accuracy, to mark the dividing line for solitonic solutions. 
Below this value of $e^2$, the solitonic branch of solutions constructed 
in this paper  encounters the Chandrashekhar limit 
at large mass. Above this value of $e^2$, smooth solitonic solutions 
exist at arbitrarily large mass. If these two different sharp changes 
do indeed occur at the same value of $e^2$, this coincidence requires an
explanation that we do not yet have. Of course it is entirely possible that 
the solitonic shift in Chandrashekhar limit does not occur at 
$e^2=\frac{32}{3}$ but at a value that is coincidentally so near to this
value that our numerics are unable to tell the two numbers apart. We leave 
a fuller consideration of this matter to future work. 

Another issue that we have not fully resolved in this paper is the precise 
nature of extremal hairy black holes. As we have 
reviewed above, the lowest mass hairy black holes appear to have zero 
temperature and finite entropy. Precisely at extremality these solutions 
appear to be singular (the curvature invariants and tidal forces grow large as we approach this configuration), however neither our perturbative techniques
nor our numerical constructions are sensitive enough to determine what
the nature of this singularity is. We leave a fuller characterization 
of these extremal solutions to future work. 

The nature of these extremal solutions also has bearing on the study of 
supersymmetric
black holes in $AdS_5 \times S_5$. As explained in 
\cite{Bhattacharyya:2010yg}, IIB supergravity on $AdS_5 \times S^5$ 
admits a consistent truncation that sets equal the three diagonal $U(1)s$ in 
the R symmetry $SO(6)$, and includes a charged scalar field under this 
$U(1)$. This consistent truncation admits hairy black hole solutions which
were constructed perturbatively at small charge in \cite{Bhattacharyya:2010yg}.
At small charge these hairy black holes reduce to a supersymmetric soliton
in the extremal limit. However the susy soliton does not exist beyond 
a critical charge of order unity. The nature of the extremal limit 
of hairy black holes beyond this charge is as yet unknown, and may 
have important bearing on our picture of the vacuum structure of 
${\cal N}=4$ Yang Mills at finite $SO(6)$ charge density. 

The plan of the paper is as follows.  Section \ref{sec:summary} provides a sharp summary of the phase diagram of the system \eqref{sysaction} in the microcanonical ensemble (omitting some details discussed in later sections). Section \ref{sec:solitons}  is entirely devoted to the study of the soliton family of solutions for all values of the scalar charge. Section \ref{sec:RNlinear} introduces the  Reissner-Nordstr\"om$-$AdS (RN AdS) black hole; studies the Klein-Gordon equation for linearized charged scalar perturbations in the RN AdS background to find the onset unstable modes; and discusses the two sources of instability, namely the scalar condensation and the superradiant instabilities. 
Section \ref{sec:middlee} addresses the solutions with scalar field charge 
$3<e^2<\frac{32}{3}$. It first uses a thermodynamic non-interacting model to find the leading order properties of the solutions and then it constructs these solutions, first using a perturbative approach and then a full numerical construction. Section \ref{sec:largee} repeats this process but this time for solutions with scalar field charge $e^2>\frac{32}{3}$. It starts with a thermodynamic non-interacting model to find the leading order properties of the solutions and then it constructs these solutions using a perturbative and a numerical construction. The details of the perturbative construction of the hairy black holes of section   \ref{sec:middlee}  and \ref{sec:largee} are left to Appendices B, C and D. 

{\em Note added:} Reference \cite{Gentle:2011kv}, which appears
simultaneously with our work on the arXiv,
discovers and studies a new branch of solitonic solutions in
a model closely related to \eqref{sysaction}. The interplay of these new
solitons with the solutions studied in this paper appears to be an interesting
topic for a future study.
We added a small section \ref{subsec:2ndbranch} in a second version of our work to address some immediate questions.

\section{Summary of the phase diagram}
\label{sec:summary}

In this we section make the assumption that the thermodynamics of 
\eqref{sysaction} is determined entirely by the three classes of static 
solutions considered in this paper. Under this assumption we present 
a summary of the phase diagram of the system \eqref{sysaction} in the 
microcanonical ensemble. This phase diagram is qualitatively different 
depending on whether $e^2 <3$, $ 3 \leq e^2 \leq \frac{32}{3}$
or $e^2 > \frac{32}{3}$. We consider these three cases in turn. The phase 
diagrams presented in this system will be justified in detail in future 
sections.

\subsection{$e^2 \leq 3$}

At these values of $e$ the phase diagram of $\eqref{sysaction}$ is always 
dominated by RN AdS black holes which are always stable. As the properties
of RN AdS black holes do not depend of $e$, their phase diagram is 
independent of $e^2$. At any given value of the charge the RN AdS phase 
exists down to a minimum mass; the phase boundary is given by extremal 
RN AdS black holes. 

Solitonic solutions exist up to a certain maximum charge, at these values 
of $e$, but they play no role in the thermodynamics of the system. 
In particular, solitons do not represent the ground state of the system at 
any given charge, as their mass is always greater than the mass of 
the extremal black hole at the same charge.

\begin{figure}
\begin{center}
\includegraphics[scale=0.4]{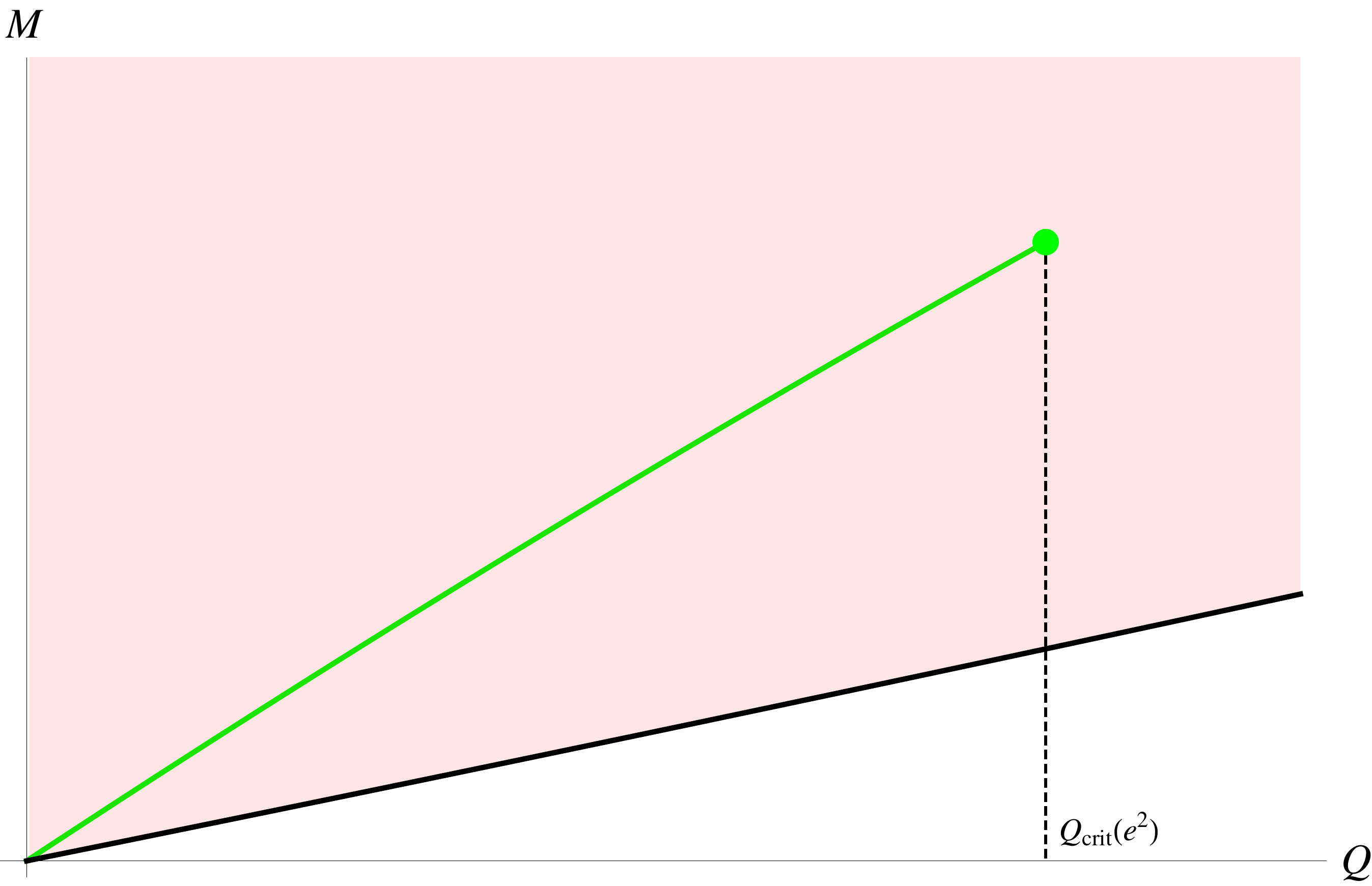}
\end{center}
\caption{Schematic phase diagram for $e^2 \leq 3$. Note that the soliton curve (green) is always above the extremal black hole (black). Pure RN AdS black holes exist for all values of masses above the extremality curve. The extremal black hole is never unstable and no hairy black hole solutions exist. The soliton exists up to a certain maximum charge, has a self-similar behaviour around a value $Q_{crit}(e^2)$ and is never the dominant phase.}
\label{esqless3}
\end{figure}

The qualitative phase diagram for $e^2 \leq 3$ is shown in Fig. \ref{esqless3}. For a numerically constructed phase diagram with full details, see Fig. \ref{fig:solitonphasesq1}.  (We find that an interesting cusp structure develops in the neighbourhood of $Q_{crit}$, whose discussion we postpone.)

\subsection{$3 \leq e^2 \leq \frac{32}{3} $}

At these values of $e^2$ RN AdS black holes are unstable in a band about 
extremality for $Q \geq Q_0(e^2)$. When they are stable, RN AdS black holes are 
always the dominant phase. New hairy black hole solutions 
are nucleated at the border of the RN AdS stability curve, and the system 
undergoes a second order phase transition to the hairy black hole phase
on this line. Hairy black hole solutions exist down to masses below the 
RN AdS extremality bound, and represent the dominant phase whenever they 
exist. The lowest mass hairy black hole phase is extremal (zero temperature) 
and appears to have nonzero entropy. This solution is likely to be singular. 

As for $e^2<3$, solitonic solutions exist up to a certain maximum charge 
but play no role in the thermodynamics of the system. 
In particular, solitons never represent the ground state of the system at 
any given charge, as their mass is always greater than the mass of 
the extremal hairy black hole at the same charge.

\begin{figure}
\begin{center}
\includegraphics[scale=0.5]{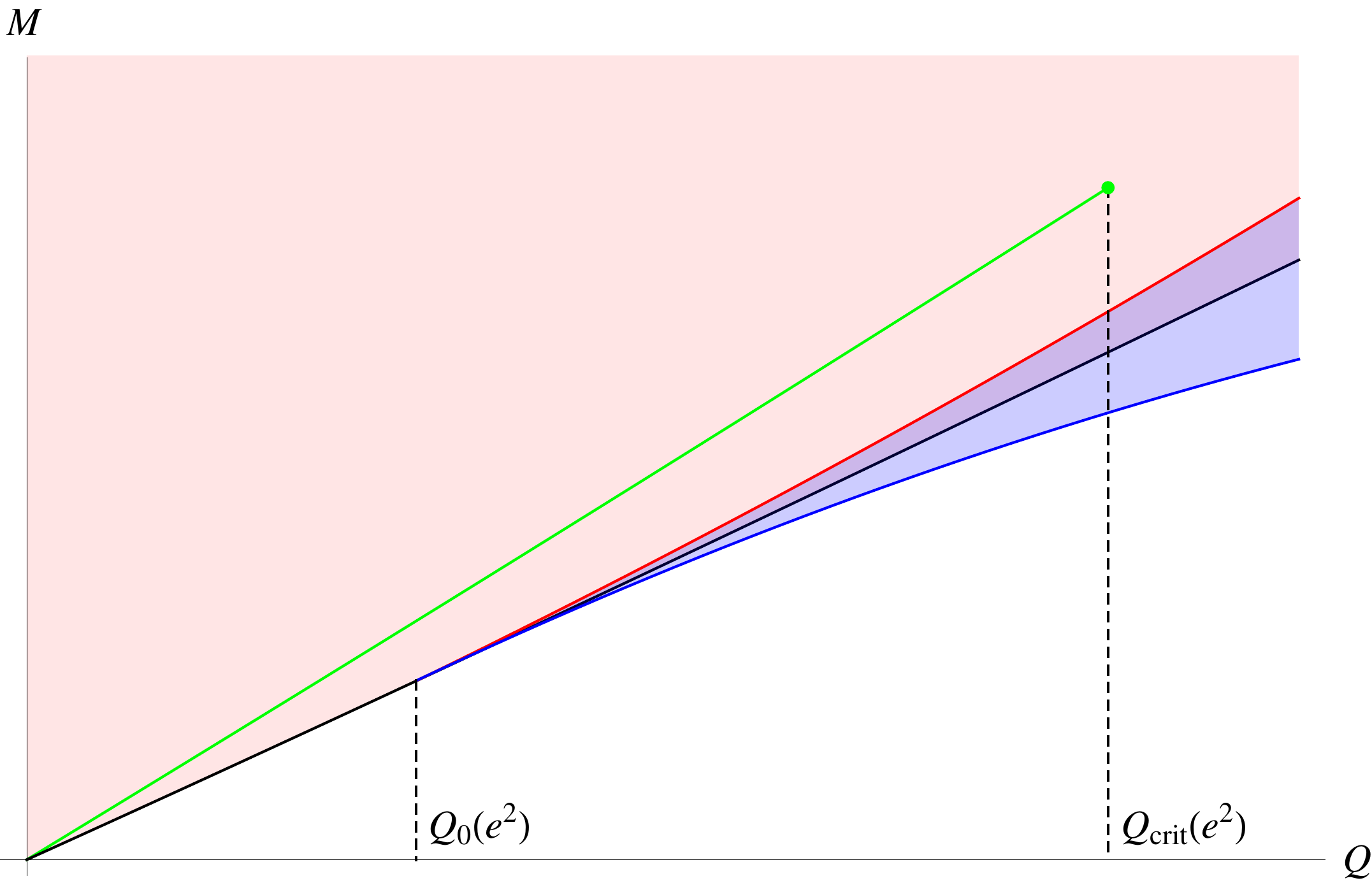}
\end{center}
\caption{Schematic phase diagram for $3 \leq e^2 \leq \frac{32}{3}$. RN AdS black holes exist (red shaded) for all values of charge and mass above the extremality curve (black). Extremal black holes are unstable for $Q\geq Q_0(e^2)$ and hairy black hole solutions exist for these values of charge. These solutions exist (blue shaded) between the curve of instability of RN AdS black holes (red) and a zero temperature hairy black hole solution (blue). Hairy black holes are the dominant phase whenever they exist. The soliton (green) exists upto a maximum charge $Q = Q_{crit}$ (where a cusp structure, to be discussed only later, appears) and is never the dominant phase. (The soliton curve can be below the extremal RN AdS line for $Q>Q_0(e^2)$ but keeps above the extremal hairy black hole).}
\label{esqgreat3}
\end{figure}

Fig. \ref{esqgreat3} gives a qualitative picture of the phase diagram at these values of $e^2$.  For a numerically constructed phase diagram, see Fig. \ref{fig:e3p2_dM_Q}.

\subsection{$e^2 \geq \frac{32}{3} $}

In this parameter regime, RN AdS black holes are unstable in a band about 
extremality, at every value of the charge.  When they are stable, RN AdS black 
holes are always the dominant phase. As in the previous subsection, 
hairy black hole solutions are nucleated at the border of the RN AdS stability 
curve, and the system 
undergoes a second order phase transition to the hairy black hole phase
at lower values of the mass. Hairy black hole solutions extend to masses 
below the RN AdS extremality bound, and represent the dominant phase whenever 
they exist. 

In this range of parameters, the lowest mass hairy black hole phase is 
an infinite temperature soliton for $Q \leq Q_{c_2}(e^2)$ 
(see Section \ref{sec:largee} for details of the function $Q_{c_2}(e^2)$). 
For $Q > Q_{c_2}(e^2)$, on the other hand, the lowest mass hairy black 
hole is extremal (i.e. has zero temperature), has finite entropy, and 
is likely singular. 

The soliton exists at all values of the charge. As explained above, 
for $Q \leq Q_{c_2}$ the soliton represents the lowest mass hairy black hole 
or the ground state of the system. For $Q> Q_{c_2}$ the soliton plays no 
thermodynamical role; in particular its mass is always larger than that 
of the lowest mass hairy black hole.

\begin{figure}
\begin{center}
\includegraphics[scale=0.5]{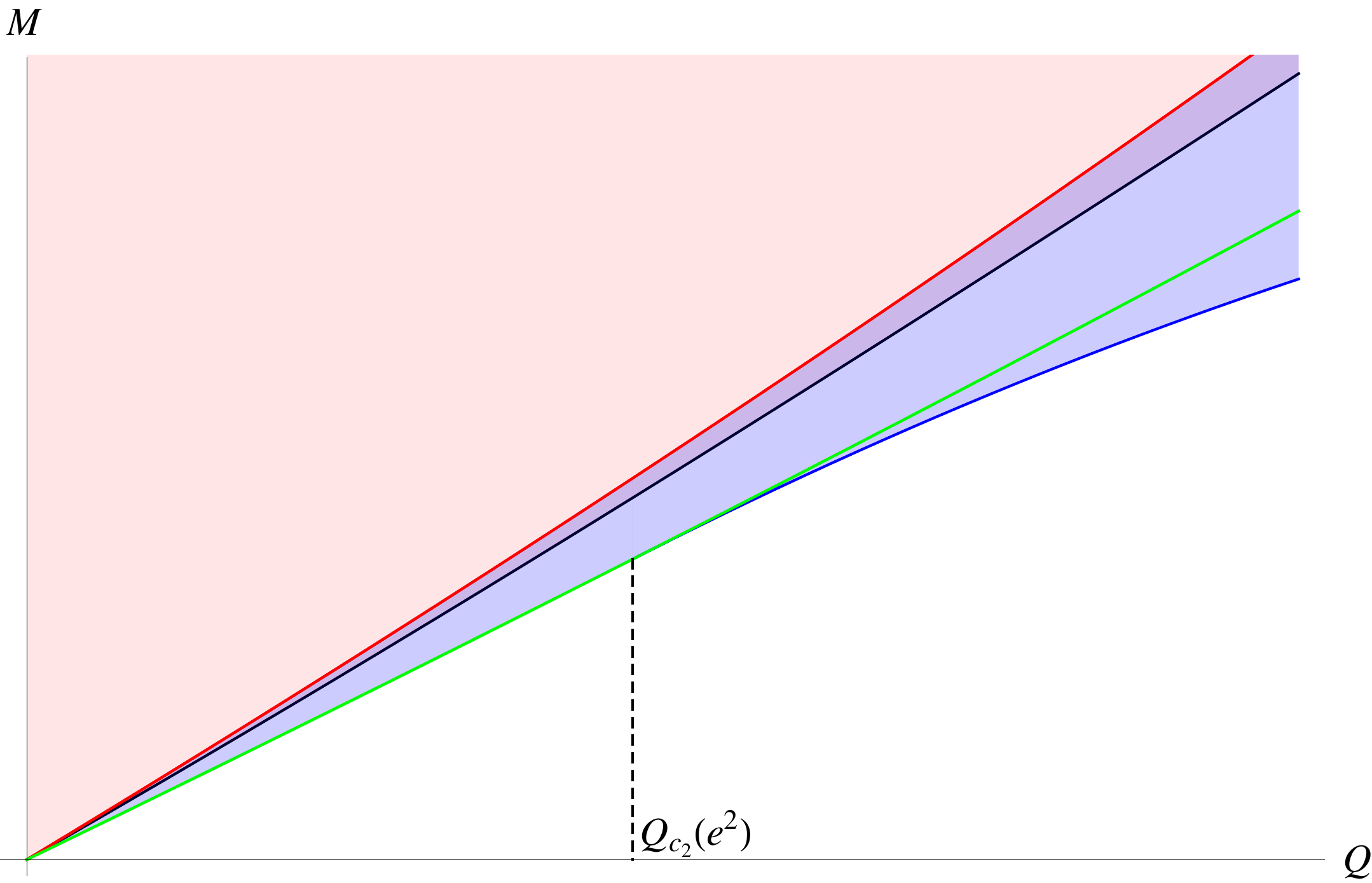}
\end{center}
\caption{Schematic phase diagram for $e^2 \geq \frac{32}{3}$. As for $e^2 \leq \frac{32}{3}$ RN AdS black holes exist (red shaded) for all charges, and all masses above extremality (black). Hairy black holes exist (blue shaded) for all values of the charge, and the masses are below the curve of instability of RN AdS black holes (red). For $Q < Q_{c_2}(e^2)$, the lowest mass hairy black hole solution is a zero entropy soliton at infinite temperature (green). For $Q > Q_{c_2}(e^2)$, the lowest mass hairy black hole is extremal with finite entropy (blue). }
\label{esqgreat32by3}
\end{figure}

Fig. \ref{esqgreat32by3} gives a qualitative picture of the phase diagram for $e^2 \geq \frac{32}{3}$. For a numerically constructed phase diagram, see Fig. \ref{fig:e4_dM_Q}.

\section{Solitons}
\label{sec:solitons}

In this section we will numerically construct solitons at the full non-linear 
level. As we have mentioned in the introduction, our system admits an 
infinite number of  solitonic solution branches. The solitons we construct in 
this paper are those that are obtained by continuously increasing the scalar 
field at the origin, starting from $AdS_5$ perturbed by the ground 
state linearized scalar fluctuation. We do not study any other branches 
of solitonic solutions in this paper.

As we have explained in the Introduction, the space of solitonic 
solutions differs qualitatively depending on whether $e^2>e^2_{solcrit}$ 
or $e^2< e^2_{solcrit}$. As we have also remarked above, 
$e^2_{solcrit} =\frac{32}{3}$ to within numerical accuracy. We first 
describe the methodology we use to construct our solitonic solutions, and 
then describe our results.

\subsection{Set up}
\label{subsec:setupsoliton}

Following \cite{Basu:2010uz}, we adopt the gauge:
\begin{equation}
ds^2=-f(r)\,dt^2+g(r)\,dr^2+r^2\,d\Omega_{(3)}^2\,,\qquad
A_\mu\,dx^\mu=A(r)\,dt\,,\qquad
\phi=\phi(r)\,,\label{eqn:ansatz}
\end{equation}
where $\phi(r)$ can be taken to be real wlog  and $d\Omega_{(3)}^2$ denotes  the standard metric on the round unit three-sphere. The equations of motion are \cite{Basu:2010uz}:
\begin{subequations}
\label{eqn:eoms}
\begin{align}
&\phi''+\frac{1}{r}\bigg[1+2\,g\bigg(1+2\,r^2\bigg)-\frac{r^2 \,A'^2}{3 f}\bigg]\phi'+\frac{g\,e^2\, A^2}{f}\,\phi=0\,,\\
&A''+\frac{1}{r}\bigg[3-\frac{2\,g\, e^2\, r^2 \,A^2\, \phi^2}{3\,f }-\frac{2}{3}\, r^2 \,\phi '^2\bigg]A'-2\,g\, e^2\, \phi^2\,A=0\,,\\
&f'+\frac{2}{r}\bigg[1-g\,(1+2\, r^2\big)-\frac{1}{3}\, r^2\, \phi '^2\bigg]f-\frac{r}{3}  \bigg(2\,g\,e^2\, A^2\, \phi^2-A'^2\bigg)=0\,,\\
&g'-\frac{1}{r}\bigg(2+\frac{r^2\, A'^2}{3\, f}+\frac{2}{3}\,r^2 \phi'^2\bigg)\,g+\frac{2\,g^2}{r}\bigg(1+2\, r^2+\frac{ e^2\, r^2\, A^2\, \phi^2}{3\,f}\bigg)=0\,.
\end{align}
\end{subequations}
where the prime $'$ denotes the derivative with respect to $r$. Notice that from the third equation in \eqref{eqn:eoms} we can eliminate $g(r)$ in terms of the other variables. Plugging this expression for $g$ into the last equation in \eqref{eqn:eoms} yields a second order equation for $f$. We will take this equation together with the first and second equations in \eqref{eqn:eoms} as our set of fundamental equations to be solved. 

The boundary conditions that we shall impose are as follows. Since we want to consider asymptotically $AdS_5$ solutions, the large $r$ behavior of the various functions is given by
\begin{equation}
\label{eqn:asymp}
\begin{aligned}
&A(r)=\mu-\frac{\tilde{q}}{r^2}+\O(1/r^4)\,,\\
&\phi(r)=\frac{\epsilon}{r^4}+\O(1/r^6)\,,\\
&f(r)=1+r^2-\frac{m}{r^2}+\O(1/r^4)\,,
\end{aligned}
\end{equation}
where $\mu$ is the chemical potential and $\epsilon$ is the expectation value of the operator dual to the scalar field $\phi$; the asymptotic charges of the solution are the mass $M=\frac{3\pi}{8} m$ and the charge $Q=\frac{\pi}{2} \tilde{q}$. At the origin $r=0$ we shall impose smoothness, which implies a Neumann condition on all functions. 

For a given value of the scalar charge $e$, solitons form a one parameter family of solutions. This parameter can be taken to be the charge, for instance, but we find it useful to use $f(0)$ instead. The reason is that, as discussed in  \cite{Basu:2010uz}, a  necessary condition in order to have smooth horizonless solution is that $f(0)>0$;  in our gauge one has that  $f(0)=1$ corresponds to pure AdS and $f(0)$ monotonically decreases as one moves along the soliton branch. Therefore, $f(0)$ uniquely labels each solution for any value of $e$; on the other hand, we find that for $e<e_{solcrit}$ the charge does \textit{not} uniquely specify the soliton solution.

The aforementioned  boundary conditions can be easily implemented introducing a new compact coordinate $y=\frac{r^2}{1+r^2}$ and redefining the functions  as follows:
\begin{equation}
f(r)=\left(1+r^2\right)p_f(y)\,,\qquad \phi(r)=\epsilon\left(1+r^4\right)^{-1}p_\phi(y)\,,\qquad A(r)=p_A(y)\,.
\end{equation}
 Then, the new functions $p_i$'s should satisfy $p_f(1)=1$, $p_A(1)=\mu$ and $p_\phi'(1)=0$ at infinity $(y=1)$, and at the origin $(y=0)$ we find:
\begin{equation}
p_f'(0)=\frac{1}{2}\,e^2\epsilon^2\,p_\phi(0)^2p_A(0)^2\,,\qquad p_\phi'(0)=-\frac{e^2\,p_\phi(0)\,p_A(0)^2}{8\,p_f(0)}\,,\qquad p_A'(0)=\frac{1}{4}\,e^2\epsilon^2\,p_\phi(0)^2\,p_A(0)\,,
\end{equation}
which follow from solving the equations of motion near the origin and imposing regularity.

We have solved the equations \eqref{eqn:eoms} numerically using Newton's method and a Chebyshev pseudospectral collocation approximation. Alternatively we have also used  shooting and the results of both methods agree. The data presented below was obtained using the pseudospectral method.

\subsection{Results: $e^2 \leq e^2_{solcrit} \approx \frac{32}{3}$}

\begin{figure}[t!]
\begin{center}
\includegraphics[scale=0.8]{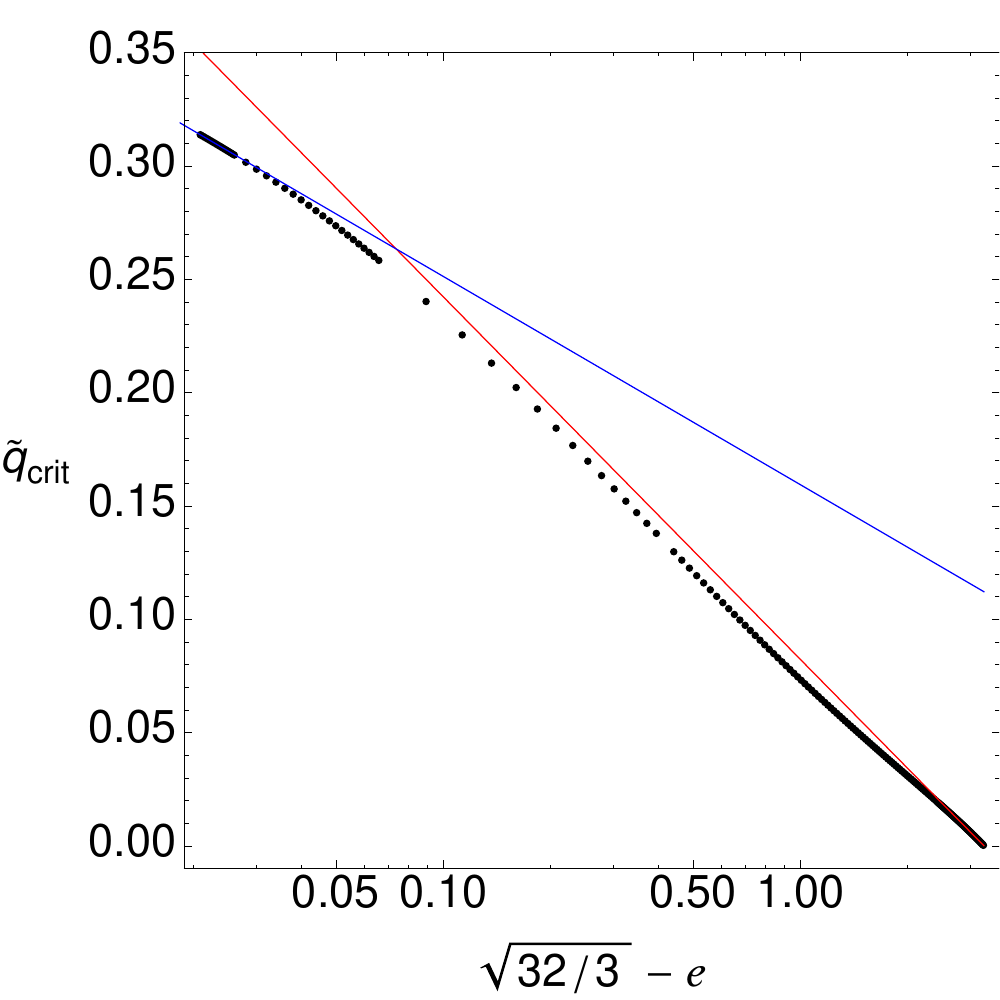}
\hspace{0.1cm}
\includegraphics[scale=0.8]{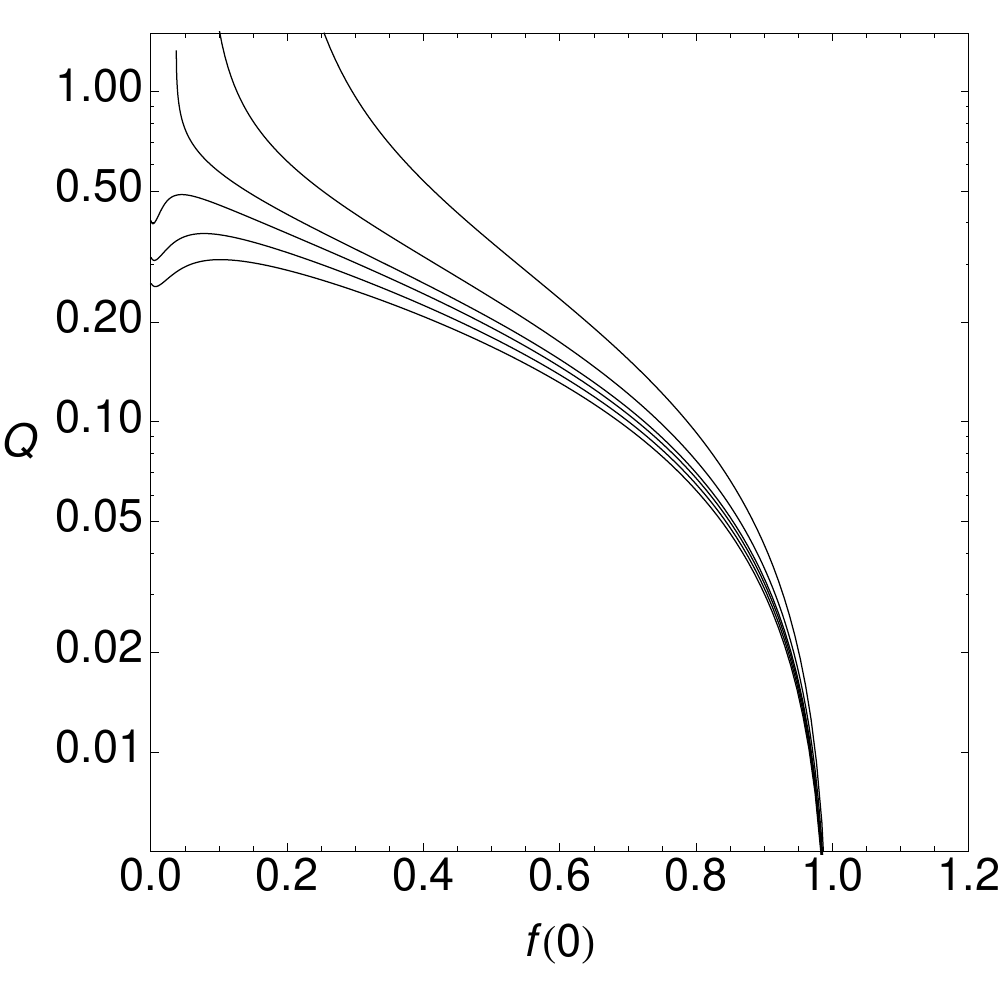}
\end{center}
\caption{\small{\textit{Left:}  $\tilde q_{crit}$ as a function of 
$\sqrt{\frac{32}{3}}-e$ for $e^2<32/3$: note that the scale on the 
$x$ axis is logarithmic. The black dots correspond to 
our data. The red line is the fit to a linear behavior as in 
\eqref{eqn:qmaxvse} in the region of small $e$. The blue line corresponds 
to a logarithmic fit of the form $\tilde q_{crit}=a\,\ln\left(\sqrt{\frac{32}{2}}-e\right)+b$ of the  data near $e \sim \sqrt{\frac{32}{3}}$. 
This diagram suggest that ${\tilde q}_{crit}$ blows up logarithmically as 
$e$ approaches $e_{solcrit}$.  \textit{Right:} $Q$ as a function of $f(0)$ for $e=3.,3.1,3.2,3.29,3.5, 4.$ (from bottom to top in the plot). For $e\leq e_{solcrit}$, $Q$ tends to a finite value as $f(0)\to 0$, and for $e>e_{solcrit}$,  $Q$ diverges as $f(0)\to 0$.}}
\label{fig:qmaxvse}
\end{figure}

Our full nonlinear numerical construction of solitons agrees well with 
the perturbative construction presented in \cite{Basu:2010uz} at small 
values of the charge (see for instance the right panel in  Fig. \ref{fig:fitsolitonq1} ). 
The most striking qualitative feature of the solitonic branch of solutions, 
at these values of $e^2$, is that it terminates in a naked singularity 
at a critical value of the charge, $\tilde{q}_{crit}$. This feature is readily 
explained at small $e$. Indeed, at very small $e$ the soliton is simply an
almost uncharged boson star. It seems intuitively reasonable that 
such a boson star hits the Chandrashekhar limit at a critical 
mass of order unity. At the first nontrivial order in $e$, the charge of
the soliton may be measured by the gauge field $A$ sourced by this 
almost uncharged boson star, and so is to order $e$.  In other words 
to leading order in $e$, 
\begin{equation}
\tilde{q}_{crit} = A\, e\,, \label{eqn:qmaxvse}
\end{equation}
Our full nonlinear numerical solutions, obtained using the methods 
of  \S\ref{subsec:setupsoliton} verify this expectation, see  
Fig. \ref{fig:qmaxvse} (left) and demonstrate that $A \approx 0.0218$.
 At larger values of $e$,  $\tilde{q}_{crit}$ continues to grow. 
For $e^2>e^2_{solcrit} \approx \frac{32}{3}$ solitons exist at every value
of the charge, i.e. ${\tilde q}_{crit}=\infty$. As we approach $e=e_{solcrit}$ 
from below, the graph of 
$\tilde{q}_{crit}$ versus $-\ln(\sqrt{\frac{32}{3}}-e)$ appears to asymptote 
to a straight line (Fig.  \ref{fig:qmaxvse} (left)) suggesting that 
the critical charge diverges like 
$$\tilde{q}_{crit} \approx - 0.0399 \ln \left( \sqrt{\frac{32}{3}} -e \right)+0.159. $$

We pause here to emphasize that we are inferring a divergence in 
${\tilde q}_{crit}$ at $e=\sqrt{\frac{32}{3}}=3.266$, even though the largest 
value of ${\tilde q}_{crit}$ we have found in a simulation is $0.314$ at $e=3.245$. 
It turns out that simulations at values of $e$ nearer to $\sqrt{\frac{32}{3}}$
are difficult; the slow growth of the logarithm prevents us from obtaining 
more direct evidence for the divergence in ${\tilde q}_{crit}$ as $e$ approaches 
$e_{solcrit}$ from below.\footnote{In particular our data is consistent 
with the possibility that the linear behaviour in Fig. \ref{fig:qmaxvse} 
levels out at very small $e-e_{solcrit}$ leading to a moderate finite value of 
${\tilde q}_{max}$ at $e=e_{solcrit}$. A definitive statement here needs 
further - preferably analytic - work. }

In order to further study the approach of $e$ to $e_{solcrit}$ from below, 
in Fig. \ref{fig:qmaxvse} (right) we have plotted $Q$ as a function of $f(0)$ 
for $e=3.,3.1,3.2,3.29, 3.5, 4.$ (from bottom to top curves in this plot). 
Recall that $f(r)$ is the coefficient of $-dt^2$ in the metric of the 
solitonic solution; $f(0)$ going to zero indicates a Chandrashekhar 
singularity in the solution. For $e<e_{solcrit}$, $f(0)$ vanishes at finite 
charge. As $e$ approaches $e_{solcrit}$ the approach of the curves in 
Fig. \ref{fig:qmaxvse} (right) presumably creep logarithmically up the 
$y$ axis. For  $e>e_{solcrit}$ the curves presumably never intersect the 
$y$ axis. 

\subsubsection{Approach to ${\tilde q}_{crit}$ at $e<e_{solcrit}$}

In the rest of this subsection we describe 
the approach of the solitonic branch of solutions to $\tilde{q}_{crit}$ at 
a fixed value of $e<e_{solcrit}$ 
in more detail. 
\begin{figure}[t]
\begin{center}
\includegraphics[scale=0.8]{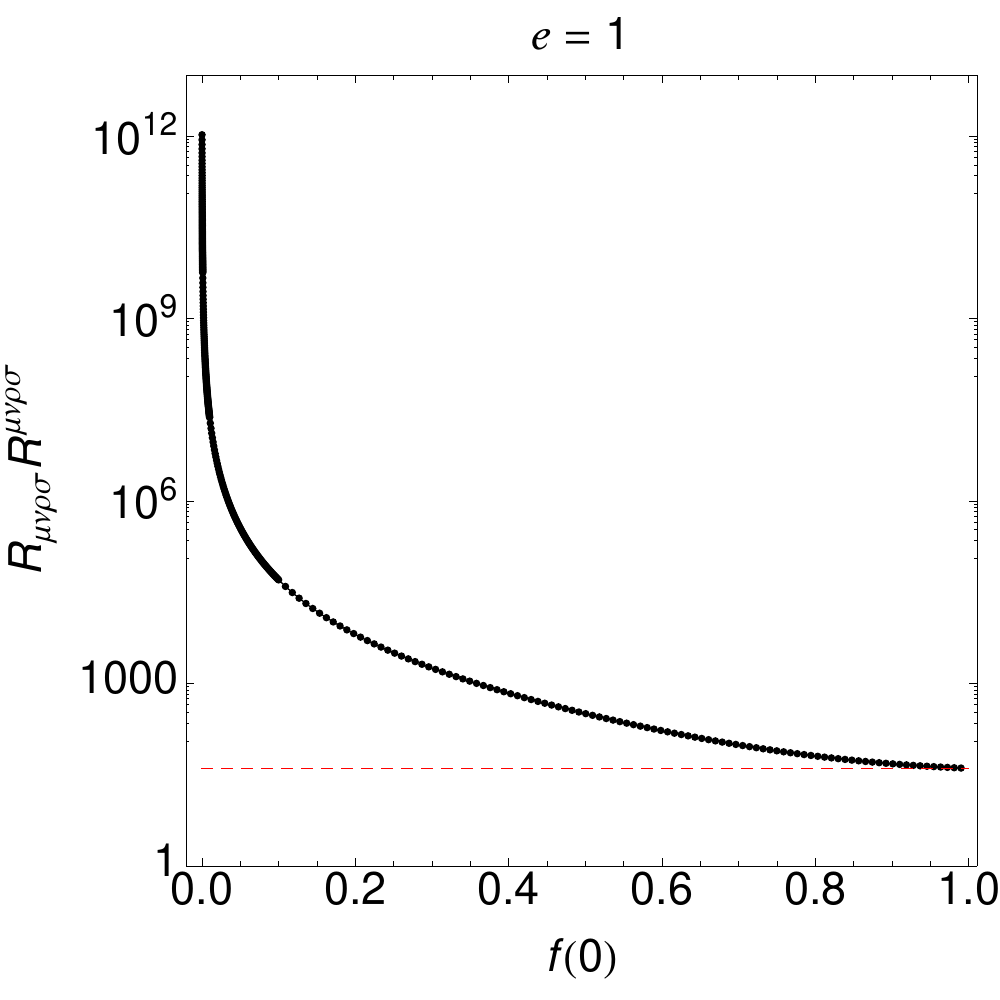}
\hspace{.5cm}
\includegraphics[scale=0.35]{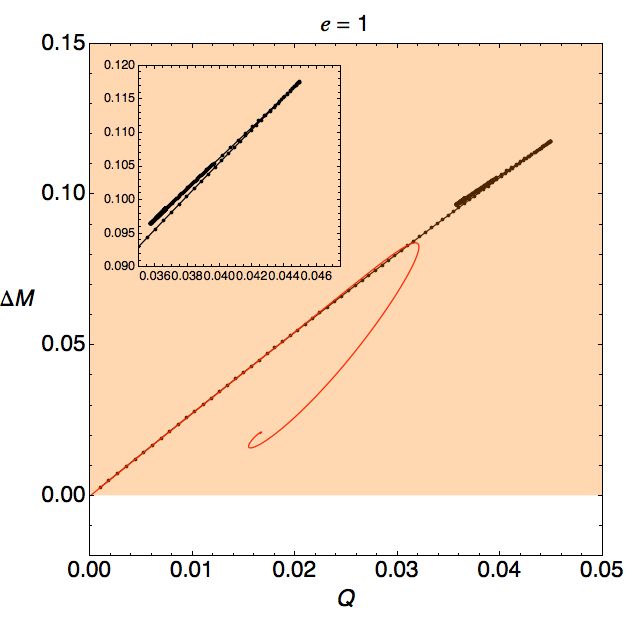}
\end{center}
\caption{\small{\textit{Left:} Kretschmann invariant evaluated at $r=0$ as a function of $f(0)$ for the $e=1$ solitons. As $f(0)\to 0$ the Kretschmann invariant diverges at the origin, which signals the appearance of a curvature singularity there. The red dashed line corresponds to the value of the Kretschmann invariant for pure $AdS_5$. \textit{Right:} Phase diagram in the microcanonical ensemble for $e=1$. On the $y$-axis of this plot we depict $\Delta M=M-M_\textrm{ext}$, where $M_\textrm{ext}$ is the mass of the extremal RN AdS black hole with the same charge $Q$. RN AdS black holes occupy the shaded region and the soliton family of solutions is given by the black curve. This curve terminates at a naked singularity at some finite $Q$. In red we show the perturbative results of \cite{Basu:2010uz}; the agreement between the perturbative calculation and our numerical results is remarkable at small values of $Q$ but they disagree at sufficiently large $Q$.}}
\label{fig:solitonphasesq1}
\end{figure} 
  
At $\tilde{q}=\tilde{q}_{crit}$ the solution becomes singular. This singularity 
is signalled by the fact that the function $f$ in \eqref{eqn:ansatz} develops a zero at the origin $r=0$ at this particular (and finite) value of 
$\tilde q_{crit}$. In turn, this implies that the curvature invariants diverge 
at this point. For instance, in Fig. \ref{fig:solitonphasesq1} (left) we 
depict the Kretschmann invariant evaluated at $r=0$ as a function of $f(0)$ 
for the $e=1$ case. Clearly, as $f(0)\to 0$ this curvature invariant diverges 
and it can be shown that the divergence is like $\sim f(0)^{-2}$. Ref. 
\cite{Basu:2010uz} had conjectured that the solitonic branch of solutions 
would cease to exist at a critical value of the charge; our results fully 
validate this conjecture in this range of parameters.  
 \begin{figure}[t]
\begin{center}
\includegraphics[scale=0.9]{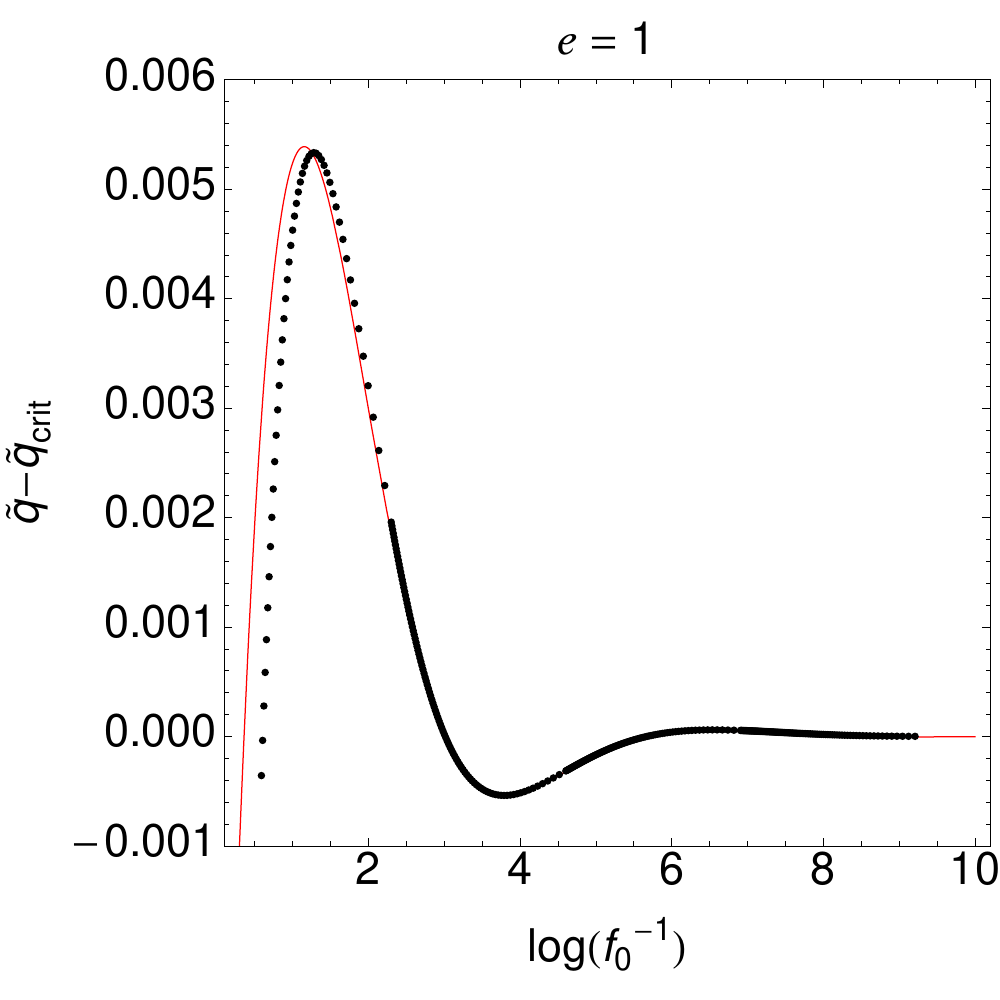}
\end{center}
\caption{\small{$\tilde q-\tilde q_{crit}$ as a function of $f_0\equiv f(0)$ for the $e=1$ case. The black dots correspond to our data and the red curve is the fit given by the functional form \eqref{eqn:solitonfit}. As we approach the singular solution, the regular solitons exhibit a (possibly) infinite set of damped self-similar oscillations  similar to the behavior observed in  \cite{Bhattacharyya:2010yg}.}}
\label{fig:fitsolitonq1}
\end{figure} 
 
As we approach the singular solution along the soliton family of solutions, the soliton exhibits a set of damped self-similar oscillations that (presumably) continue \textit{ad infinitum}. More precisely, we find that near the singular solution the dependence of charge $q$ of the solitons on $f_0\equiv f(0)$ obeys a relation of the form,
\begin{equation}
\tilde q-\tilde q_{crit}\approx \alpha\,e^{-\beta\log f_0^{-1}}\cos\left(\gamma\,\log f_0^{-1}+\delta\right)\,,
\label{eqn:solitonfit}
\end{equation}
 for some constants $(\alpha,\beta,\gamma,\delta)$ that can be determined numerically from the data, and $\tilde q_{crit}$ is the charge of the critical solution. We have analyzed different values of $e$ (always with $e^2<\frac{32}{3}$) and we have found that $\alpha$ and $\delta$ do depend on $e$ but $\beta$ and $\gamma$ seem to be roughly independent of $e$. More specifically, we find $\beta\sim 0.90\pm 0.05 $ and $\gamma\sim 1.20\pm0.05$ for the different values of $e$ that we have checked.  In Fig. \ref{fig:fitsolitonq1} we illustrate this behavior for the $e=1$ case.    Therefore, for $e^2<\frac{32}{3}$  the phase diagram for the solitons (in the microcanonical ensemble) is the same as in the truncation of $\mathcal N=8$ gauge supergravity of \cite{Bhattacharyya:2010yg}. We should emphasize that our analysis is fully numerical and it would be interesting to get some analytical results  along the lines of \cite{Bhattacharyya:2010yg}; in particular, it would be nice to calculate some of the constants in \eqref{eqn:solitonfit} analytically, get a better understanding of the singular solution and spell out the dependence on $e$ (if any).
 
 \begin{figure}[t]
 \begin{center}
 \includegraphics[scale=0.6]{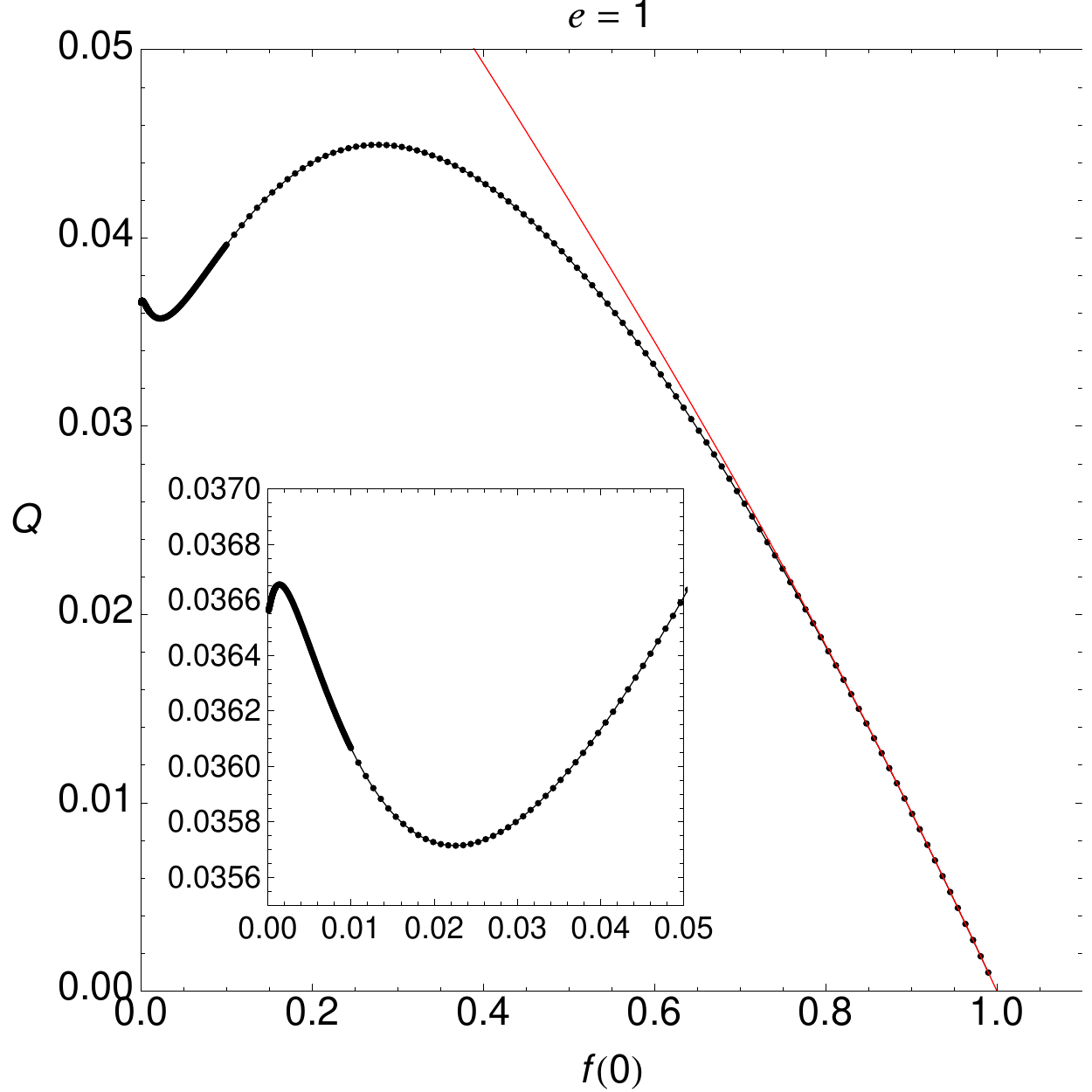}
 \hspace{.5cm}
 \includegraphics[scale=0.6]{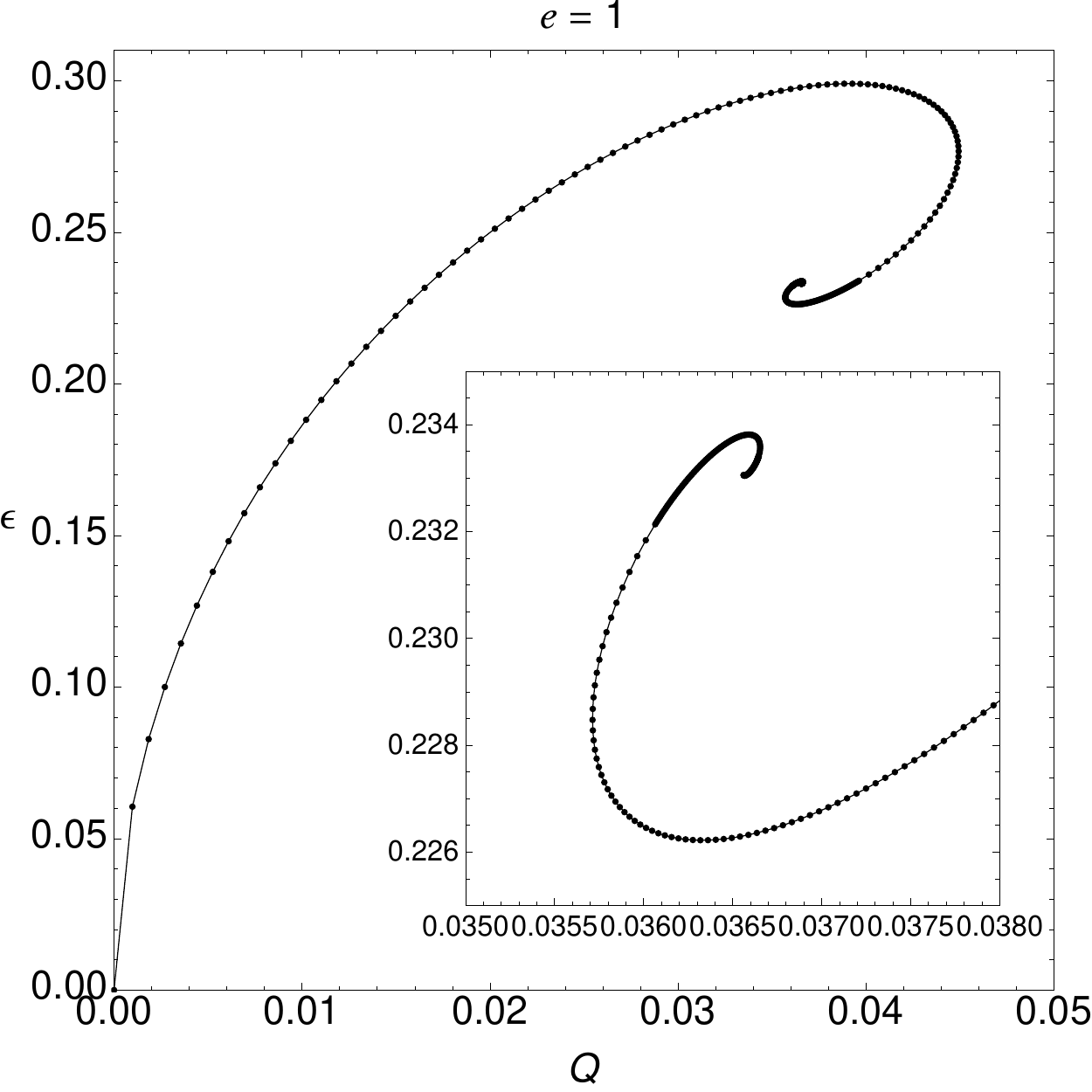}
 \end{center}
 \caption{\small{\textit{Left:} $Q$ as a function of $f(0)$. The charge exhibits a (possibly infinite) series of damped oscillations around the critical value. \textit{Right} $\epsilon$ vs. $Q$. The expectation value of the scalar field as a function of the charge $Q$ forms a spiral near the critical point.}}
 \label{fig:solitonq1}
 \end{figure}
  
  In Fig. \ref{fig:solitonq1} (left) we depict the ADM charge $Q$ vs. $f(0)$ for the $e=1$ case. We can observe that for small charges the perturbative results of \cite{Basu:2010uz} (in red) agree very well with our numerics but they disagree at sufficiently large charges. In addition, this plot clearly shows that $Q$ does \textit{not} uniquely parametrize the family of solutions. Instead, as discussed above,  this quantity (and the other physical quantities too) exhibits a (possibly infinite) series of self-similar damped oscillations as we approach the singular solution, which corresponds to $f(0)=0$. In the right panel of Fig. \ref{fig:solitonq1} we depict $\epsilon$ vs. $Q$, which shows a spiralling behavior towards the singular solution. 
  
We close the discussion by noting that we have checked that the behaviour 
described in this subsection applies to values of $e^2$ that differ from 
$\frac{32}{3}$ by  less  than $1\%$. We have also checked, on the other 
hand, that solitons exist for arbitrarily large charge for $e^2$ that 
exceed $\frac{32}{3}$ by  $1\%$. This is the basis of our 
claim that $e^2_{solcrit} \approx \frac{32}{3}$. 

\subsubsection{Approach to the singularity}

\begin{figure}[t]
\begin{center}
\includegraphics[scale=0.8]{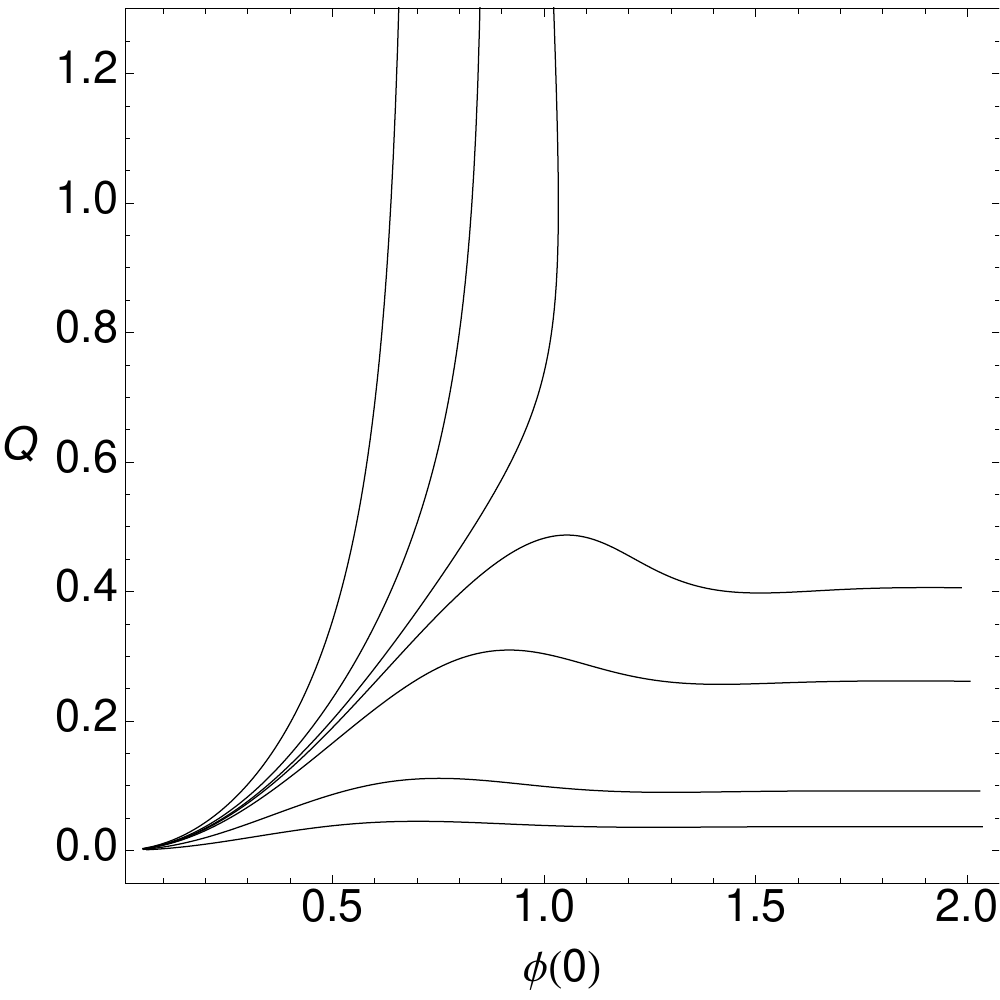}
\end{center}
\caption{\small{Charge of the soliton $Q$ vs. $\phi(0)$ for $e=1.,2.,3.,3.2,3.29,3.5,4$ (from bottom to top curves). The behavior of $Q$ as a function of $\phi(0)$ changes discontinuously as we vary $e^2$ across $e^2_{solcrit}\approx 32/3$. The amplitude of the oscillations increases as we approach  $e^2_{solcrit}$ from below and they completely disappear for  $e^2>e^2_{solcrit}$ .}}
\label{fig:Qvsphizvarye}
\end{figure}

In this brief subsection we study, from yet another angle, 
how the solitonic branch 
of solutions behaves as $e$ is increased above $e=e_{solcrit}$. 
In Fig. \ref{fig:Qvsphizvarye}  we display 
a set of graphs for the solitonic charge as a function of the value of 
the scalar field at zero, $\phi(0)$, for different values of $e^2$ that go through
the phase transition. Note that for $e^2< e_{solcrit}^2$ the amplitude of the oscillations becomes larger as we approach $e^2_{solcrit}$, and the oscillations  completely disappear
for $e^2 > e^2_{solcrit}$. In other words, if we plotted $\phi(0)$ vs. $Q$ we would see that the spiral unwraps for $e^2>e^2_{solcrit}$. For $e^2>e^2_{solcrit}$ we note that  $\phi(0)$ approaches a constant for large $Q$ and this constant is not too far from $\frac{2}{e}$. This is consistent with a singular limit as in  \cite{Horowitz:2009ij} and the results in Appendix \ref{sec:AppPlanarSol}. Similarly, the graph of $Q$ vs. $f(0)$ (see the right panel in Fig. \ref{fig:qmaxvse})   is monotonic for
$e^2 > e^2_{solcrit}$.


\subsection{Results: $e^2 \geq e^2_{solcrit} \approx \frac{32}{3}$ }
\label{subsec:respsoliton}

 \begin{figure}[t]
 \begin{center}
 \includegraphics[scale=0.7]{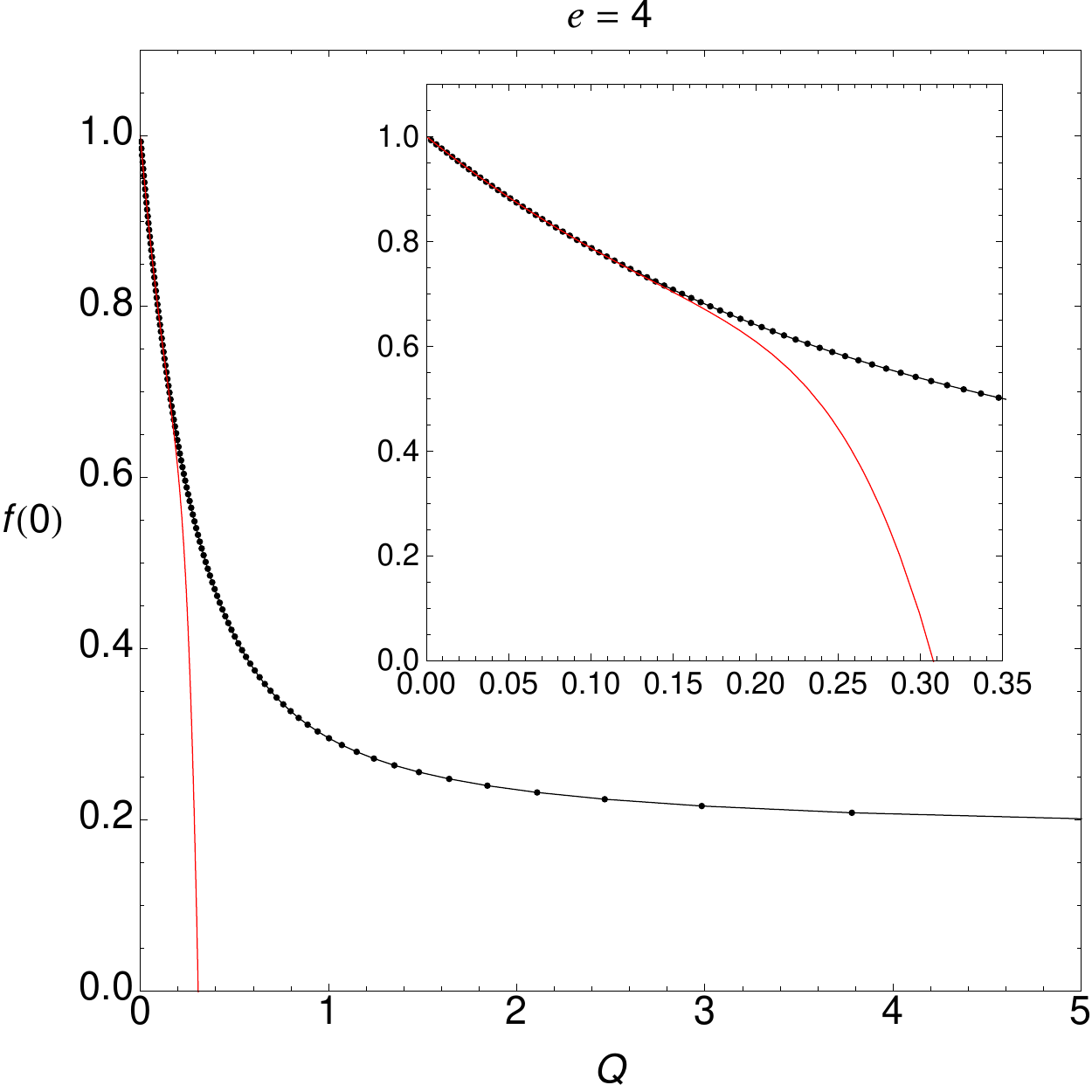}
 \hspace{.5cm}
 \includegraphics[scale=0.70]{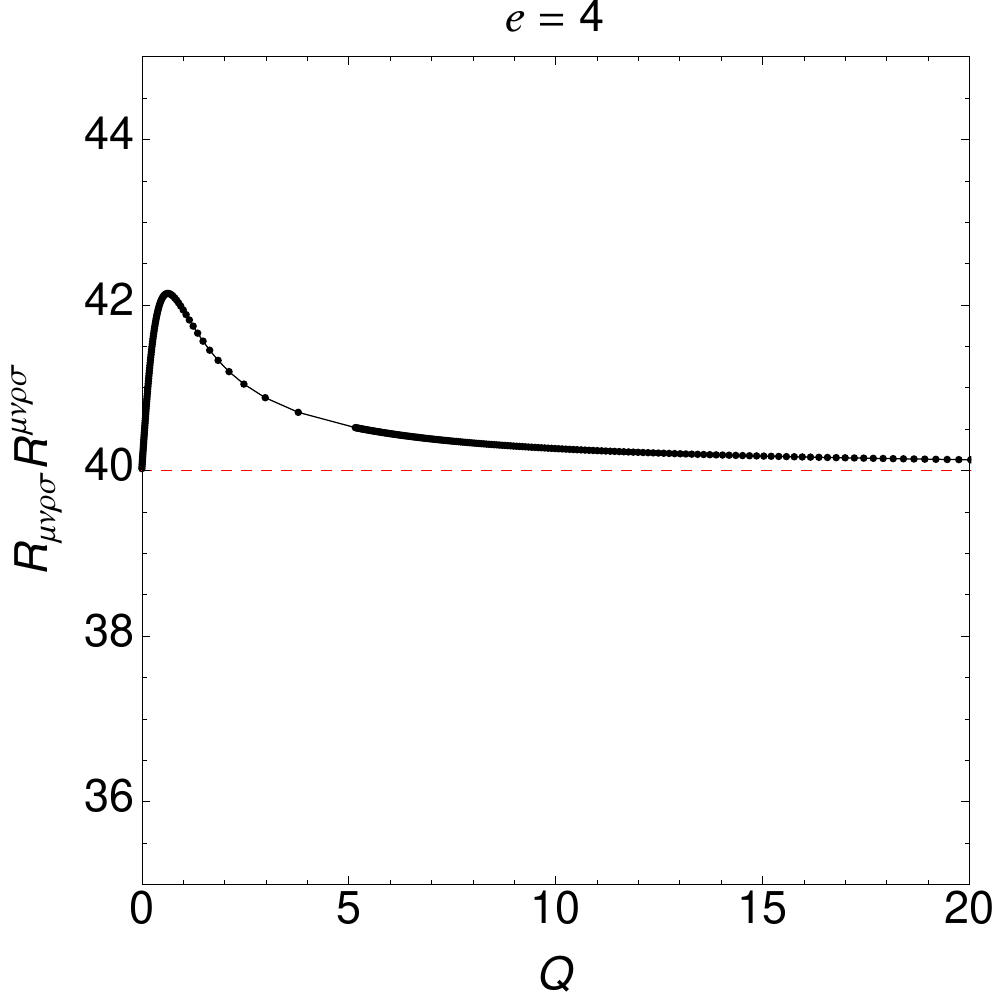}
 \end{center}
 \caption{\textit{Left:} $f(0)$ as a function of $Q$ for the $e=4$ case. In black we depict our numerical data and the red curve corresponds to the perturbative results of \cite{Basu:2010uz}. The agreement between the two is excellent for small $Q$. An unjustified extrapolation of the perturbative results suggest, 
however,  that $f(0)$ should develops a zero at a finite $Q$. 
 On the other hand our nonlinear results indicate that $f(0)$ remains finite 
for any finite value of $Q$. \textit{Right:} Kretschmann invariant at $r=0$ as  a function of $Q$. This curvature invariant remains finite for any finite value of $Q$, and it approaches $40$, (which is the value for pure $AdS_5$) for large $Q$. }
 \label{fig:solitonq4}
 \end{figure}

For $e^2>\frac{32}{3}$   the behavior of the soliton family in the phase 
diagram changes completely. Whilst an unjustified extrapolation of 
the perturbative results of \cite{Basu:2010uz} to large charge 
 suggest that solitons should have a Chandrasekhar bound for all values of 
$e$, our numerical results demonstrate that for 
$e^2>e_{solcrit} \approx  \frac{32}{3}$ solitons can exist for arbitrarily 
large values of the charge. 
In Fig. \ref{fig:solitonq4}   (left) we plot $f(0)$ as a function of the charge $Q$ for the $e=4$ case. The black curve corresponds to our data and the red curve corresponds to the perturbative results of \cite{Basu:2010uz}. Note that for small values of the charge the agreement between the perturbative results and our non-linear calculation is remarkable. On the other hand, an 
unjustified extrapolation of the perturbative results to large charge 
 suggest  \cite{Basu:2010uz} that $f(0)$ should become zero at some finite 
value of $Q$, but our non-linear results show that this is not the case and in fact the soliton family of solutions seems to exist for arbitrarily large values of $Q$. The explanation for this  is that for any $e^2>\frac{32}{3}$ the electrostatic repulsion is always strong enough to counter-balance the gravitational self-attraction, no matter how massive the soliton is. To check the regularity of the spacetime, we have computed the value of the Kretschmann invariant at the origin as a function of $Q$ (see Fig. \ref{fig:solitonq4} right). Interestingly, for large values of the charge (and hence well in the full non-linear regime!)  the value of the Kretschmann invariant at the origin reduces to that of pure $AdS_5$ (red dashed line in the right panel of Fig. \ref{fig:solitonq4}), which implies that the curvature there is not large.

\begin{figure}[t]
 \begin{center}
 \includegraphics[scale=0.8]{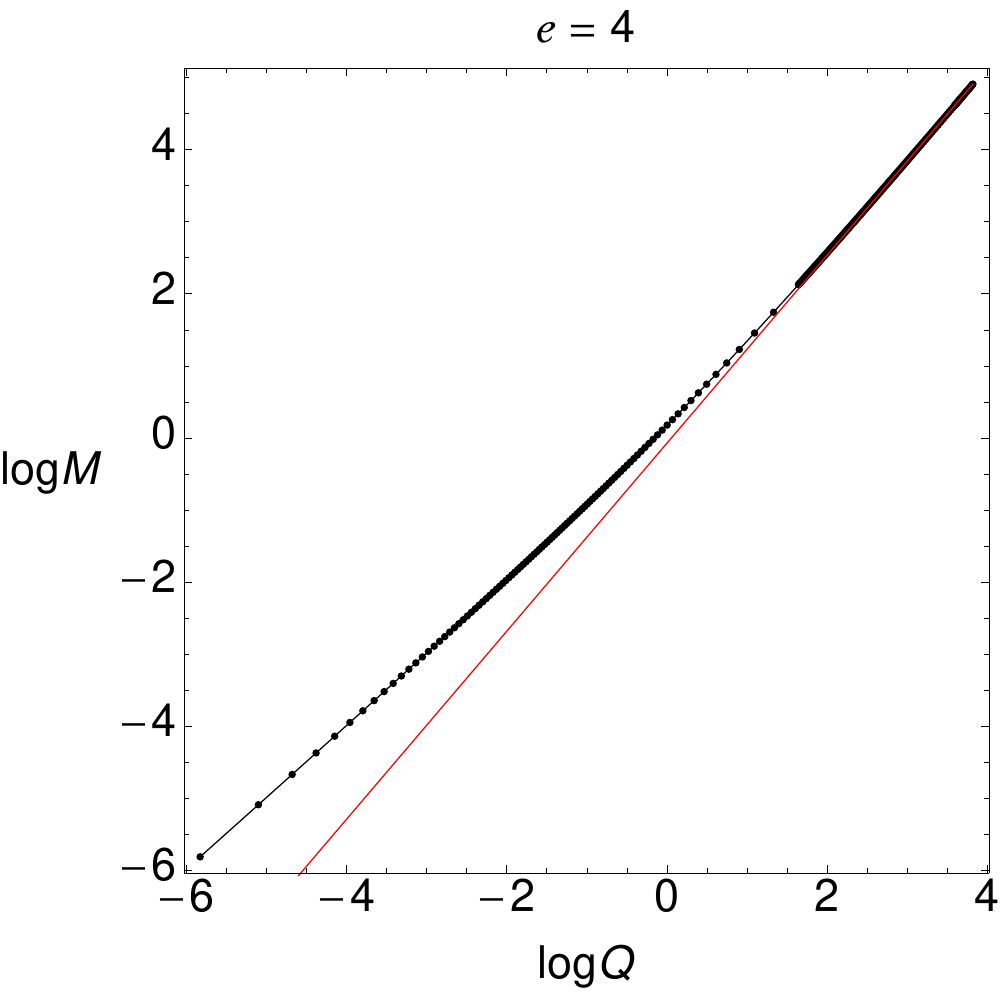}
 \hspace{.5cm}
 \includegraphics[scale=0.8]{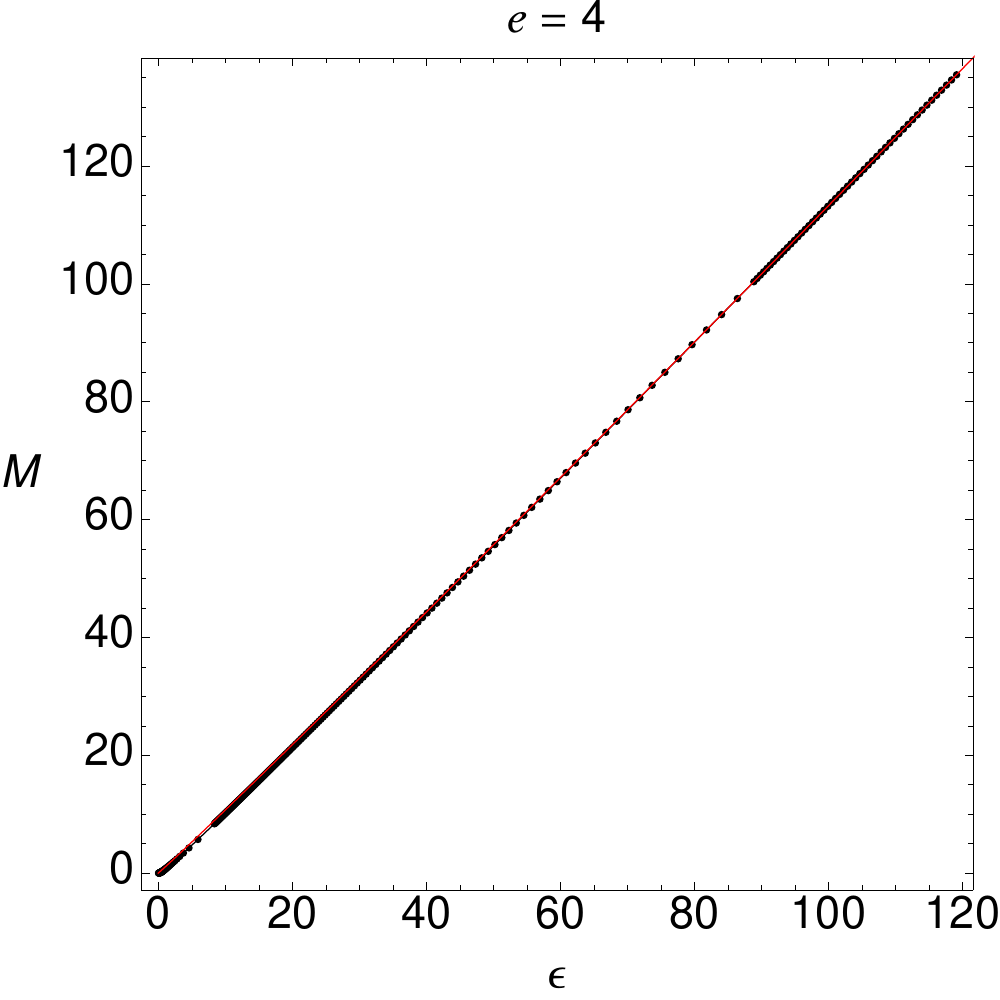}
 \end{center}
 \caption{\small{\textit{Left:} $\log M$ vs $\log Q$ for the $e=4$ case. The black dots correspond to our data and the red curve corresponds to the best fit 
of our data to a function of the of form $M= B Q^\alpha$ in the large charges regime. We find that $\alpha\simeq 1.3$, which is close to  the scaling \eqref{scnn} required by conformal invariance, and $B=0.9$. \textit{Right:}. $M$ vs. $\epsilon$. As this plot shows, for large values of these quantities the relation between them is approximately linear, in accordance with \eqref{scnn}. We find $M=C \epsilon^{\alpha}$ with $C \simeq \alpha \simeq 1.0$.}}
 \label{fig:scalingsolitonq4}
 \end{figure}

As solitonic solutions exist at all values of the charge in this range of 
parameters, it is interesting to study the infinite charge limit of these 
solutions. In this limit the solitonic solutions should tend to solitonic 
black branes. Indeed we have numerically verified the following. 
Let $L={\tilde q}^\frac{1}{3}$. Then the limits 
\begin{equation}\label{sf}
\begin{split}
g_P(\rho)&= \lim_{L \to \infty} L^2\,g(\rho L)\\
f_P(\rho)&=\lim_{L \to \infty} \frac{f(\rho L)}{L^2}\\
A_P(\rho)&=\lim_{L \to \infty} \frac{A(\rho L)}{L}\\
\phi_P(\rho)&=\lim_{L \to \infty} \phi(\rho L)\\
\end{split}
\end{equation}
all appear to exist, implying that the large charge hairy black hole 
may be rewritten in new coordinates $r=L\,\rho$, $t =\tau/L $ in the 
black brane form
\begin{equation} \label{scso}
\begin{split}
ds^2& =-f_P(\rho) \,d\tau^2+ g_P(\rho)\,d\rho^2 + \rho^2\, (dx^i)^2\\
A&= A_P(\rho)\, d\tau 
\end{split}
\end{equation}
In particular this implies that the mass and charge of our solitonic 
solutions scale, at large charge, like 
\begin{equation} \label{scnn}
M \approx B\, Q^{4/3}\,,\qquad M \approx C\, \epsilon\,, 
\end{equation}
by conformal invariance. Indeed, as  Fig. \ref{fig:scalingsolitonq4} shows, our data approximately satisfies the aforementioned scaling behavior and 
yields $B \approx 0.9$ and $C \approx 1.0$ at $e=4$. The coefficients 
$B$ and $C$ are, of course, functions of $e$. 

\begin{figure}[t]
\begin{center}
\includegraphics[scale=0.8]{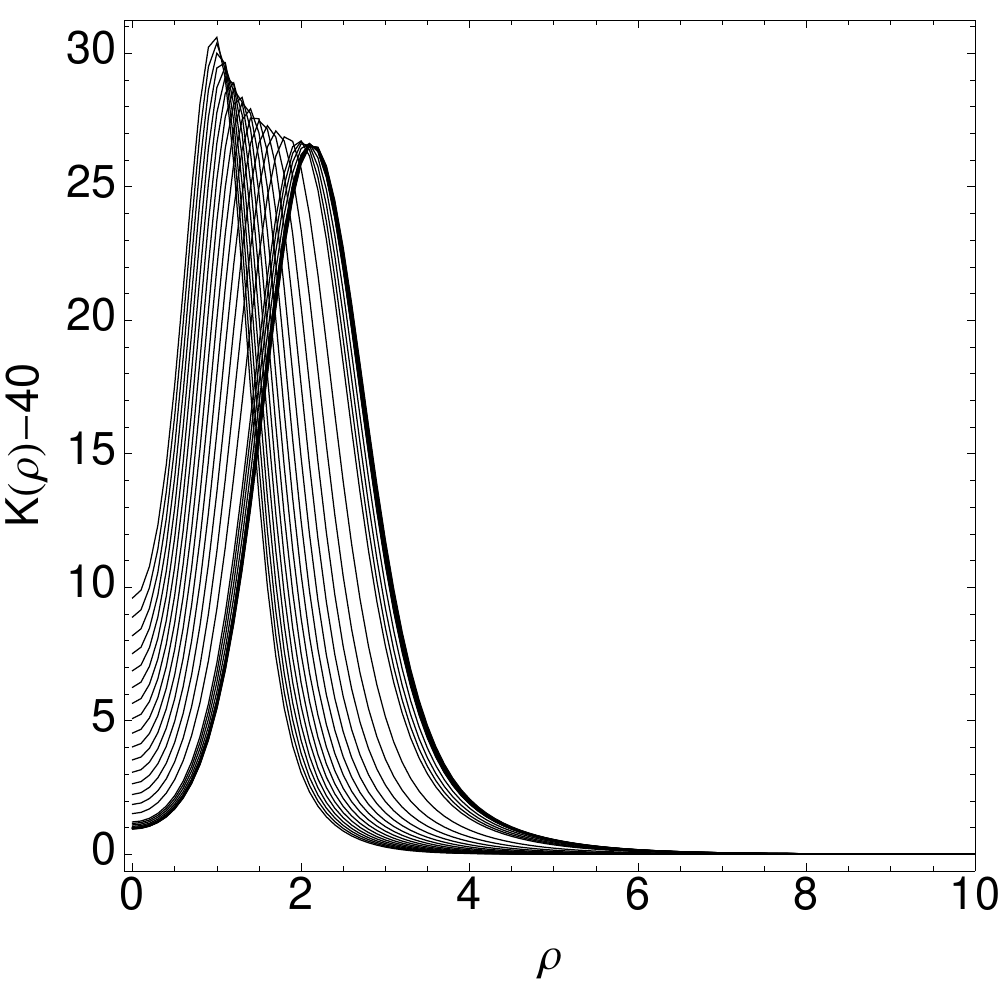}
\includegraphics[scale=0.8]{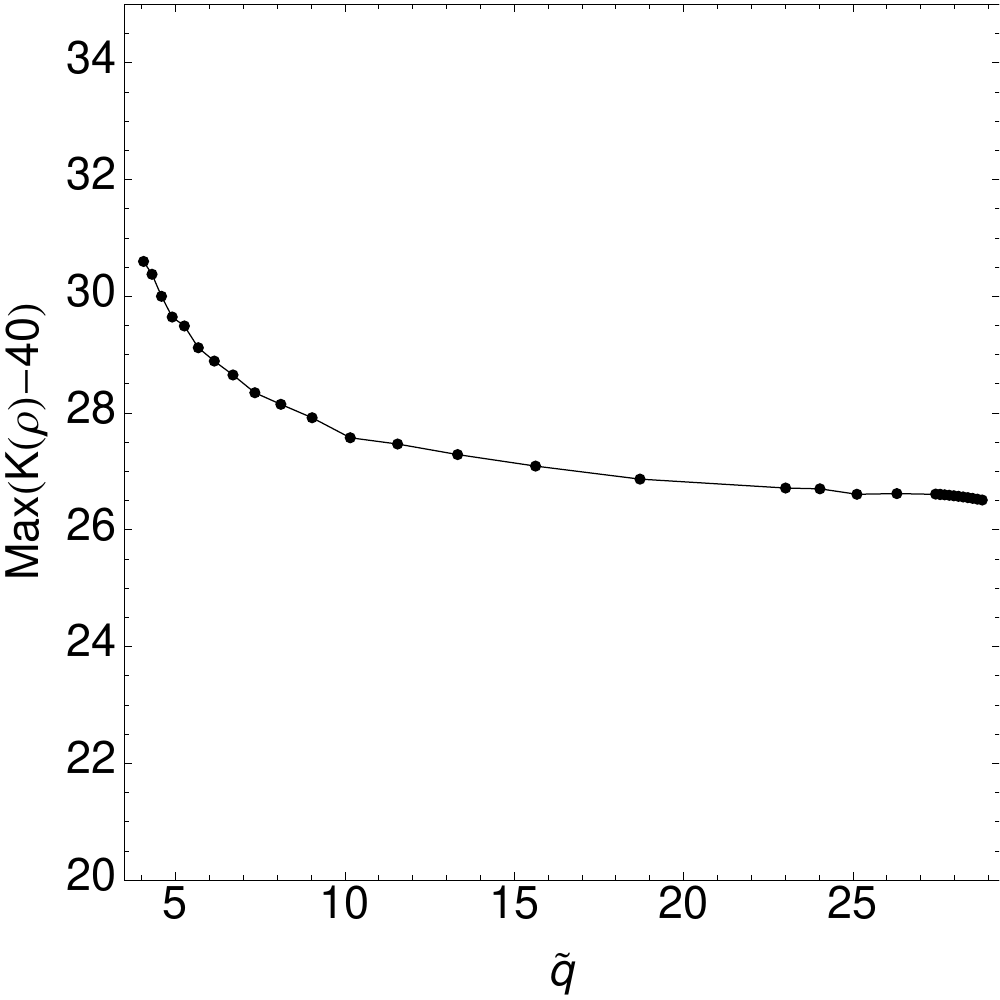}
\end{center}
\caption{\small{\textit{Right:} The Kretschmann invariant, $K(\rho)$ 
for the scaled solution \eqref{scso} may obtained from the Kretschmann 
invariant $K_s(r)$ of the soliton as $K(\rho)=\lim_{L \to \infty} 
K_s(r/L)$, where $L=\tilde q^{\frac{1}{3}}$. We plot the RHS of this equation for $5\lesssim \tilde q \lesssim 30$. For the largest values of $\tilde q$ shown in these plots the mass of the soliton is larger than $100$ (in AdS units) and therefore these solitons are ``large" and the considerations in \S\ref{subsec:respsoliton} should apply.  
Note that the curves fall nicely on top of each other (towards the right in the plot) and show no sign of a 
blow up at small $\rho$. \textit{Right:} Note that the maximum value of the Kretschmann 
scalar asymptotes to a constant at large $\tilde q$, \textit{i.e.,} large $Q$.}}
\label{fig:scaledK}
\end{figure}

While the solitonic solution is regular at every finite value of the charge, 
we have demonstrated in Appendix \ref{sec:AppPlanarSol} that scaled solution \eqref{scso} is not regular  but instead has a singularity at $r=0$ at least at 
generic values of $e^2$. This singularity appears, however, to be rather mild in nature. Neither the Kretschmann invariant nor  $(\nabla \textrm{Riem})^2$
of the scaled solution appear to blow up anywhere
(see Fig. \ref{fig:scaledK}).  It would be nice to better understand
the near horizon behaviour of this planar solitonic solution along the lines of 
\cite{Horowitz:2009ij}; however we leave this to future work. 

Later in this paper we study hairy black holes. As we demonstrate below, 
at finite but large values of the charge extremal black holes have lower mass 
than solitons. In the planar scaling limit described above, the solitonic mass approaches 
that of the hairy black brane, and it appears that the planar 
limit of the solitonic solution in fact coincides with the extremal hairy 
black brane. We provide numerical evidence for this later, when we discuss hairy black holes.

\subsection{Second soliton branch for $e^2 < e^2_{solcrit}$}
\label{subsec:2ndbranch}

\begin{figure}[t]
 \begin{center}
 \includegraphics[scale=0.8]{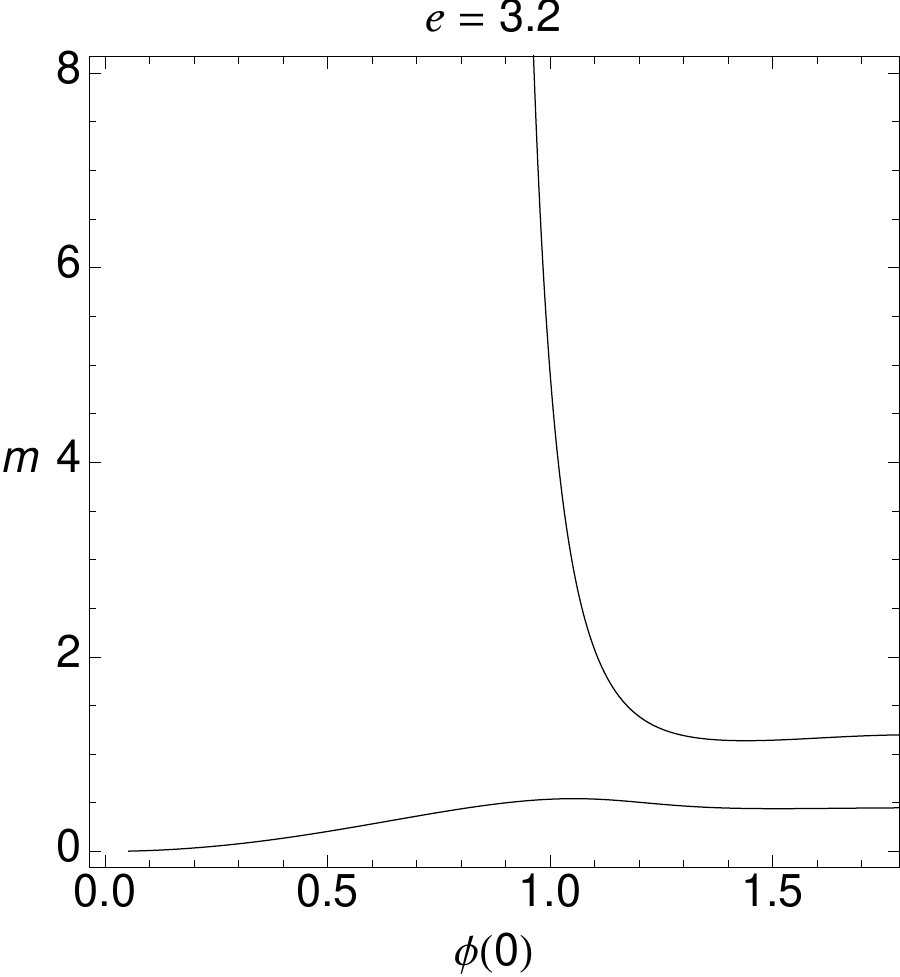}
 \end{center}
 \caption{\small{Mass versus scalar field at the origin, with two soliton branches represented for $e=3.2<e_{solcrit}$. The upper branch, which exists for arbitrarily large charges, was found in \cite{Gentle:2011kv}. The two branches merge for $e^2 = e^2_{solcrit}$.}}
 \label{fig:mvsphi0e3p2}
 \end{figure}

\begin{figure}[t]
 \begin{center}
 \includegraphics[scale=0.8]{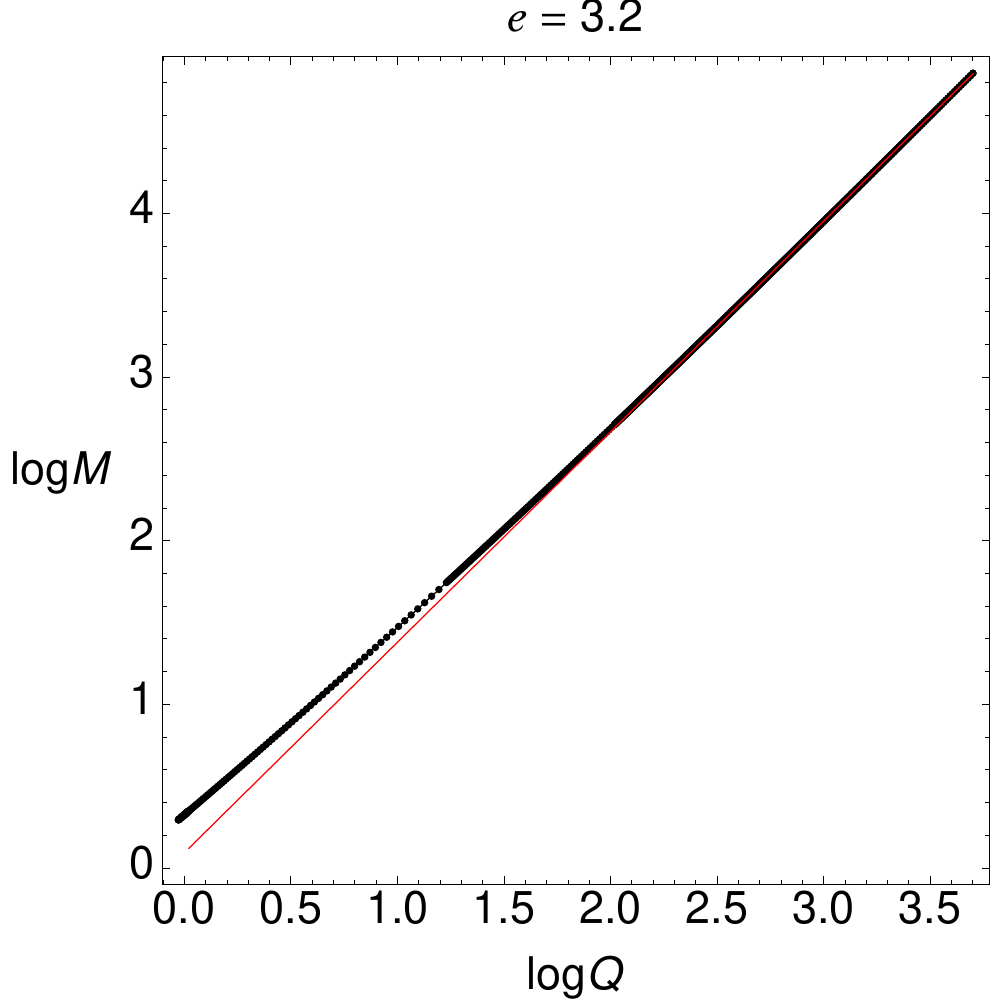}
 \hspace{.5cm}
 \includegraphics[scale=0.8]{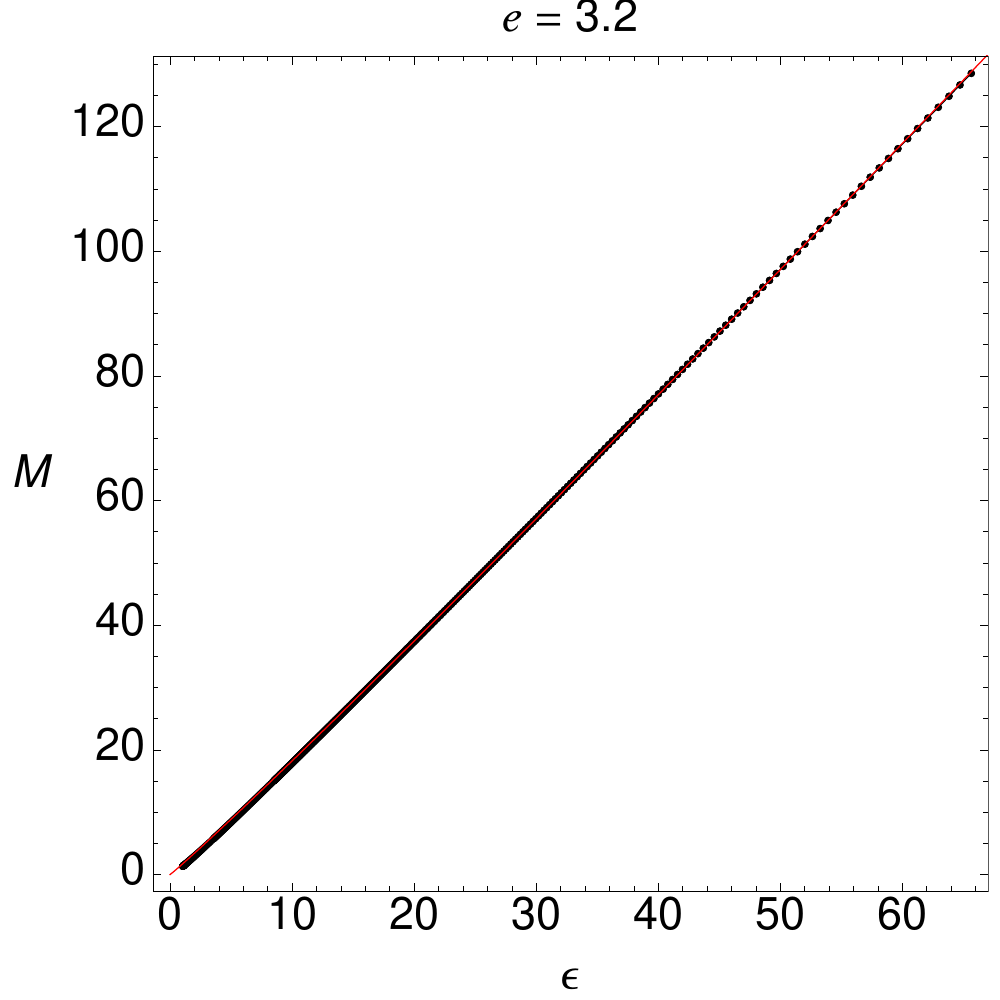}
 \end{center}
 \caption{\small{\textit{Left:} $\log M$ vs $\log Q$ for the second soliton branch, in the case $e=3.2$. The black dots correspond to our numerical data and the red curve corresponds to the best fit of our data to a function of the of form $M= A Q^\alpha$ in the large charges regime. We find that $\alpha \simeq 1.287$, which is close to  the scaling \eqref{scnn} required by conformal invariance, and $A \simeq 1.095$. \textit{Right:}. $M$ vs. $\epsilon$. As this plot shows, for large values of these quantities the relation between them is approximately linear, in accordance with \eqref{scnn}. The best fit to $M= A \epsilon^{\alpha}$ is $\alpha \approx 1.030$ and $A\simeq1.727$.}}
 \label{fig:scalingsolitonq3p2}
 \end{figure}

Reference \cite{Gentle:2011kv}, which appears simultaneously with the first version of our work on the arXiv, discovers a new branch of solitonic solutions in a model closely related to \eqref{sysaction}. We confirmed with our numerical methods that this second branch exists for $e^2 < e^2_{solcrit}$ in our case, and that it merges with the branch described above for $e^2 = e^2_{solcrit}$, which agrees with the expectations from \cite{Gentle:2011kv}.\footnote{Ref.~\cite{Gentle:2011kv} also finds that a second branch of solitons exists for $e^2 > e^2_{solcrit}$. Unlike the branch studied in section~\ref{subsec:respsoliton}, which exists for all values of the charge, that second branch exists only for a finite range $0<Q_{min}(e^2) \leq Q \leq Q_{max}(e^2)$.}
Figure~\ref{fig:mvsphi0e3p2} represents the two branches, for the particular value $e=3.2<e_{solcrit}$. It is clear that one branch is continuously connected to small solitons, studied above, while the second branch exists for arbitrarily large charge/mass. We can study the large charge limit of the latter branch, as we did previously for solitons in the range $e^2 > e^2_{solcrit}$. In particular, the scaling behaviour \eqref{scnn}, which follows from the expressions \eqref{sf} and \eqref{scso}, should apply. We confirm this expectation with our numerical analysis in Figure~\ref{fig:scalingsolitonq3p2}.

\section{Instabilities of the Reissner-Nordstr\"om$-$AdS black hole}
\label{sec:RNlinear}

We start this subsection with a review of the $d=5$ Reissner-Nordstr\"om$-$AdS (RN AdS) black hole. We then study its instability with respect to the condensation of a minimally coupled massless scalar, which signals the bifurcation to a hairy black hole.

The RN AdS black hole is a two-parameter solution to the equations of motion that follow from \eqref{sysaction}, given by:
\begin{equation}
\label{RN AdS}
\begin{aligned}
ds^2&=-V(r)\,dt^2+\frac{dr^2}{V(r)}+r^2\,d\Omega_3^2\,,\quad V(r)=\left(1-\frac{R^2}{r^2}\right)\left(1+r^2+R^2-\frac{2}{3}\frac{\mu^2R^2}{r^2}\right)\,,\\
A_t&=\mu\left(1-\frac{R^2}{r^2}\right)\,,\\
\phi&=0\,,
\end{aligned}
\end{equation}
where $r=R$ is the location of the event horizon and $\mu$ is the chemical potential. The existence of an event horizon requires that
\begin{equation}
\mu^2\leq \frac{3}{2}(1+2\,R^2)\,,
\end{equation}
with the equality saturated by the extremal (zero temperature) black hole. 

Here we record the basic thermodynamic quantities of the RN AdS black hole for later use:
\begin{equation}
\begin{aligned}
M&\equiv\frac{3\,\pi}{8}\,m=\frac{3\,\pi}{8}\,R^2\left(1+R^2+\frac{2}{3}\,\frac{\tilde{q}^2}{R^4}\right)\,,\\
Q&\equiv\frac{\pi}{2}\,{\tilde q}=\frac{\pi}{2}\,\mu\,R^2\,,\\
S&=\frac{\pi^2}{2}\,R^3\,,\\
T&=\frac{1}{2\pi\,R}\left(1+2\,R^2-\frac{2}{3}\,\mu^2\right)\,,\\
\end{aligned}
\end{equation}
where $M$ is the mass, $Q$ is the charge, $S$ is the entropy and $T$ is the temperature of the black hole. The rescaled parameters $m$ and ${\tilde q}$ will be used later for convenience.


\subsection{Linear instability}
\label{subsec:RNstab}

In this subsection, we consider the stability of the RN AdS black hole \eqref{RN AdS} under scalar condensation. Let us focus on the extremal solutions,
\begin{equation}\label{bhext}
 \begin{split}
V(r) &= \left(1-\frac{R^2}{r^2}\right)^2(1+r^2+2R^2) = 4\left(3+\frac{1}{R^2}\right)(r-R)^2+\Or\left((r-R)^3\right),\\
A_t &= \sqrt{\frac{3}{2}(1+2R^2)}\left(1-\frac{R^2}{r^2}\right) = \sqrt{6\left(2+\frac{1}{R^2}\right)}(r-R)+\Or\left((r-R)^2\right), 
 \end{split}
\end{equation}
which are parameterized by the horizon radius $R$. 
The charge of the extremal black hole is given as a function of its 
radius by  
\begin{equation}\label{chR}
{\tilde q}= R^2\sqrt{\frac{3}{2}(1+2R^2)}.
\end{equation} 

\subsubsection{Near horizon instability for large black holes}

In the near horizon limit $r-R \ll R$, the metric \eqref{bhext} reduces to 
$AdS_2 \times S^3$. In this region, the linearized equation for the charged 
scalar field $\phi$ (see \eqref{sysaction}) about this background
reduces to the equation for a massive minimally coupled 
scalar with 
$$m_s^2 l_{AdS_2}^2= -\frac{3e^2 R^2}{8} \frac{1+2R^2}{(1+3R^2)^2},$$ 
where $l_{AdS_2}$ is the radius of the $AdS_2$ region. As is well known \cite{Breitenlohner:1982jf}, 
a minimally coupled scalar in $AdS_2$ is unstable whenever
$$m_s^2 l_{AdS_2}^2< -\frac{1}{4}.$$
It follows that the extremal black hole of horizon radius $R$ is unstable
whenever
\begin{equation} \label{instab} 
e^2\geq \frac{2 (1+3 R^2)^2}{3 R^2(1+2R^2)}.
\end{equation}
The RHS of \eqref{instab} is a monotonically decreasing function of $R$. 
At large $R$, \eqref{instab}  reduces to 
\begin{equation} \label{instabl} 
e^2\geq 3+ \frac{1}{2 R^2} + {\cal O}(1/R^4).
\end{equation}
It follows that very large extremal RN AdS black holes are unstable when 
$e^2>3$. The end point of the instability involves a condensate of the 
scalar field. By the Hawking area increase theorem it also
has a horizon. Consequently, the end point of this instability is 
a hairy black hole.

It follows from the previous analysis that 
hairy black holes of charge ${\tilde q}$ exist in the system \eqref{sysaction} whenever\footnote{In order to obtain this equation, we solved \eqref{instab} to obtain $R^2$ as a function of $e^2$, and then plugged the solution to this equation into \eqref{chR}.}
\begin{equation}\label{qineq}
{\tilde q} \geq \frac{\left(e \left(\sqrt{9 e^2-24}-3 e\right)+12\right) \sqrt{e \left(3 e+\sqrt{9 e^2-24}\right)-6}}{24
   \left(e^2-3\right)^{3/2}} = {\tilde q}_{BF}(e^2).
\end{equation}
This condition is not only sufficient, but also necessary for the instability of extremal black holes if $R\geq1$ (${\tilde q}\geq\frac{3}{\sqrt{2}}$), as shown in \cite{Dias:2010ma}. 

\subsubsection{Superradiant instability for small black holes}

For $R<1$, i.e for $e^2 \geq \frac{32}{9}$, 
 the stability is not ensured by an analysis of the near-horizon region. Indeed, the leading instability of small extremal black holes is of the superradiant rather than Breitenl\"ohner-Freedman variety\footnote{This picture of superradiance can be reconciled with the more traditional version of an instability $\hat{\phi}\sim e^{-i w t}$ with onset mode $w=e\mu$ by a change of gauge. Consider the gauge $\hat{A}=A-\mu dt$, for which the potential vanishes at infinity. A static scalar field in the gauge $A$ (corresponding to the bifurcation to a hairy black hole) transforms to $\hat{\phi}\sim e^{-i e\mu t}$ in the gauge $\hat{A}$.}
(see \cite{Basu:2010uz}); the corresponding unstable modes are not localized entirely within the near horizon region. At small values of the black hole charge the extremal RN AdS black hole undergoes the superradiant instability provided
\begin{equation}\label{srinst}
{\tilde q} \geq {\tilde q}_{sr}(e^2)= \frac{1}{2}\sqrt{\frac{3}{2}}\left(1 - \frac{3e^2}{32}\right) + {\cal O}\left(\left(1- \frac{3e^2}{32}\right)^2\right).
\end{equation}

In summary we expect 
that hairy black holes exist in our system whenever 
\begin{equation}\label{srinst2}
\tilde{q} \geq \tilde{q}_0(e^2),  \qquad  \textrm{with} \quad  \begin{cases} \tilde{q}_0(e^2)={\tilde q}_{BF}(e^2)  &  \textrm{for}\quad
e^2 < \frac{32}{9}~(R>1),  \\ \tilde{q}_0(e^2)\leq{\tilde q}_{sr}(e^2) 
& \textrm{for}\quad e^2 \geq \frac{32}{9} ~(R<1),  \end{cases}
\end{equation}
where the function ${\tilde q}_o(e^2)$ represents the smallest charge at which an extremal
RN AdS black hole is unstable.\footnote{We assume here that all black holes of a given charge $\tilde{q}$ are
stable if the extremal black hole at that charge is stable. This 
expectation is intuitively reasonable, and has been borne out by all
explicit computations to date.} 
It was demonstrated in \cite{Basu:2010uz} that arbitrarily small extremal 
black holes in \eqref{sysaction}
suffer from the superradiant instability when $e^2>\frac{32}{3}$. 
It follows that ${\tilde q}_o(\frac{32}{3}) =0$. Therefore it is natural to consider two different regions in the parameter space of hairy black holes,  $3<e^2<\frac{32}{3}$ and $e^2>\frac{32}{3}$. We will analyse these regions separately in later sections.


\subsection{Numerical results}
\label{subsec:linear}

We now present a numerical analysis of the linearized instability of the RN AdS black hole, which complements the considerations in the previous subsection. We consider the wave equation for a massless scalar field of charge $e$ in the background of the RN AdS black hole and we look for marginally stable (that is, time-independent) spherically symmetric modes. The existence of such modes signals a bifurcation point in the phase diagram, and a new family of black holes with a non-zero scalar condensate (i.e. a hairy black hole) should emerge from the RN AdS family.  We find that such modes exist provided that $e^2>3$, which indicates that hairy black holes will exist if that the scalar charge $e$ satisfies this condition, as predicted above.

The equation that we are going to solve is
\begin{equation}
D^2\phi=0
\label{scalareqB}
\end{equation}
for $\phi(r)$ real and where $D_\mu=\nabla_\mu-i\,e\,A_\mu$ is the gauge covariant derivative on the RN AdS background.  As discussed above, we will only consider time-independent and spherically symmetric modes and, furthermore, we will impose the following asymptotic behavior, 
\begin{equation}
\phi(r)\sim \frac{\epsilon}{r^4}\,\quad \textrm{for}\quad r\to \infty\,,
\end{equation}
so that $\epsilon$ is the vacuum expectation value of the operator dual to $\phi$. In addition, we will also require that $\phi$ is regular at the horizon. These boundary conditions can be easily implemented redefining the field $\phi(r)$ as
\begin{equation}
\phi(r)=p_\phi(r)\left(\frac{R}{r}\right)^4\,,
\end{equation}
with
\begin{equation}
p'_\phi\big|_{r\to\infty}=0\,,\qquad \left(p'_\phi-4\,p_\phi\right)\big|_{r=R}=0\,.
\end{equation}
With these boundary conditions we can solve \eqref{scalareqB} as a boundary value problem using shooting as in \cite{Gubser:2008px}. However, in this paper we will follow a slightly different route. Following  \cite{Dias:2010ma}, we cast \eqref{scalareqB} as a generalised eigenvalue problem,
\begin{equation}
L(r)\phi(r)=e^2\,\Lambda(r)\,\phi(r)\,,
\end{equation}
where the scalar charge $e$ appears as the generalised eigenvalue. Here $L(r)$ is a second order linear differential operator. Then, for a given RN AdS background uniquely specified by $(R,\mu)$, our strategy consists in finding the eigenvalue $e$ for which there exists a mode that satisfies the above boundary conditions.

\begin{figure}[t]
\begin{center}
\includegraphics[scale=0.35]{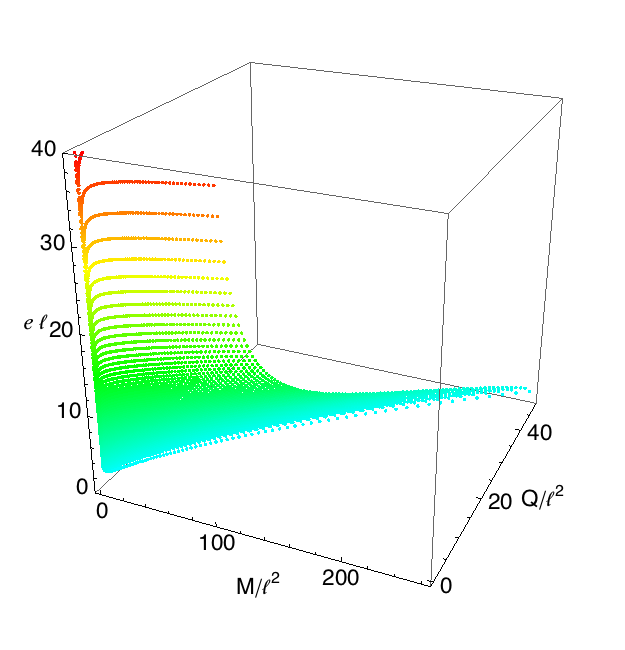}
\hspace{0.1cm}
\includegraphics[scale=0.575]{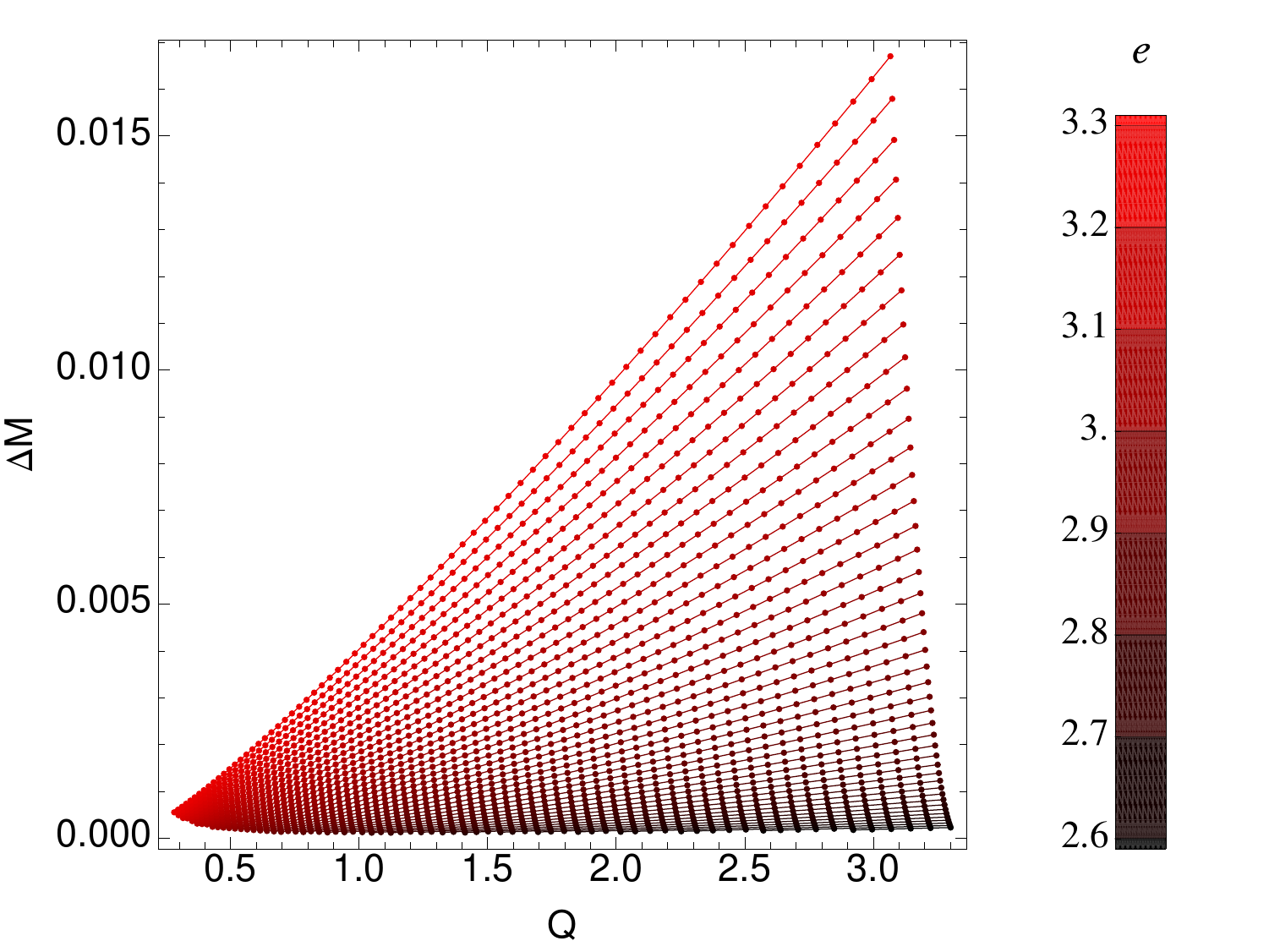}
\end{center}
\caption{\small{ Critical value of the scalar charge $e$ for the existence of an instability as a function of the mass $M$ and charge $Q$ of the RN AdS background. The minimum value of $e^2$ monotonically decreases from $\frac{32}{3}$ to $3$ as the size of the black hole increases. $\Delta M$ is the mass difference with respect to the extremal RN AdS black hole of the same charge.}}
\label{fig:e_linear}
\end{figure}

The results are depicted in Fig.~\ref{fig:e_linear}. For a given black hole size $R$, we find that the minimum value of $e^2$ for instability is obtained for black holes in the extremal limit. Also, our results indicate that the minimum value of $e^2$ monotonically decreases from $\frac{32}{3}$ to $3$ as the black hole size increases, and for $e^2<3$ all black holes are stable under scalar condensation. On the other hand, for $e^2>\frac{32}{3}$ all extremal black holes are unstable.

\begin{figure}[t]
\begin{center}
\includegraphics[scale=0.7]{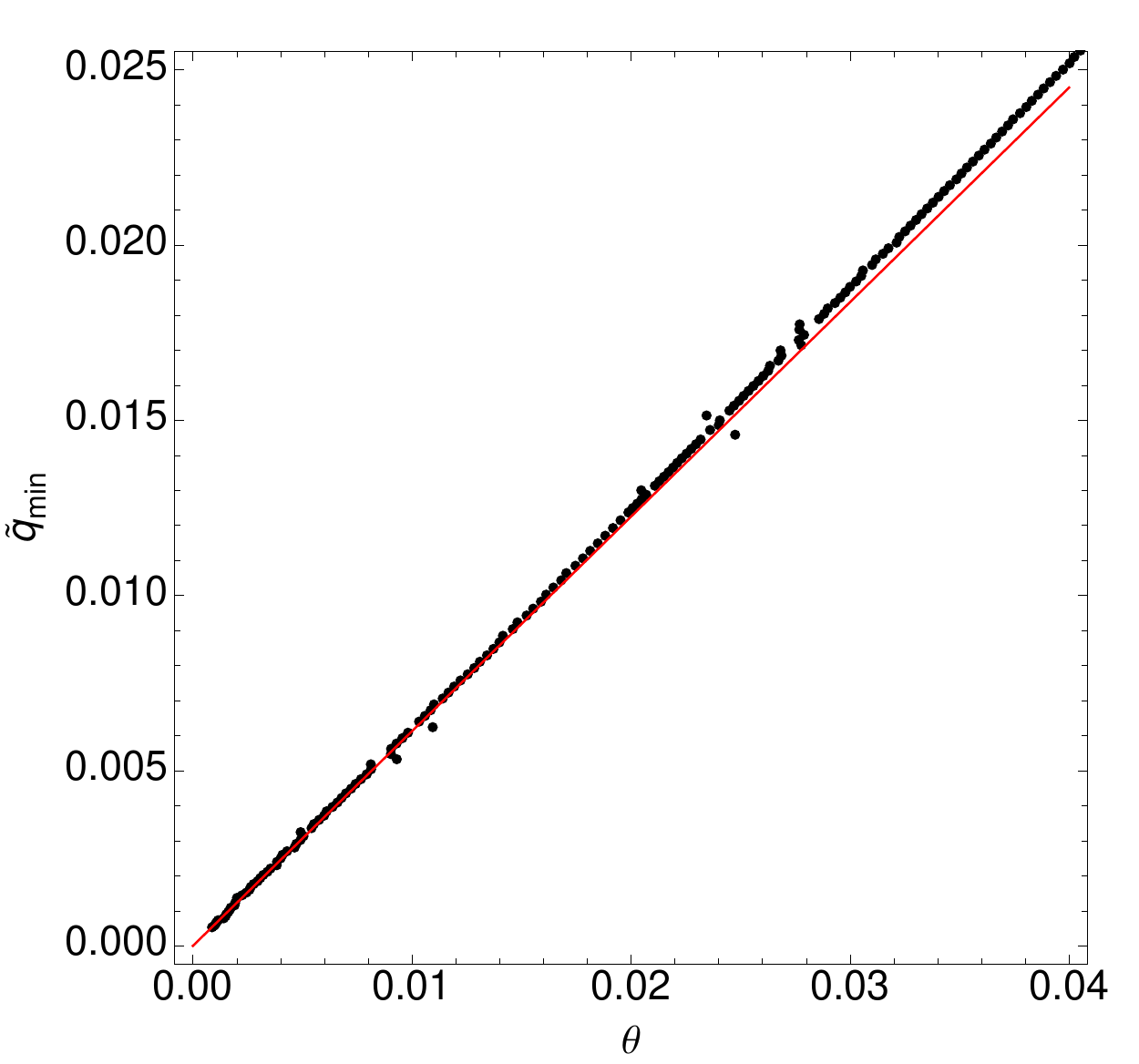}
\end{center}
\caption{\small{Minimum charge of unstable RN AdS black holes vs. scalar charge parametrized by $\theta = 1-\frac{3e^2}{32}$. The numerical results agree with the analytical prediction for small $\theta$.}}
\label{fig:merger_linear}
\end{figure}

These results agree with the analytical predictions. In Fig.~\ref{fig:merger_linear}, we plot the minimum charge of unstable RN AdS black holes for values of $e^2$ close to (but below) $\frac{32}{3}$; the prediction of \eqref{srinst2} is confirmed.

\section{Hairy black holes for $3<e^2<\frac{32}{3}$}
\label{sec:middlee}

When $e^2\leq3$, solitons and RN AdS black holes 
are the only static charged solutions of the Lagrangian
\eqref{sysaction}.
When $e^2>3$, however,  the spectrum 
of charged static solutions also includes hairy black holes as the linearized analysis in \S\ref{subsec:RNstab} and \S\ref{subsec:linear} suggests.

In this section we will investigate the spectrum of hairy black
hole solutions 
in the parameter range
\begin{equation}\label{range}
3 \leq e^2 \leq \frac{32}{3}\,. 
\end{equation} 
While our investigations are partly numerical, we have also
been able
to obtain some analytic results at the upper end of the
parameter range \eqref{range} as we now explain.

Whenever \eqref{sysaction} hosts hairy black hole solutions of
parametrically
small charge, these solutions may be constructed analytically
using perturbative techniques of \cite{Basu:2010uz}. 
As $\tilde{q}_0(e^2)$ is of order $\theta$ when 
$e^2=\frac{32}{3}\left(1-\theta\right)$, it is possible to use 
the perturbative techniques described above to completely characterize 
the spectrum of hairy black holes with charges of order $\theta$ at these 
values of $e^2$. In the first two subsections below we present an analytic
construction of
hairy black holes, perturbatively in $\theta$. We use the
results of our
analysis to conjecture the qualitative properties of 
hairy black holes in the full range $3<e^2<\frac{32}{3}$. In the
rest of this subsection we present the results of a detailed
numerical analysis that give evidence for our conjectures and fill out
several details.

\subsection{Non-interacting model for small hairy black holes at
small $\theta$}
\label{subsec:middleenonint}

In this section and the next, we study small hairy black holes
for values
of scalar charge $e^2$ close to but less than $\frac{32}{3}$. We
set
\begin{equation}\label{deftheta}
e^2=\frac{32}{3}\left(1-\theta \right) 
\end{equation}
where $\theta$ is a small positive number. 
In \S\ref{subsec:middleepert} we demonstrate that our
system admits a spectrum of hairy black hole
solutions at small $\theta$ by explicitly constructing the 
hairy black holes in a perturbative expansion in 
$\theta$. This construction is rather 
involved but the final result 
for the thermodynamics is rather simple at leading order in
$\theta$.
It turns out that the leading order 
thermodynamics of these small hairy black holes is reproduced by
modelling them
as a non interacting mix of a RN AdS black hole and a 
soliton (this is in accord with previous experience in
\cite{Basu:2010uz}
and \cite{Bhattacharyya:2010yg}). As a prelude of, and
motivation for
our perturbative construction, in this subsection we 
work out this simple thermodynamical model of small hairy black
holes.
The formulas obtained in this subsection will be derived more 
systematically (as the 
first terms in a series expansion) in the next subsection.

\subsubsection{Thermodynamics of small RN AdS black
holes}\label{subsubsec:bhtherm}
Small RN AdS black holes appear in a two parameter family labeled
by
their radius $R$ and their chemical potential $\mu_{BH}$. In this
paper we will
find it convenient to parameterize $R$ and $\mu_{BH}$ in terms of
auxiliary
variables $\alpha$ and $a$ defined by 
\begin{equation} \label{mudef} \begin{split}
R^2&=a \theta\,, \\
\mu_{BH}^2&=\frac{3}{2}(1+\alpha R^2)=\frac{3}{2}(1+\alpha a
\theta)\,,\\
\end{split}
\end{equation}
where $\alpha$ and $a$ are positive numbers. In this subsection
and the next
we are interested in the small $\theta$ limit 
with $\alpha$ and $a$ of order unity. $a$ parametrizes the
radius $R$ of the black hole
in units of 
$\theta$ and $\alpha$ measures the deviation, in units of $\theta$,
of the chemical potential squared of the black hole from a
critical value
$\frac{3}{2}$.  The usual formulas of black hole 
thermodynamics determine all thermodynamical properties of the
black hole
as a function of $a$ and $\alpha$; 
\begin{equation} \begin{split}\label{vactherm}
\mu_{BH}&= \left[\frac{3}{2}\left(1+\alpha a
\theta\right)\right]^{1/2} = 2q_c \left( 1+ \frac{\alpha a
\theta}{2} -\frac{\alpha^2 a^2 \theta^2}{8} \right) + {\cal
O}(\theta^3)\,,\\
m_{BH}&=R^2\left(1+R^2+\frac{2}{3}\mu_{BH}^2\right) =2 a \theta
+ a^2(1+\alpha) \theta^2\,, \\
\tilde{q}_{BH}&= \mu_{BH}R^2 = 2a q_c \theta + \alpha a^2
q_c\theta^2
-\frac{1 }{4}\alpha^2 a^3 q_c\theta^3 + {\cal O}(\theta^4)\,,\\
T&=\frac{1}{2\pi R}\left(1+2R^2-\frac{2}{3}\mu_{BH}^2\right) =
\frac{(2-\alpha)}{2 \pi} \sqrt{a \theta}\,,\\
S_{BH}&=\frac{\pi^2R^3}{2} = \frac{\pi^2 (a \theta)^{3/2} }{2}\,,
\end{split}
\end{equation}
where 
\begin{equation} \label{defqc}
q_c=\frac{1}{2} \sqrt{\frac{3}{2}}\,.
\end{equation} 
Note that all nonsingular black holes have $\alpha \leq 2$.
Black holes
with $\alpha=2$ are extremal. Black holes
with $\alpha >2$ formally have negative temperature and are
unphysical (they
have naked singularities). Note also that when $a$ and $\alpha$
are of order
unity, the black hole mass and charge is of order $\theta$.

\subsubsection{Thermodynamics of small charge solitons}
\label{subsubsec:middleesoltherm}

As we have explained above, in addition to black holes, our
system admits
regular solitonic solutions at all values of $e^2$. In
\cite{Basu:2010uz}
these solutions were constructed in a perturbative expansion in
their charge.
This construction allowed a determination of the energy
$m_{sol}$
and the chemical potential $\mu_{sol}$ of the these solitons
as a function of their charge $\tilde{q}_{sol}$ at small charge.
In this section we are interested in
solitons 
whose charge is of order $\theta$, and so we set 
$\tilde{q}_{sol}=b \theta$ where the small number $\theta$ was
defined in \eqref{deftheta}
and $b$ is a positive number of order unity.\footnote{Note that
while the small number 
$\theta$ is a parameter of the theory, $b$ parametrizes the
solution we study.} Plugging into the formulas of \cite{Basu:2010uz} we
find
\begin{equation} \begin{split}\label{soltherm}
\tilde{q}_{sol}&= b \theta\,,\\
\mu_{sol}&=
\frac{4}{e}+\left(\frac{9}{7}-\frac{64}{7e^2}\right)q_{sol} +
\Or (q_{sol}^2) = 2 q_c + \theta \left( q_c + \frac{3b}{7}
\right) +{\cal O}(\theta^2)\,,\\
m_{sol}&=
\frac{16}{3e}q_{sol}+\frac{2}{21}\left(9-\frac{64}{e^2}\right)q_{sol}^2+\Or(q_{sol}^3)
\\
&=\frac{b}{q_c} \theta + \left( \frac{b}{2q_c}+ \frac{2b^2}{7}
\right)
\theta^2 + \left( q_c b - \frac{4 b^2 }{7}
-\frac{1541b^3}{339570q_c}\right) \theta^3 +
{\cal O}(\theta^4)\,.
\end{split}
\end{equation}
As the solitonic solutions have no event horizon they have zero
entropy. As these solutions can be continued to Euclidean space
with a thermal circle of arbitrary size, their temperature is
indefinite.

\subsubsection{Hairy black holes as a non-interacting mix}
\label{subsubsec:middleehbhnonint}

Consider a system with net charge
\begin{equation}\label{qdef}
{\tilde q} = q \theta \,,
\end{equation}
and net mass 
\begin{equation}\label{dmdef}
m = \frac{q}{q_c} \theta + \left( \frac{2}{3} q^2 + \Delta m
\right)
 \theta^2 \,.
\end{equation}
The parameters $q$ and $\Delta m$ parameterize the hairy black
hole
we wish to study. In this section we assume that $q$ is held
fixed
as $\theta$ is taken to zero. It will turn out that  hairy black holes
with charge $q \theta$ occur over values of $\Delta m$ that
range from
approximately $0$ to a negative number independent of $\theta$.
In other
words $\Delta m$ is always order unity or smaller.

In this subsection we will model hairy 
black holes as a non interacting mix of solitons and RN AdS black
holes. A hairy black hole can partition its charge and mass arbitrarily
between the black hole and soliton phases, and chooses to do so in the
manner that maximises its entropy. 

How does a system maximise its entropy? As the soliton carries
no entropy, all of the entropy lies in its black hole component. Let us
suppose that the total mass $m$ is partitioned up between black hole and
soliton as $m=m_{sol}+m_{BH}$ and a similar partitioning for its total
charge $\tilde{q}=\tilde{q}_{sol}+\tilde{q}_{BH}$)
\begin{equation}
S_T = S_{BH}(\tilde{q}_{BH},m_{BH}) +
S_{sol}(\tilde{q}_{sol},m_{sol}) =S_{BH}(\tilde{q} -
\tilde{q}_{sol},m- m_{sol})\,. \nonumber
\end{equation}
Maximising $S_T$ w.r.t. $\tilde{q}_{sol}$ and use of the first law gives
\begin{equation}\label{mueq}
-\frac{\left(\frac{\partial{S_{BH}}}{\partial{\tilde{q}_{BH}}}\right)}{\left(\frac{\partial{S_{BH}}}{\partial{m_{BH}}}\right)}
= \frac{\partial{m_{sol}}}{\partial{\tilde{q}_{sol}}} \qquad\Rightarrow \qquad  \mu_{BH} = \mu_{sol} \,. 
\end{equation}
In other words we must partition up charges so as to ensure that
the chemical
potential of the black hole equals that of the soliton. 
Assuming the model spelt out above, we will now describe the
spectrum
of hairy black holes in our system. We will first work this out
at leading
order in $\theta$. It turns out that the leading order result
has a
degeneracy that is lifted at next order in $\theta$; for this
reason we will
also compute the next correction in $\theta$ to a particular
aspect of the
thermodynamics of the non interacting model. 

\subsubsection{Hairy black holes at leading order in
$\theta$}\label{subsubsec:middleeleadord}
Let us suppose that the system divides itself into an RN AdS
black hole
parameterized by $a$ and $\alpha$ and a soliton parameterized by
$b$. We
will now determine $a$, $b$ and $\alpha$ in terms of $q$ and 
$\Delta m$ (as defined in \eqref{qdef} and \eqref{dmdef}). 
To the order of interest in this subsubsection
\begin{equation}
 \begin{split}\label{bhthermod}
\mu_{BH}&=2q_c+ \alpha a q_c \theta + {\cal O}(\theta^2)\,,\\
m_{BH}&=2 a \theta +  a^2(1+\alpha) \theta^2\,, \\
\tilde{q}_{BH}&=2a q_c \theta + \alpha a^2 q_c\theta^2+ {\cal
O}(\theta^3)\,,\\
\end{split}
\end{equation}
while for the soliton one has
\begin{equation} \begin{split}\label{musol}
\mu_{sol}&= 2 q_c + \theta \left( q_c + \frac{3b}{7} \right)
+{\cal O}(\theta^2)\,,\\
m_{sol}&= \frac{b}{q_c} \theta + \left( \frac{b}{2q_c}+
\frac{2b^2}{7} \right) \theta^2 +{\cal O}(\theta^3)\,,\\
\tilde{q}_{sol}&= b \theta \,.\\
\end{split}
\end{equation}
Charge and mass conservation imply that 
\begin{align}
\label{qeq}
{\tilde q} &= q \theta = \tilde{q}_{sol} + \tilde{q}_{BH}
\Rightarrow q = \left(2aq_c+b\right) +{\cal O}(\theta)\,, \\
\label{meq}
m &= \frac{q}{q_c} \theta + \left( \frac{2}{3} q^2 + \Delta m
\right)
\theta^2 = m_{sol} + m_{BH} \Rightarrow \Delta m =
\left(\frac{b}{2q_c}-\frac{a b}{q_c}-\frac{8b^2}{21}\right)
+{\cal O}(\theta)\,.
\end{align}
Eq. \eqref{qeq} gives
\begin{equation}\label{solfora}
 a = \frac{q-b}{2q_c}+{\cal O}(\theta)\,.
\end{equation}
The requirement of chemical potential matching \eqref{mueq}
gives
(using  the first equations of 
\eqref{bhthermod} and \eqref{musol}) 
\begin{equation}
\label{alpha}
\alpha = \frac{1}{a}\left(1+\frac{3b}{7q_c}\right)+{\cal
O}(\theta)
=\frac{2}{q-b}\left(q_c+\frac{3b}{7}\right)+{\cal O}(\theta)\,,
\end{equation}
where we have used \eqref{solfora} in the last step. Plugging 
\eqref{solfora} in \eqref{meq} gives an equation for $b$.
Solving this, we find
\begin{equation}
\label{bsol} \begin{split} 
b = \left(\frac{7}{10}(q-q_c) \pm
\sqrt{\left[\frac{7}{10}(q-q_c) \right]^2 + \frac{21}{20}\Delta
m}\right)+{\cal O}(\theta)\,.
\end{split}
\end{equation}
Recall that the black hole component of the mix must have
$\alpha \leq 2$.
From \eqref{alpha}, however, this implies that 
$ b\leq\frac{7}{10}(q-q_c)+{\cal O}(\theta)$. 
As $b\geq0$, it follows immediately that no solution exists 
for $q<q_c+{\cal O}(\theta)$. 
It also follows that acceptable roots for $b$ (to leading order)in \eqref{bsol} are given by 
\begin{equation}
\label{bsoln} \begin{split} 
b &= \left(\frac{7}{10}(q-q_c) -\sqrt{\left[\frac{7}{10}(q-q_c)
\right]^2 + \frac{21}{20}\Delta m}\right)+{\cal O}(\theta)
 ~~~~{\rm when}~~~q\geq q_c+{\cal O}(\theta)\,.\\
\end{split}
\end{equation}
This solution
is physical only when $b\geq 0$; this requires 
$$\Delta m \leq{\cal O}(\theta).$$
 Moreover $b$ must be real; this requires 
$$ \Delta m \geq -\frac{7}{15}(q-q_c)^2+{\cal O}(\theta)\,.$$
Note that $\alpha=2$ when the last inequality is saturated. It
follows that
the black hole component of a hairy black hole is extremal at
the lowest
allowed value of $\Delta m$ at any $q$. 

\begin{figure}
\begin{center}
\includegraphics[height=74mm]{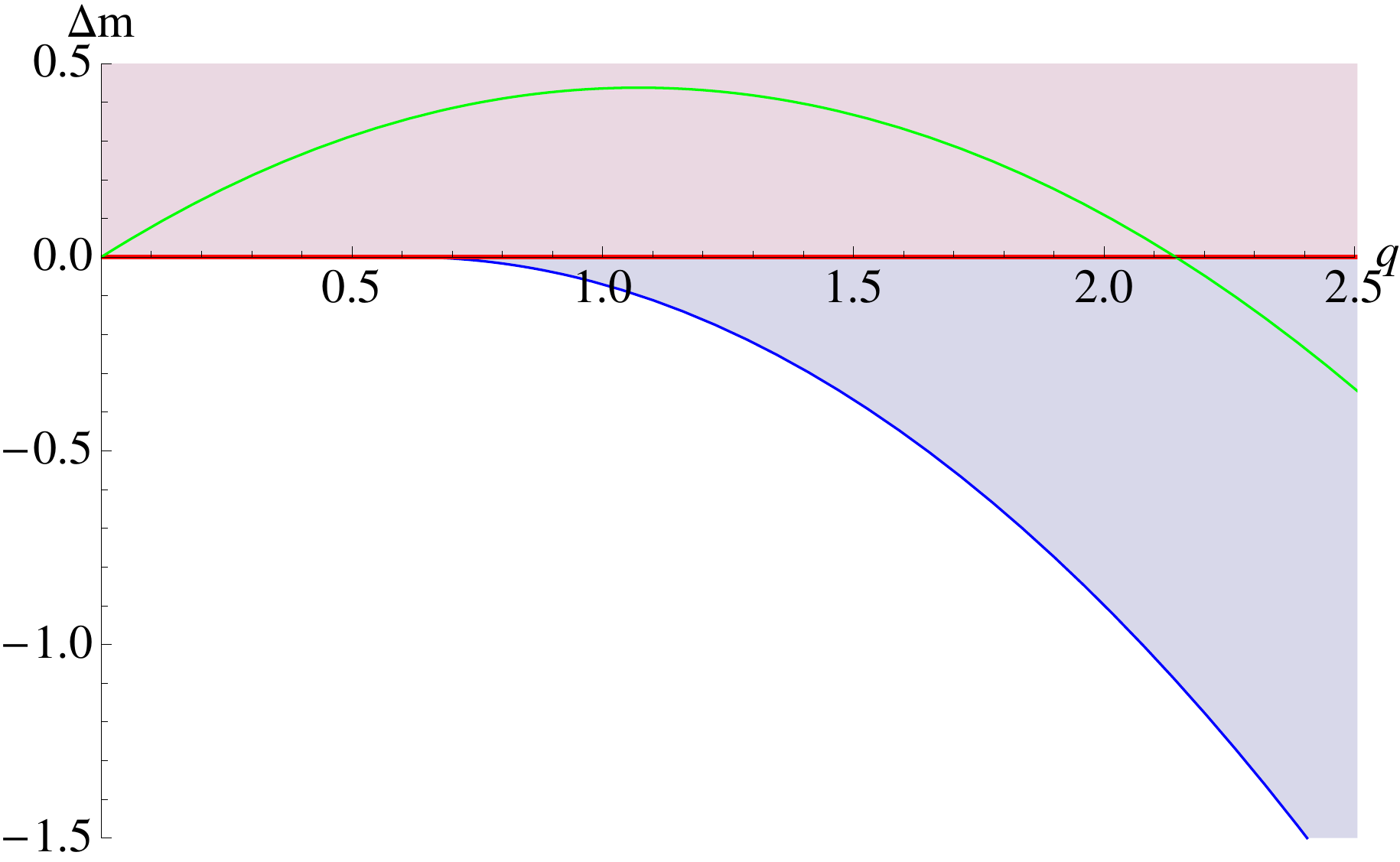} 
\end{center}
\caption{Range of allowed values of $\Delta m$ for hairy
black holes
(blue shaded). Note that hairy solutions exist only 
for $q\geq q_c$. The upper limit, $\Delta m=0$ denotes 
the onset of superradiant instabilities and corresponds to $b=0$
(red line).
This is the also the extremality line for the pure RN AdS black
hole.
The lower limit, $\Delta m = -\frac{7}{15}(q-q_c)^2$
corresponds to the extremal hairy black hole with $\alpha = 2$
(blue line).
RN AdS black holes exist for all values of $q$ and $\Delta m >0$
(pink shaded).
The soliton lies on $\Delta m_{sol} = \frac{q}{2q_c} -
\frac{8q^2}{21}$
(green curve).}
  \label{figure1}
\end{figure}

In summary, we have a solution whenever 
\begin{equation}\label{qrange}
q \geq  q_c +{\cal O}(\theta)\,,
\end{equation}
and 
\begin{equation}\label{mrange}
{\cal O}(\theta) \geq \Delta m \geq -\frac{7}{15}(q-q_c)^2+{\cal
O}(\theta).
\end{equation}
This range is plotted in Fig. \ref{figure1}.
In this range the black hole component of the mix has 
\begin{equation} \begin{split}
\label{sola}
 a &= \frac{1}{2q_c}\left(\frac{3q}{10}+\frac{7q_c}{10} + 
\sqrt{\left[\frac{7}{10}(q-q_c) \right]^2 + \frac{21}{20}\Delta
m}\right)+{\cal O}(\theta)\,,\\
\alpha &=\frac{1}{a}\left(\frac{3q}{10q_c}+\frac{7}{10}
-\frac{3}{7q_c}\sqrt{\left[\frac{7}{10}(q-q_c) \right]^2 +
\frac{21}{20}\Delta m}\right)+{\cal O}(\theta)\,.\\
\end{split}
\end{equation}
The entropy, 
temperature and chemical potential of the hairy black hole are
given by
the formulas \eqref{vactherm} with the values of $a$ and
$\alpha$ in
\eqref{sola}. Note that 
\begin{equation}\label{arange}
 \begin{split}
a \geq
\frac{1}{2q_c}\left(\frac{3q}{10}+\frac{7q_c}{10}\right)+{\cal
O}(\theta) \geq \frac{1}{2}+{\cal O}(\theta)\,,
 \end{split}
\end{equation}
where we have used \eqref{qrange} in the last inequality. As we
have explained
above, at a fixed value of $q$, $\Delta m$ varies in the range
\eqref{mrange}.
As $\Delta m$ is lowered within this range the hairy black hole
horizon
area - parameterized by $a$ - decreases while its chemical
potential -
parameterized by $\alpha$ increases. The temperature of the
hairy black hole
also decreases upon lowering $\Delta m$, reaching zero at the
lowest allowed
value of $\Delta m$.

\subsubsection{Splitting the degeneracy at 
$\mathcal{O}(\theta)$}\label{subsubsec:nextord}
The phase diagram depicted in Fig. \ref{figure1} has the property
that
RN AdS black holes and hairy black holes coexist only on a single
line,
namely $\Delta m=0$. This odd feature is an artefact of working
at
leading order in the $\theta$ expansion. The actual phase
diagram
includes a region of coexistence of these two phases. It turns
out
that the height of this region is of ${\cal O}(\theta)$ 
(in the variable $\Delta m$) as we will now explain. 

\begin{figure}[ht]
 \begin{center}
\includegraphics[height=75mm]{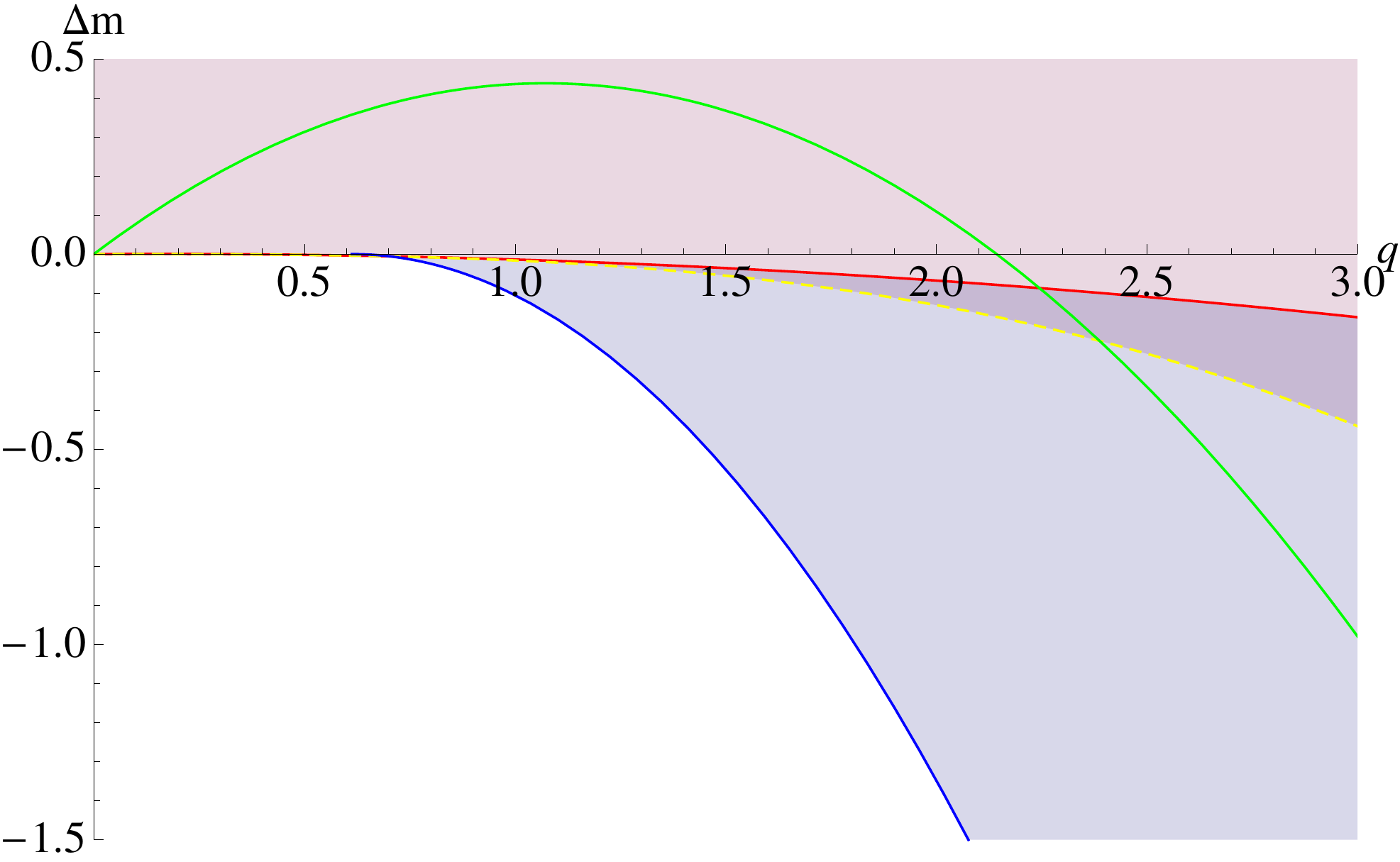} 
\end{center}
\caption{Plot of $\Delta m_{Sr} =
-\frac{2q}{3}\left(q-\frac{q_c}{2}\right)\theta$(red line),
$\Delta m_{ext} =
-\frac{2}{3}\sqrt{\frac{2}{3}}q^3\theta$(yellow dashed line)
and the allowed range $\Delta m$ for hairy black holes(blue
shaded)
for $\theta=0.03$. Hairy black holes exist for all $q\geq q_c$
and $\Delta m$
between the red curve (onset of superradiance) 
and the blue curve (extremal hairy black hole). 
The yellow dashed curve denotes the extremality 
line for RN AdS black holes. Pure black hole solutions exist
above
 this line for all values of $q$. RN AdS and hairy black holes 
coexist in the region between the solid red curve and dashed 
yellow curve.}
\label{figure2}
\end{figure}

The extremality curve for extremal RN AdS black holes is given by
(see \cite{Basu:2010uz})
\begin{align}
\Delta m_{ext} = -\frac{2}{3}\sqrt{\frac{2}{3}}q^3\theta+{\cal
O}(\theta^2)\,.
\end{align}
We will now show that hairy black holes start existing at
larger
values of $\Delta m$, demonstrating that hairy black holes and
RN AdS black
holes coexist over a range of charges and masses. The
instability curve
for RN AdS black holes is given by equating the chemical
potential of these
black holes to $\frac{4}{e}$($= 2q_c+q_c \theta + \Or
(\theta^2)$)
\footnote{This is the minimum value of chemical 
potential of the soliton.}. This is the condition 
(from \eqref{mueq} and \eqref{vactherm})
\begin{equation}\label{alphasr}
\alpha = \frac{1}{a}+\Or(\theta)\,.
\end{equation}
Plugging \eqref{alphasr} into the second and third 
equations of \eqref{vactherm}, we get
\begin{align}
\label{dmeq1}
\Delta m_{Sr} = &\frac{1}{4}\alpha(\alpha-4)a^3\theta+{\cal
O}(\theta^2) = \left(\frac{1}{4}-a\right)a\theta+
\Or(\theta^2)\,,\\
\label{qeq1}
q = &2aq_c+{\cal O}(\theta)\,.
\end{align}
Eliminating $a$ from \eqref{dmeq1} and \eqref{qeq1}, we get
\begin{align}\label{mrange1}
\Delta m_{Sr} = &-\frac{2}{3}q\left(q-\frac{q_c}{2}\right)\theta
+{\cal O}(\theta^2)\,.
\end{align}
A plot of $\Delta m_{Sr}$ and $\Delta m_{ext}$ is shown in Fig
\ref{figure2}.
Note that for $q<q_c$, $\Delta m_{Sr} > \Delta m_{ext}$. In other words,
the RN AdS black holes reach their extremal limit before the
onset of superradiant instabilities and therefore, hairy black holes
do not exist for these values of charges. 

For $q\geq q_c$, $\Delta m_{Sr} \geq \Delta m_{ext}$\footnote{The
inequality is saturated at $q=q_c$.}. For these values 
of charge $q$, the RN AdS black holes suffer from a superradiant instability before (i.e. at a higher mass than) 
they go extremal; the region of the phase
diagram between these two masses therefore 
hosts both RN AdS and hairy black holes.

\subsection{Perturbative analysis of hairy black holes}
\label{subsec:middleepert}

In \S\ref{subsec:middleenonint}, we have presented 
a thermodynamic model that makes 
the following predictions for small hairy black holes in the
theory with
$$e^2=\frac{32}{3}(1-\theta).$$
\begin{itemize}
\item[1.] Hairy black holes at charge ${\tilde q}$ exist only
when
$${\tilde q} \geq q_c \theta 
+{\cal O}(\theta^2),~~~~~q_c=\frac{1}{2} \sqrt{\frac{3}{2}}\,.$$ 
\item[2.] We study hairy black holes with charge ${\tilde q} 
\sim {\cal O}(\theta)$, and so define 
$q = \frac{{\tilde q}}{ \theta}$ (so that $q$ is of unit order
at small
$\theta$) . We also parameterize the mass $m$ of hairy black
holes by
\begin{equation}\label{mass} 
m= \theta \frac {q}{q_c} + \theta^2 \left( 
\frac{2}{3} q^2 + \Delta m \right).
\end{equation}
The thermodynamic model predicts that hairy black holes exist
only in the range
$$- \frac{2}{3}\left( q-\frac{q_c}{2} \right) \theta  + 
{\cal O}(\theta^2)\geq \Delta m \geq -\frac{7}{15} 
\left(q - q_c \right)^2 + {\cal O}(\theta).$$ 
Note in particular that $\Delta m$ is always of order unity or smaller
at small $\theta$. 
\item[3.]The lowest mass hairy black hole is roughly 
approximated as a non-interacting mix of an extremal RN AdS black hole
and a soliton. It has zero temperature, but non-zero entropy.
\item[4.] The value of the scalar field in this solution is 
is of ${\cal O}(\sqrt{\theta})$.
Hairy black holes are approximately a non interacting mix 
of a RN AdS black hole of radius $R\sim \sqrt{\theta}$, charge 
${\tilde q}\sim \theta$ 
and energy above extremality $\sim \theta^2$.
\end{itemize}

In this section we justify all these predictions by presenting a
perturbative construction of hairy black holes at small $\theta$ 
and charges of order $\theta$. 

\subsubsection{The mathematical problem} \label{subsubsec:mp}

In this subsection we describe the mathematical problem we
require to
solve in order to construct hairy black holes in our model. 

The equations of motion for the Lagrangian \eqref{sysaction} are\begin{align}
\label{scalareq}
D^2\phi &= 0\,,  \\
\label{maxwelleq}
\partial_\mu {F^{\mu}}_{\nu} &= ie\left(\phi^* D_\nu \phi - \phi
(D_\nu \phi)^*\right)\,, \\
\label{einsteineq}
R_{\mu\nu} - \frac{1}{2}g_{\mu\nu}R - 6g_{\mu\nu} &=
T^{EM}_{\mu\nu} + T^{mat}_{\mu\nu}\,, 
\end{align}
where 
\begin{equation}
\begin{split}
T^{EM}_{\mu\nu} &=
-\left(F_{\mu\sigma}{F^{\sigma}}_\nu\right)+\frac{1}{4}g_{\mu\nu}F_{\rho\sigma}F^{\rho\sigma}\,, 
\\
T^{mat}_{\mu\nu} &=
(D_\mu\phi)(D_\nu\phi)^*+(D_\mu\phi)^*(D_\nu\phi) -
g_{\mu\nu}|D_\mu\phi|^2\,. \nonumber
\end{split}
\end{equation}
We are interested in stationary, spherically symmetric 
solutions of the system described by \eqref{sysaction}. As explained in \S\ref{subsec:setupsoliton} we adopt
a
Schwarzschild-like gauge \eqref{eqn:ansatz}. The four unknown functions $f(r)$, $g(r)$, $A(r)$ and $\phi(r)$
are
constrained by Einstein's equations, the Maxwell equation and
the
minimally coupled scalar equation. With this ansatz, the
equations of motion boil down to the equations \eqref{eqn:eoms}. It was shown in \cite{Basu:2010uz}
that these equations admit a 6 parameter set 
of solutions. One of the solutions is empty $AdS_5$  space, 
given by $f(r)=r^2+1$, $g(r)=\frac{1}{1+r^2}$, $A(r)=\phi(r)=0$.
We are interested in those solutions that asymptote 
to AdS spacetime, i.e. solutions whose large $r$ behavior is as in eq. \eqref{eqn:asymp},
\begin{equation}\label{gCondns} \begin{split}
f(r)&=r^2+1+{\cal O}(1/r^2)\,, \\
g(r)&=\frac{1}{1+r^2} + {\cal O}(1/r^6)\,, \\
A(r)&={\cal O}(1) + {\cal O}(1/r^2)\,,  \\
\phi(r)&= {\cal O} (1/r^4)\,.\\
\end{split} 
\end{equation}
It turns out imposing these conditions on $f$ and $\phi$ 
eliminates 2 of the 6 solutions so that the system of equations
admits a
four parameter set of asymptotically AdS  solutions. 
We will also be interested in solutions that are regular in the
interior.
In particular we demand that the functions $f$ and $\phi$ do not
blow up anywhere apart from at the singularity of the solution
(so
that our solution has a smooth event horizon).
These two requirements generically cut down solution 
space to distinct classes of two parameter solutions 
In this section we will determine the two parameter family of
hairy black hole solutions at small $\theta$. 

\subsubsection{Parametrization of solutions}

In the next subsection we will describe a perturbative
construction
of hairy black hole solutions at small $\theta$. In this
subsection
we will specify how the perturbative solutions we construct are
parameterized.
We use the parameters $k$ and $a$ to label solutions, as we now
explain.

$k$ is defined by the requirement that
\begin{equation}\label{kdef}
\phi(r)= \frac{\sqrt{k \theta}}{r^4} + {\cal O}(1/r^6)
\end{equation} at large $r$.

$a$ parametrizes the radius of our hairy black holes. Let $R$
denote
the radial location (in the coordinate system of
\eqref{eqn:ansatz}) of the
event horizon 
of the hairy black hole\footnote{More invariantly $R$ is defined
so that the area of the event horizon is given by $A_H
=2\pi^2R^3$.}.
The parameter $a$ is then defined by the relation
\begin{equation}
 \label{adef}
R^2 = a\theta\,.
\end{equation}

We wish to construct hairy black holes with $k$ and $a$ taken 
to be of order unity, in a perturbative expansion in
$\theta$. As is
clear from the previous section (and as we 
shall see below), the solutions we will construct are
approximately a
non interacting mix of an RN AdS black hole and a soliton. 
The RN AdS component of this mix has radius of order
$\sqrt{\theta}$,
charge and mass of order $\theta$. 
The soliton component of this mix 
has charge and mass both of order $\theta$. 

The hairy black hole is constructed in a perturbation expansion
in
$\sqrt{\theta}$. The starting point of this expansion is an
RN AdS black
hole labeled by $a$ and $\alpha$ (see subsection
\ref{subsubsec:bhtherm}).
This solution is 
then perturbed by a scalar condensate in order to meet the
requirement
\eqref{kdef}. It turns out that the three parameters 
$k$, $a$ and $\alpha$ cannot be independently varied. The
requirement of
regularity of our solutions fixes $\alpha$ as a 
function of $a$ and $k$. To leading order, the relation so determined 
is in precise agreement with the results of the non interacting 
model of \S\ref{subsec:middleenonint}, justifying the non interacting 
model.

\subsubsection{Nature of the perturbative
expansion}\label{subsubsec:natureofpert}

In this subsection we briefly explain the nature of the 
perturbation theory we employ in order to construct hairy black 
hole solutions in a perturbative expansion in $\theta$. Many 
more details of our procedure, together with a detailed listing 
of all our results are presented in Appendix \ref{appert}.

$\sqrt{\theta}$ parametrizes
the amplitude of the scalar field perturbation about the RN AdS
black
hole that constitutes the starting point of our perturbative
expansion.
However $\theta$ also appears as a parameter that governs the
size of the
starting RN AdS solution. 

 In the `far field region' $r \gg R=\sqrt{a \theta}$ 
where the RN AdS solution is a small perturbation about AdS
space, the expansion
of our solution in $\theta$ is a standard expansion 
in the amplitude of the scalar perturbation about unperturbed
AdS space.
When $r \sim \sqrt{\theta}$ the starting solution of perturbation
theory, i.e. the RN AdS black hole, 
depends on $\theta$ in a crucial way. However it turns 
out that the dependence of the RN AdS solution on $\theta$ in
this region is
rather simple, and may be scaled out of a problem with an
appropriate
choice of coordinates and variables. As a consequence, in this
region as
well, the expansion of the hairy black hole solution in $\theta$
is
simply an expansion in the amplitude of the scalar perturbation,
albeit in
the appropriate coordinates.

In summary, the perturbative 
expansion of the hairy black hole solution in $\theta$ is
everywhere an expansion in the
amplitude of the scalar field, though we need to employ
different variables
in different regions in order to make this apparent and useful.
In order
to practically implement our perturbative expansion, we found it
useful to
work separately in three regions. In the far field region $r \gg
R$ we
worked with the usual AdS radial variable $r$. In order to
describe the
other regions of interest to the perturbative analysis, we need
to introduce
some new notations. Let $R_{in}$ denote the inner horizon radius
of our RN AdS
black hole, while $R$ is its outer horizon radius. In the
parameter regime of
interest to this paper it turns out that 
$R-R_{in} \sim \frac{1}{2}\zeta R^3 \sim \theta^{3/2}$
\footnote{\label{zetadef}where $\zeta = 2-\alpha_0$ is independent of $\theta$.}
(i.e. the black holes we study are very near extremality). 
The intermediate field region is
defined by
the condition $r \ll 1$ but $r-R \gg\zeta R^3$. In this region our
solution
admits an expansion in $\theta$ when expressed as a function of
the scaled
variable $y=\frac{r}{R}$. In the so called near field region, $r-R
\ll R-R_{in}$,
on the other hand, our solution admits an amplitude expansion in
$\theta$
when expressed as a function of the variable
$z=\frac{r-R}{R-R_{in}} = \frac{r-R}{\zeta R^3}$.
In the small $\theta$ limit of interest to this subsection, the
far field
and intermediate field regions share a substantial overlap.
Similarly
the intermediate and near field regions overlap substantially.
For
this reason we are able to construct the full solution by
working individually
in the three region and then patching the full solution up using
a matching
procedure (following \cite{Basu:2010uz} and
\cite{Bhattacharyya:2010yg}).

In Appendix \ref{appert} we have discussed the general
structure of our perturbative
expansion in some detail. In subsection \ref{ap:setup} of that
Appendix we have
described in detail the general method employed in our
perturbative expansion.
In subsections \ref{ap:leadordpertmethod} and
\ref{ap:ordonepertmethod}
 we have described the explicit 
implementation of this procedure at order $\sqrt{\theta}$ and
$\theta$
respectively. Finally in subsection \ref{ap:middlee} we present
a detailed listings of the final solutions of our hairy black
holes to
${\cal O}(\theta^{3/2})$ in each of the far field, intermediate
field
and near field regions. In the rest of this subsection we
compute the
thermodynamics of the solution presented in subsection
\ref{ap:middlee}.

\subsubsection{Thermodynamics}\label{subsubsec:therminmeless}

It is a straightforward exercise to compute the 
thermodynamical charges and potentials of the black holes
presented in
Appendix \ref{ap:middlee}. For convenience we work with the
rescaled mass and charge 
variables $\tilde{q}$ and $m$ defined and employed  in \cite{Basu:2010uz}
\begin{equation}\label{mqdef}
 \begin{split}
  Q &= \frac{\pi}{2} \tilde{q}\,,  \\
M &= \frac{3\pi }{8}m\,.
 \end{split}
\end{equation}
The formulas we have employed to compute the mass and charge and
other
thermodynamic quantities from the solutions of Appendix
\ref{ap:middlee} are
\begin{equation}\label{thermform}
\begin{split}
\mu &= \lim \limits_{r\rightarrow\infty}\left[ A^{out}(r)\right],
\\
m &= \lim \limits_{r\rightarrow\infty}\left[
r^2\left(1+r^2-f^{out}(r)\right)\right]\,,  \\
\tilde{q} &= \lim \limits_{r\rightarrow\infty}\left[
r^2\left(\mu-A^{out}(r)\right)\right]\,,  \\
T &= \frac{1}{4\pi\zeta
\left(a\theta\right)^{3/2}}\frac{d}{dz}f^{in}(z)\bigg|_{z=0}\,,  \\
S &= \frac{\pi^2\left(a\theta\right)^{3/2}}{2}\,.
\end{split}
 \end{equation}
(see Appendix \ref{ap:setup} for a definition of the functions
$f^{out}$, $A^{out}$ and $f^{in}$). Employing these formulae we find 
\footnote{Throughout this paper, we follow \cite{Basu:2010uz}
and
consistently omit the factor 
of $G_5^{-1}$ from all our extensive quantities}
\begin{equation}\label{actualtherm}
 \begin{split}
M &= \frac{3\pi}{8}\bigg[\theta \left(2 a+\frac{8
k}{9}\right)+\theta ^2 \left(a^2+\frac{1024 a
k}{189}+a+\frac{9232
   k^2}{59535}\right) \\
&+\theta ^3 \left(3 a^3-\frac{32}{3} a^2 k 
\log \left(a\theta^2\left(2 a-\frac{8k}{21}-1\right)\right)+a^2
\left(-\frac{332888 k}{19845}-3\right) \right. \\
&\left.+a\left(\left(\frac{331396792}{22920975}-\frac{512 \pi
^2}{567}\right) k^2+\frac{680 k}{63}+1\right)+\frac{16
k^2 (7554439 k+200607715)}{1764915075}\right)\bigg]+{\cal
O}\left(\theta ^4\right)\,, 
\\
Q = &\frac{\pi}{4}\sqrt{\frac{3}{2}}\bigg[\theta \left(2
a+\frac{8 k}{9}\right)+\theta ^2 \left(\frac{4}{189} (256 a-21)
k+a+\frac{4192
   k^2}{59535}\right) \\
&+\theta ^3 \left(3 a^3-\frac{32}{3} a^2 k \log
\left(a\theta^2\left(2 a-\frac{8k}{21}-1\right)\right)+a^2
\left(-\frac{359348 k}{19845}-3\right)+\right. \\
&\left.a \left(\frac{4
k \left(\left(76099378-5174400 \pi ^2\right)
k+53967375\right)}{22920975}+\frac{3}{4}\right)\right. \\
   &\left.+\frac{893786896
k^3}{15884235675}+\frac{5776
k^2}{2835}-\frac{k}{9}\right)\bigg]+{\cal O}\left(\theta
^4\right)\,, \\
\mu = &\sqrt{\frac{3}{2}}+\frac{(8 k+21) \theta }{14
\sqrt{6}}+\frac{\theta ^2 \left(11088 (1642 a-525) k+4584195
(2a-1)^2+135136 k^2\right)}{4074840 \sqrt{6}}+O\left(\theta
^3\right)\,,  \\
T = &\frac{\sqrt{\theta } (42 a-8 k-21)}{42 \pi  \sqrt{a}} \\
&+\frac{\theta ^{3/2} \left(-2772 (172 a+315) k-1528065 (3
(a-1) a+1)+189208 k^2\right)}{3056130 \pi
\sqrt{a}}+O\left(\theta ^{5/2}\right)\,, \\
S = &\frac{\pi^2(a\theta)^{3/2}}{2}
 \end{split}
\end{equation}
We have verified that these quantities obey the 
first law of thermodynamics
$$dM = T dS + \mu dQ\,.$$

\subsubsection{Comparison with the thermodynamics of the non-interacting model}
\label{subsubsec:leadordtherm}

In this subsection, we demonstrate 
that the thermodynamical formulae \eqref{actualtherm}
reduce to \eqref{qeq} and \eqref{meq} at leading order. 
Using \eqref{actualtherm}
and definitions \eqref{qdef} and \eqref{dmdef}, we can calculate
$q$ and $\Delta m$ for this system. Upto the order of interest in this
section,
\begin{equation}\label{qandm0}
\begin{split}
q &= \frac{(9 a+4 k)}{3 \sqrt{6}}
+{\cal O}\left(\theta\right)\,,  \\
\Delta m & =-\frac{4k}{567} (126 a+16 k-63)+\Or(\theta)\,,  \\
T &=\frac{ (42 a-8 k-21)}{42 \pi \sqrt{a}}\sqrt{\theta } +{\cal
O}(\theta^{3/2}).
\end{split}
\end{equation}
In order to make contact with the results of \S\ref{subsec:middleenonint},
we need to identify the parameter $k$ defined in \eqref{kdef} with the 
parameter $b$ of \S\ref{subsec:middleenonint}. We do this as follows.

The pure soliton with 
$$\phi(r) = \frac{\sqrt{k\theta}}{r^4}+\Or(1/r^6),~e^2 = \frac{32}{3}(1-\theta)$$
has charge $$\tilde{q}_{sol} = \frac{k}{3q_c}\theta+\Or(\theta^2)$$ (see \cite{Basu:2010uz}).
From the definition of $b$ (see \eqref{soltherm}), we immediately conclude $b = \frac{k}{3q_c}$. 
Written in terms of $a$ and $b$ the formulae \eqref{qandm0} are modified to
\begin{equation}
 \begin{split}
  q &= (2aq_c + b)+\Or(\theta)\,,  \\
\Delta m &= \left(\frac{b}{2q_c} - \frac{a b}{q_c} - \frac{8b^2}{21}\right)+\Or(\theta)\,.
 \end{split}
\end{equation}
These are exactly the formulae \eqref{qeq} and \eqref{meq} of 
the non-interacting model.

At leading order, the hairy black hole can therefore, be thought 
of as a non-interacting mix of an RN AdS black hole and a soliton.

We now compute the equation for the line denoting the onset of 
the superradiant instability at one higher order in $\theta$. 
This line is obtained by setting $k=0$ in \eqref{actualtherm}. 
To the order of interest (in terms of $q$ and $\Delta m$) we have
\begin{equation}\label{qanddmatnextorder}
 \begin{split}
q &=  2aq_c+\Or(\theta^2)\,,  \\
\Delta m &= \left(\frac{1}{4}-a\right)a\theta+\Or(\theta^2)\,.
\end{split}
\end{equation}
Eliminating $a$ from \eqref{qanddmatnextorder}, we get
\begin{equation}
\Delta m = -\frac{2}{3}q\left(q-\frac{q_c}{2}\right)\theta
+{\cal O}(\theta^2)
\end{equation}
as predicted by \eqref{mrange1} in \S\ref{subsubsec:nextord}.

\subsubsection{Break down of perturbation theory very near
extremality}

Note that our spectrum of hairy black holes includes a one
parameter set of extremal hairy black holes; at leading order this occurs
when
$42 a-8 k-21= 21a\zeta = 0$ (see \eqref{actualtherm})
\footnote{$\zeta = 2-\alpha_0 = \frac{42a-8k-21}{21a}$
(see footnote \ref{zetadef}).} 
and in particular when $\zeta =0$.
While all thermodynamical formulae in
\eqref{actualtherm}
are smooth in this extremal limit, the solutions themselves
develop
a singularity, as we now explain. 

The small hairy black hole solutions we have constructed in
Appendix \ref{appert}
are regular (away from the black hole singularity). In
particular the value
of the scalar field $\phi$ is finite everywhere 
outside the outer event horizon; the value of the scalar field
at
the event horizon is of order $\sqrt{\theta}$ Strictly at
extremality,
however, the scalar field 
$\phi$ diverges at the event horizon (see \eqref{nfexpr}). 
As our perturbative
expansion is essentially an expansion in the amplitude of the
scalar field,
this divergence indicates a breakdown of our perturbative
techniques
(and so in particular is not a self-consistent and reliable prediction of our
perturbation theory).
It really indicates only that our perturbative construction of
hairy black
holes does not apply all the way down to extremality.

How near to extremaliity does perturbation theory break down? 
We can examine the breakdown in perturbation theory in more
detail
as follows. As is clear from \eqref{actualtherm}, 
all hairy black holes constructed in this paper have $T \sim
\sqrt{\theta}$.
The parameter $t=\frac{T}{\sqrt{\theta}}$ characterizes the
departure of our
hairy black holes from extremality. 
$$t = \frac{T}{\sqrt{\theta}} =
\frac{\sqrt{a}\zeta}{2\pi}\,.$$
It turns out (see  \eqref{nfexpr}), that at very small 
values of $t$ the value of the scalar field at the horizon is
proportional
(at ${\cal O}(\theta^{3/2})$ in the perturbative expansion)  to
$$-\frac{4a}{k}\theta^{3/2}\ln t .$$ 
This term becomes comparable to the scalar field at leading
order (signalling
a breakdown of the perturbative expansion) when 
\begin{equation}\label{val}
 \begin{split}
-\frac{4a}{k}\theta^{3/2}\ln t &\sim \theta^{1/2} \\
t &\sim  e^{-k/(4a\theta)}  \,.
 \end{split}
\end{equation}
Consequently perturbation theory breaks down at exponentially
low values of
the temperature. Our perturbative construction of hairy black
holes, and
so the thermodynamical formulae \eqref{actualtherm} and the
results of the
non interacting model, all strictly apply only above the
temperature
\eqref{val}. In particular our perturbative results cannot
reliably be
used to study extremal hairy black holes, although we can get
exponentially near
to these solutions at small $\theta$.

\subsection{Numerical results}
\label{subsec:middleenum}

\subsubsection{Method}

In this subsection we construct the hairy black holes at the full non-linear level by numerically solving the equations of motion. Our ansatz for the metric, gauge field and scalar field are the same as in Eq.\eqref{eqn:ansatz},  and therefore the equations of motion are given by \eqref{eqn:eoms}. We shall require that the asymptotic behaviour of the various fields is the same as in the soliton case, which is given in Eq.\eqref{eqn:asymp}. Of course, the main difference now is that we have to require that the spacetime has a (regular and \textit{non-degenerate}\footnote{As mentioned in the Introduction, regular extremal hairy black holes are not allowed in the system under study \cite{FernandezGracia:2009em}.}) event horizon; this implies that $f(r)$ has a simple zero at $r=R$. The parameter $R$ then  determines the radius of the horizon. Furthermore, we choose a regular gauge for the gauge field $A$ on the horizon, such that $A_t(r=R)=0$. 

To implement the boundary conditions we find it useful to introduce a compact radial coordinate, $y=1-\frac{R^2}{r^2}$ and redefine the functions $f$, $\phi$ and $A$ as follows,
\begin{equation}
\begin{aligned}
f(r)&=\left(1+r^2\right)\left(1-\frac{R^2}{r^2}\right)p_f(y)\,, \\
\phi(r)&=\epsilon \left(1+r^4\right)^{-1}p_\phi(y)\,,\\
A(r)&=\left(1-\frac{R^2}{r^2}\right)p_A(y)\,,
\end{aligned}
\end{equation}
with $r=\frac{R}{\sqrt{1-y}}$. The boundary conditions that we shall impose are:
\begin{equation}
p_f(1)=1\,,\qquad p_\phi(1)=1\,,\qquad p_\phi'(1)-\frac{1}{12}\left(8+e^2\,p_A(1)^2\right)=0\,,
\end{equation}
at infinity and 
\begin{equation}
\begin{aligned}
p_f'(0)&=\Big[-6 p_f(0) \left(1+R^2\right) \left(1+R^4\right)^2 \left(3 p_f(0) R^4-2 p_A(0)^2 \left(1+3 R^2\right)\right)\\
&~~~~~+ e^2\epsilon^2p_\phi(0)^2 p_A(0)^2 R^2 \left(2 p_A(0)^2+3 p_f(0) \left(1+R^2\right)\right) \Big]\\
&~~~\times\left[18 \left(1+R^2\right)^2 \left(p_f(0)+2 p_f(0) R^2\right) \left(1+R^4\right)^2\right]^{-1}\,,\\
p_\phi'(0)&=\frac{2 p_\phi(0) R^4}{1+R^4}\,,\\
p_A'(0)&=\frac{e^2 \epsilon^2p_\phi(0)^2 p_A(0)  R^2 \left(2 p_A(0)^2+3 p_f(0) \left(1+R^2\right)\right)^2}{36 p_f(0)^2 \left(1+2 R^2\right) \left(1+R^2+R^4+R^6\right)^2}\,.
\end{aligned}
\end{equation}
The latter boundary conditions follow from simply solving the equations of motion in a near-horizon expansion. 

We construct, in the present range $3<e^2<\frac{32}{3}$, hairy black holes whose radius is not too small compared to the radius of $AdS_5$, but we find no evidence for the existence of very small hairy black holes. This is in agreement with the previous perturbative analysis. Instead we find that the extremal limit of the hairy black holes is singular. It does \textit{not} coincide with the soliton branch studied, as we also predicted from the perturbative treatment.

\subsubsection{Results}

\begin{figure}[h]
\begin{center}
\includegraphics[scale=0.46]{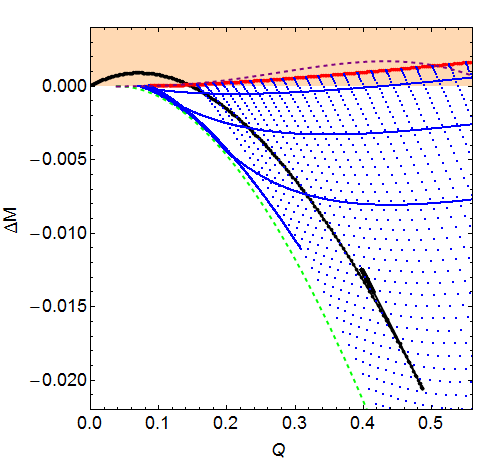}
\includegraphics[scale=0.46]{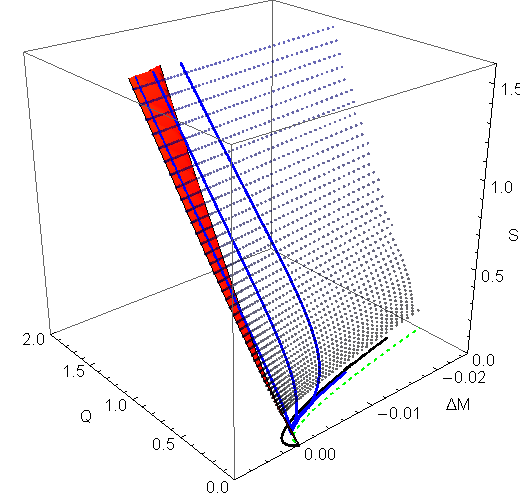}
\end{center}
\caption{\small{ {\it Left:} $\Delta M$ vs. $Q$ for $e=3.2$. The shaded region is the area occupied by RN AdS black holes, the black curve corresponds to a soliton branch 
and the blue region corresponds to hairy black holes (the `horizontal' lines of the blue grid correspond to fixed values of $\epsilon$, and $R$ decreases to the left; the `diagonal' blue lines near the green curve are segments of a hairy black hole with fixed $R$, and $\epsilon$ grows to the right). The lower mass bound of hairy black holes is well described at small charge by the dashed green line, which is the perturbative prediction. The red curve is the line of marginal modes of the linear problem discussed in section \ref{subsec:linear}; it agrees with the dashed magenta line for small charge, which is the perturbative prediction. It is clear that the soliton curve and the hairy black holes surface are not related in the range $3<e^2<32/3$. {\it Right:} same data with the entropy as third axis. The red surface is the strip of RN AdS black holes between the marginal line of stability and extremality.}}
\label{fig:e3p2_dM_Q}
\end{figure}

\begin{figure}[t]
\begin{center}
\includegraphics[scale=0.6]{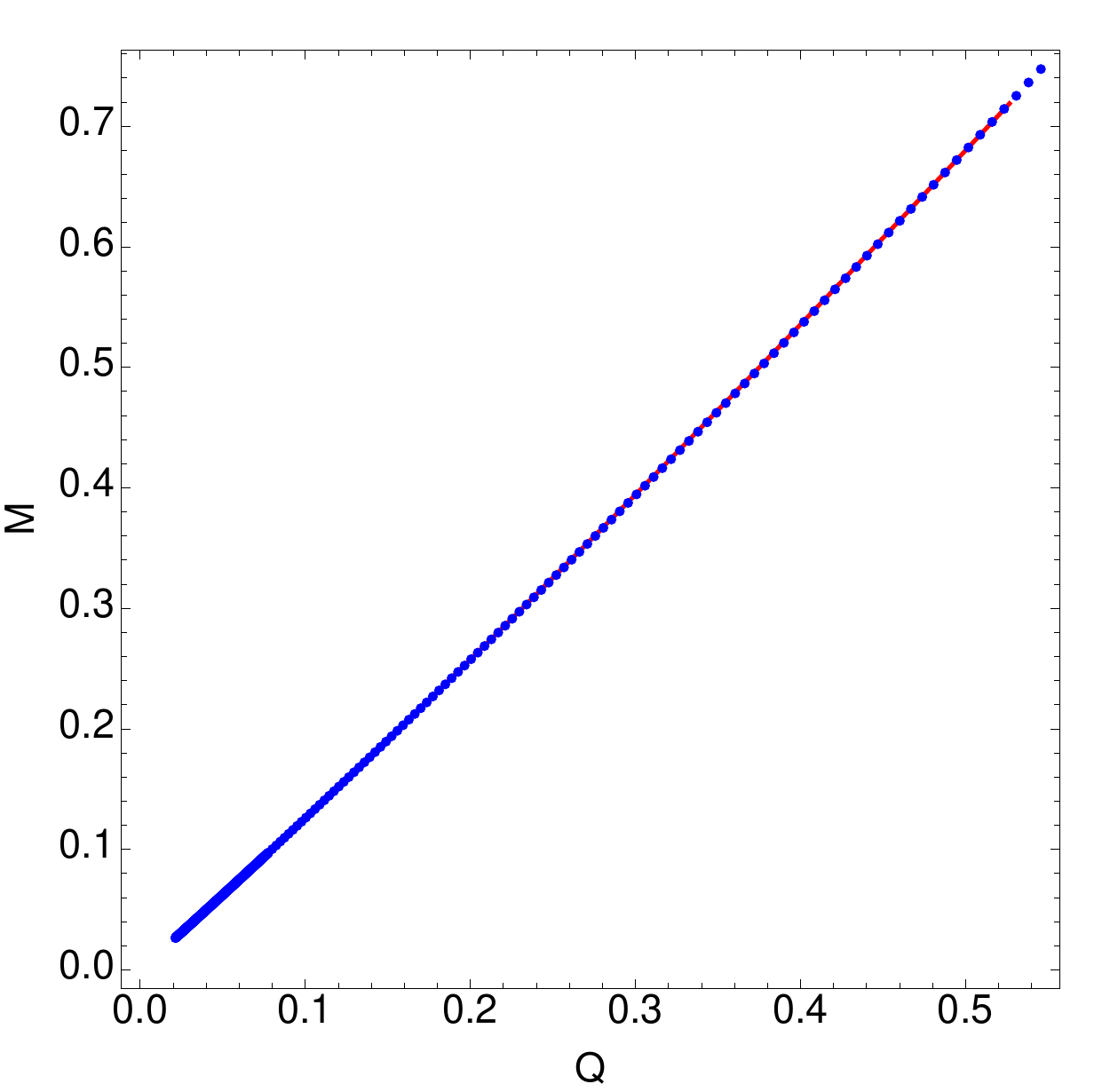}
\hspace{0.5cm}
\includegraphics[scale=0.6]{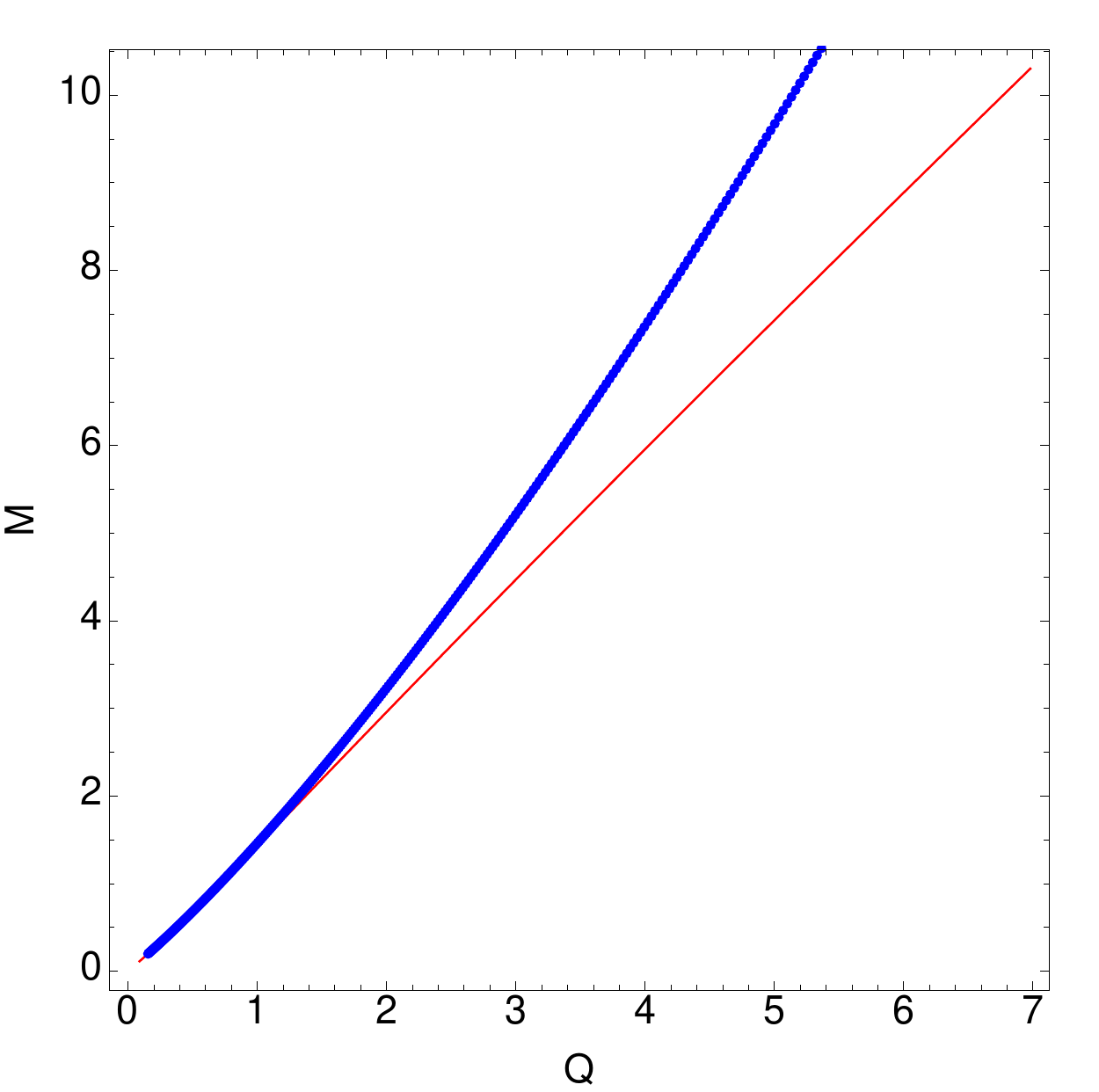}
\end{center}
\caption{\small{These plots $M$ vs. $Q$ represent the agreement of the numerical data (blue lines, made of data points) with the perturbative analysis (red lines) for small charge. \textit{Left:} Merger between hairy black holes and RN AdS black holes along the line of marginal stability for $\theta=0.01$. \textit{Right:} Hairy black holes along a line of constant boundary condensate $\epsilon=0.1$ for $e=3.2$.}}
\label{fig:nonlinearcomparasionthetam}
\end{figure}

The results are presented in Fig.~\ref{fig:e3p2_dM_Q}. On the left, we have plotted $\Delta M$ vs. $Q$ for $e= 3.2$ ($\theta=0.04$) for both the soliton and the hairy black holes; $\Delta M$ is the mass difference with respect to the extremal RN AdS black hole of the same charge. We have added the analytical predictions from the perturbation theory, based on \eqref{mass} and \eqref{actualtherm}, and there is good agreement for small charge as it should. On the right, we have also plotted the same data with the entropy as third axis. Other values of $e$ within the range $3<e^2<\frac{32}{3}$ give a qualitatively similar picture. As this Fig. shows, there are no hairy black holes with arbitrarily small charge. As an extra check of our results, we represent in Fig.~\ref{fig:nonlinearcomparasionthetam} hairy black holes along the merger line (upper mass bound) and along a line of constant boundary condensate $\epsilon=0.1$, comparing the numerical data with the prediction of \eqref{actualtherm} for small charge.

\begin{figure}[t]
\begin{center}
\includegraphics[scale=0.55]{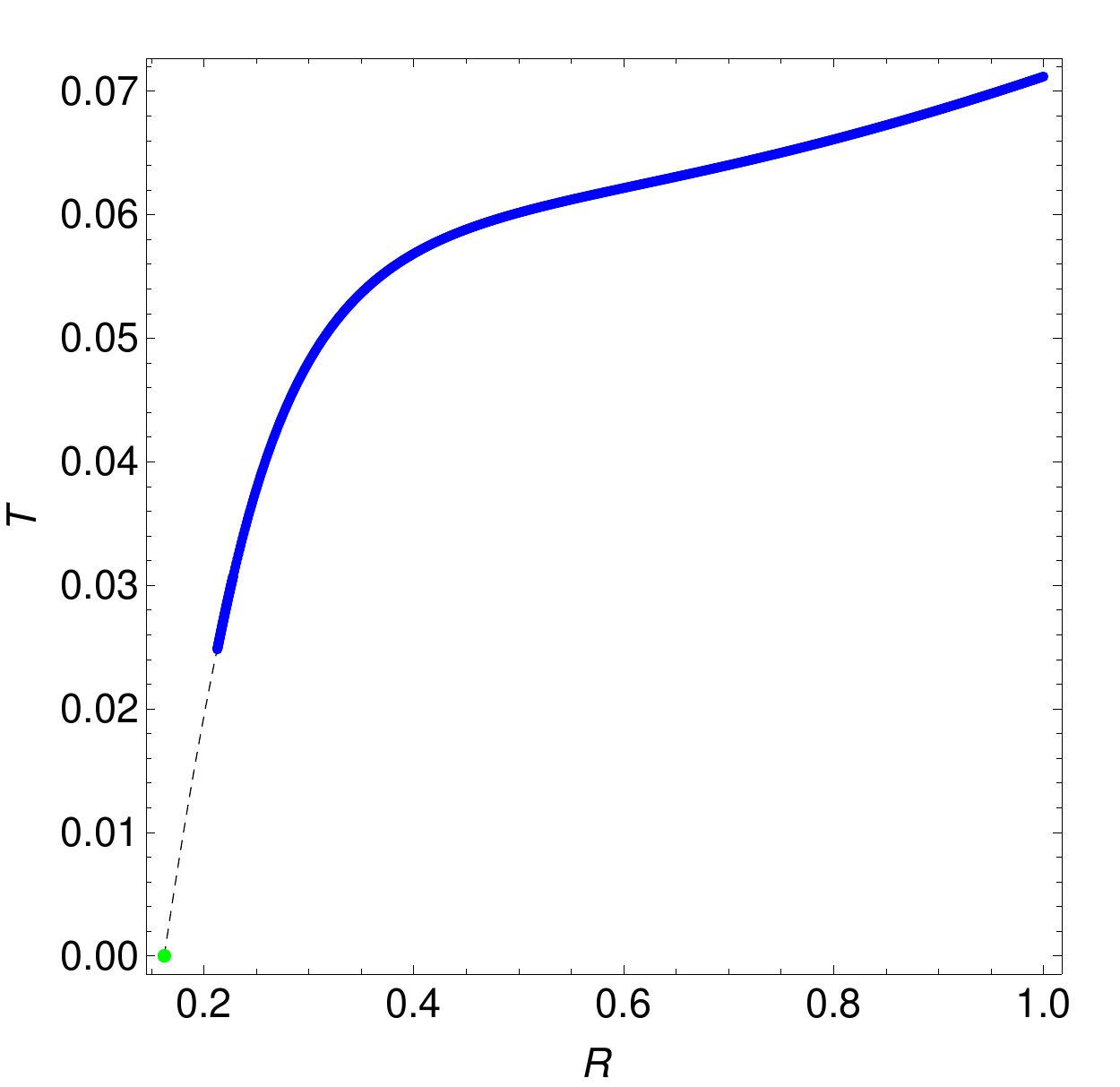}
\hspace{0.5cm}
\includegraphics[scale=0.55]{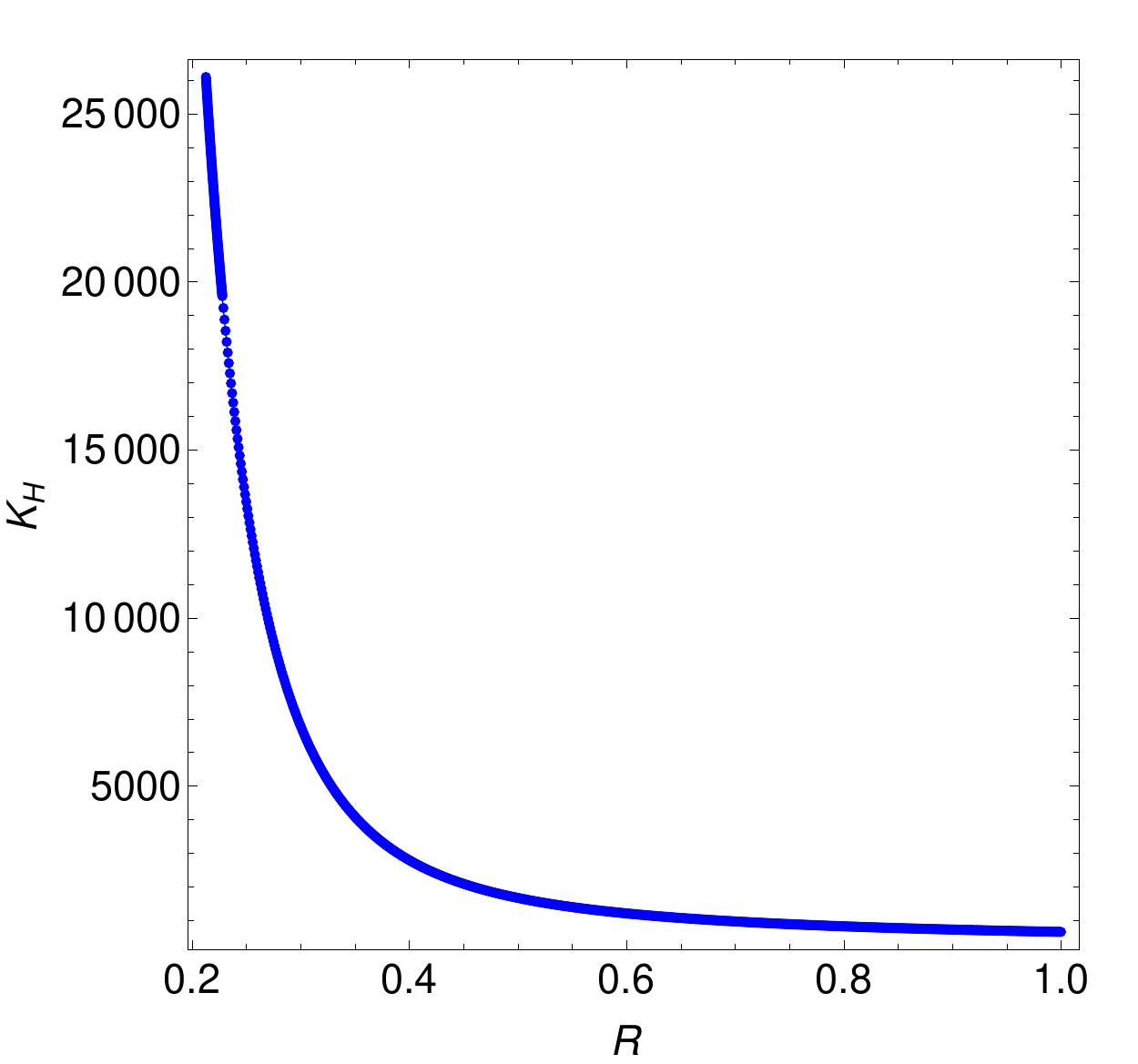}
\end{center}
\caption{\small{Temperature vs. $R$ \textit{(left)} and Kretschmann invariant at the horizon vs. $R$ \textit{(right)} for $e=3.2$ and a fixed $\epsilon=0.5$.  As the size of the black hole becomes smaller, the temperature decreases and it appears to become zero for $R\approx 0.162$. Simultaneously, the Kretschmann invariant seems to blow up as the black hole becomes smaller. Extrapolation of our data indicates that the blow up should occur for $R\approx 0.154$, which is close enough given the uncertainty. These results suggest that for $3<e^2<32/3$ hairy black holes posses an extremal \textit{singular} limit.}}
\label{fig:e3p2zero_rp}
\end{figure}

It is clear that the soliton family does not arise as a zero size limit of the hairy black hole. To investigate this, we plotted the temperature $T$ and the Kretschmann invariant evaluated at the horizon, as a function of the horizon size, keeping $\epsilon=0.5$ fixed; see Fig.~\ref{fig:e3p2zero_rp}. The plots suggest that the minimum size of hairy black holes corresponds to an extremal (i.e. zero temperature) limit which is singular. Since very close to extremality our numerics are not reliable, we have used extrapolation to estimate the value of $R$ for which the black hole would have zero temperature, and we find that $T=0$ for $R\approx 0.162$. Similarly we find that the Kretschmann invariant would blow up for $R\approx 0.154$; which is close enough, given the extrapolation. Finally, as predicted before, notice that the solitons are more massive than the extremal hairy black holes of the same charge.

For other values of the scalar condensate $\epsilon$ in this intermediate $e$ range, the picture is qualitatively the same.

In this section we have described the construction of a hairy black hole 
solutions labelled, for instance, by their mass $Q$ and charge $M$. 
Let us now set $M=\zeta Q^{\frac{4}{3}}$. If $Q=L^3$ is taken to infinity at 
fixed $\zeta$ then we expect, on physical grounds that the limits  
\begin{equation}\label{sfp}
\begin{split}
g_P(\rho, \zeta)&= \lim_{L \to \infty} L^2\,g(\rho L)\\
f_P(\rho, \zeta)&=\lim_{L \to \infty} \frac{f(\rho L)}{L^2}\\
A_P(\rho, \zeta)&=\lim_{L \to \infty} \frac{A(\rho L)}{L}\\
\phi_P(\rho,\zeta)&=\lim_{L \to \infty} \phi(\rho L)\\
\end{split}
\end{equation}
exist (see \eqref{sf}) 
all appear to exist, so that the large charge hairy black hole 
may be rewritten in new coordinates $r=L\,\rho$, $t =\tau/L $ in the 
black brane form
\begin{equation} \label{scsop}
\begin{split}
ds^2& =-f_P(\rho, \zeta) \,d\tau^2+ g_P(\rho, \zeta)\,d\rho^2 + \rho^2\, 
(dx^i)^2\\
A&= A_P(\rho, \zeta)\, d\tau 
\end{split}
\end{equation}

\begin{figure}[t]
\begin{center}
\includegraphics[scale=0.55]{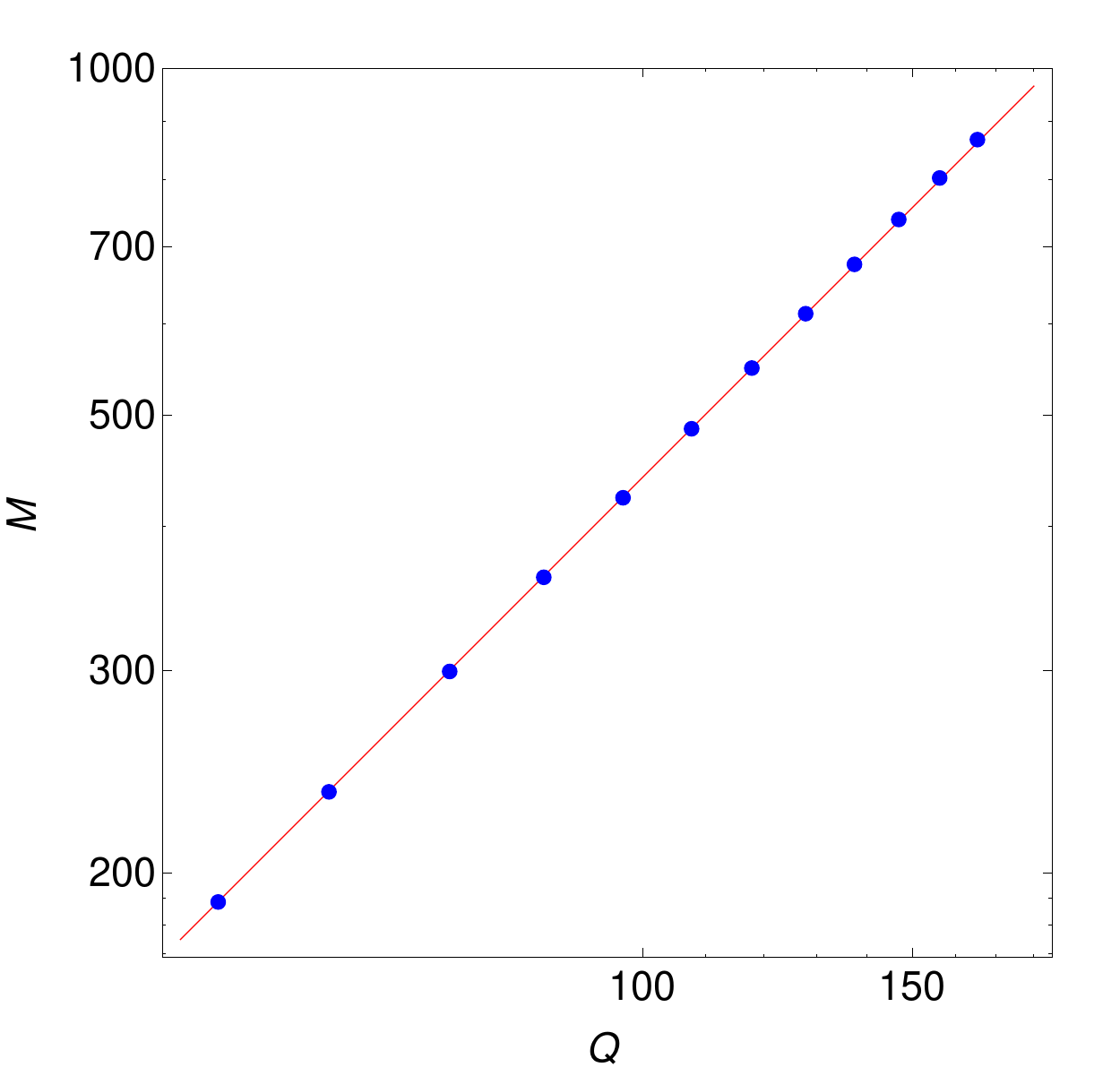}
\hspace{0.5cm}
\includegraphics[scale=0.55]{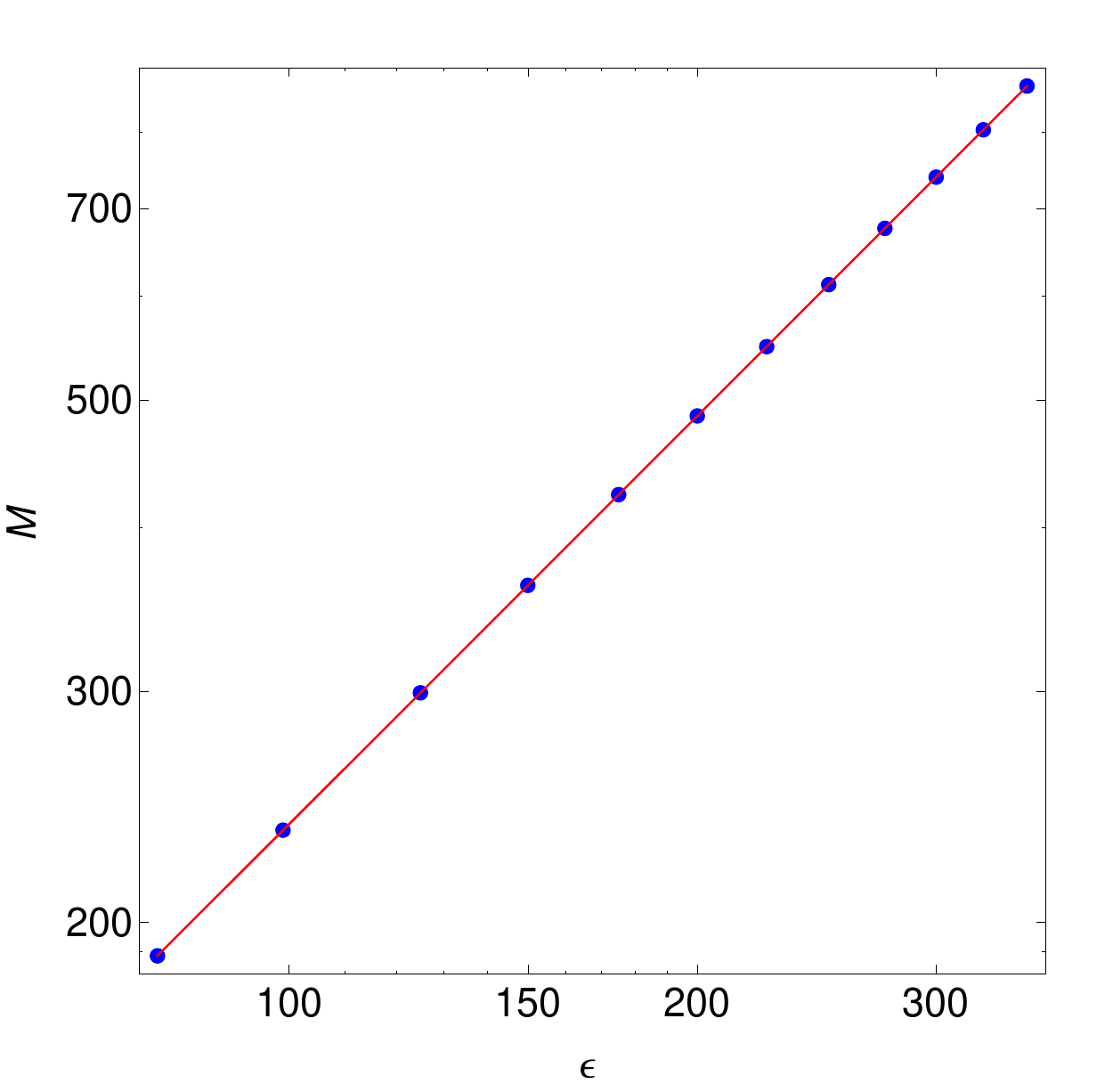}
\end{center}
\caption{\small{These plots $M$ vs. $Q$ \textit{(left)} and $M$ vs. $\epsilon$ \textit{(right)}, for $e=3.2$, represent our best estimate for where the extremal singular curve lies in the phase diagram, for large charges. In order to estimate the location of this curve, we take the points of smallest temperature for each line of constant $\epsilon$. The blue dots correspond to our numerical data and the red curves correspond to the best fit of our data to functions of the form $M = A \,Q^{4/3}$ and $M = B\,\epsilon^{C}$, respectively. We find that $A \simeq 1.0$, $B \simeq 2.0$ and $C \simeq 1.0$. These results roughly agree with the scaling found in Fig.~\ref{fig:scalingsolitonq3p2} for the second branch of the soliton.}}
\label{fig:e3p2scalingextre}
\end{figure}

We have numerical evidence that the limits \eqref{sfp} exist. We will leave the direct evidence for the next Section, where we will compare the planar limit of solitons and hairy black holes. For now, we verify the corresponding scaling of the mass with the charge and with the asymptotic scalar field. Working at the particular value $e=3.2$, in Fig 
\ref{fig:e3p2scalingextre} (left) we present a graph of $M$ versus 
$Q^{\frac{4}{3}}$, 
where $M$ represents the mass of the lightest black hole we have been able 
to construct at charge $Q$. In Fig. \ref{fig:e3p2scalingextre} (right) we 
present the plot of $M$ verus the scalar vev, $\epsilon$, for the same 
solutions. These curves are both rather close to straight lines, 
demonstrating that the scaling described above works, 
at least for extremal black holes. 

Assuming that the limits \eqref{sfp} do exist in general, 
\eqref{scsop} yield the one parameter set
of hairy black brane solutions labeled by $\zeta$. The entropy $S$ of 
these solutions must scale like $S=sL^3$, so that the ratio $S/L^3$ has 
a good limit. These solutions should reduce to the black branes 
studied by Hartnoll \emph{et al.} \cite{Hartnoll:2008vx,Hartnoll:2008kx}. We argued before that the same should be valid for the scaling limit of the second branch of the soliton; see Fig.~\ref{fig:scalingsolitonq3p2}. Indeed, the scaling of the mass, charge and scalar vev match roughly (notice that the large charges required to find the scaling become problematic for the numerics).

\section{Hairy black holes for  $e^2>32/3$}
\label{sec:largee}

In this section we will study the solitonic and hairy black hole
solutions
of \eqref{sysaction} in the parameter range $e^2>\frac{32}{3}$.
Let us
first describe solitonic solutions. These solutions were
constructed
analytically at small charge (in a power series expansion in the
charge)
in \cite{Basu:2010uz}. In this paper we use numerical techniques
to construct the
solitonic solutions at all values of the charge. It appears to
be the case
that at all values of $e^2>\frac{32}{3}$, solitonic solutions
exist at arbitrary charge (i.e. they do not hit a Chandrashekhar
instability
at finite charge). Intuitively, for this range of parameters,
Maxwell
repulsion balances gravitational attraction without leading to
collapse
at the core. 

Let us now turn to the study of hairy black holes. In this range
of parameters
hairy black holes of charge $\tilde{q}$ exist at all values of
$\tilde{q}$. In \cite{Basu:2010uz},
the small $\tilde{q}$ limit of these black holes was studied at
fixed values of
$e^2$. In this section we supplement 
the study of \cite{Basu:2010uz} with a related but distinct
perturbative construction
of hairy black holes. We construct hairy black holes 
with charge $\sim \theta$, in the theory with
$e^2=\frac{32}{3}(1+\theta)$.
In other words we construct hairy black holes whose charge is
scaled to
zero in a manner coordinated with the scaling of $e^2$ to
$\frac{32}{3}$,
in close analogy to our construction of hairy black holes in the
previous
section. 

It turns out that the nature of hairy black holes in this range
of parameters
undergoes a qualitative change when their charge exceeds a
critical value. 
Below this critical value of charge $\tilde{q}_{c_2}$, the hairy black hole at its lowest 
mass ends on the soliton, as was predicted by \cite{Basu:2010uz}.
For charges greater than $\tilde{q}_{c_2}$, hairy black holes at lowest mass
are extremal (similar to the behaviour of hairy black holes
when $3<e^2<\frac{32}{3}$).
When $e^2=\frac{32}{3}(1+\theta)$ this critical value 
turns out to be of order $\theta$, and so is reliably captured
by our new
small $\theta$ perturbation theory, though it lies outside the
range of
the perturbative expansion of \cite{Basu:2010uz}.

The results of our new perturbative expansion, together with the
results of
\cite{Basu:2010uz} and certain other considerations, suggest a
qualitative picture of
the spectrum of hairy black holes at all values of ${\tilde q}$. In the
rest of this
section we  then verify and fill in the details of 
this picture with the aid of extensive numerical simulations.

\subsection{Non-interacting model for small hairy black holes
with
$e^2=\frac{32}{3}(1+\theta)$}
\label{subsec:largeenonint}

In this section, we predict the leading order thermodynamics of
small hairy black holes for
\begin{equation}
e^2 = \frac{32}{3}(1+\theta)
\end{equation}
where $\theta$ is a small positive parameter.

We follow the methods of \S\ref{subsec:middleenonint} to analyze
these hairy black holes and define
\begin{equation}
\begin{split}
 R^2 &= a\theta \,\\
\mu^2 &= \frac{3}{2}(1+\alpha a\theta)\,\\
\tilde{q}_{sol} &= b\theta\,  
\end{split}
\end{equation}
where $a$, $\alpha$ and $b$ are positive numbers of order unity.
In terms of these
quantities, the thermodynamical expressions for the black hole
and soliton are given by
(modified versions of \eqref{vactherm} and \eqref{soltherm})
\begin{equation} \begin{split}\label{vactherm1}
\mu_{BH}&=2q_c \left( 1+ \frac{\alpha a \theta}{2} 
-\frac{\alpha^2 a^2 \theta^2}{8} \right) + {\cal O}(\theta^3)\,,\\
m_{BH}&=a \theta \left( 2+ (1+\alpha) a \theta \right)\,, \\
\tilde{q}_{BH}&= 2a q_c \left( 1+ \frac{\alpha a \theta}{2} 
-\frac{\alpha^2 a^2 \theta^2 }{8} \right)\theta + {\cal
O}(\theta^4)\,,\\
T&=\frac{(2-\alpha)}{2 \pi} \sqrt{a \theta}\,,\\
S_{BH}&=\frac{\pi^2 (a \theta)^{3/2} }{2}\,, \\
\end{split}
\end{equation}
and 
\begin{equation} \begin{split}\label{soltherm1}
\mu_{sol}&= 2 q_c + \theta \left(-q_c + \frac{3b}{7} \right)
+{\cal O}(\theta^2)\,,\\
m_{sol}&= \frac{b}{q_c} \theta + \left( -\frac{b}{2q_c}+
\frac{2b^2}{7} \right)
\theta^2+{\cal O}(\theta^3)\,.
\end{split}
\end{equation}
Note again that $\alpha \leq 2$ (black holes
with $\alpha >2$ formally have negative temperature and are
unphysical;
black holes with $\alpha=2$ have zero temperature and are
extremal).

\begin{figure}[ht]
 \begin{center}
\includegraphics[height=75mm]{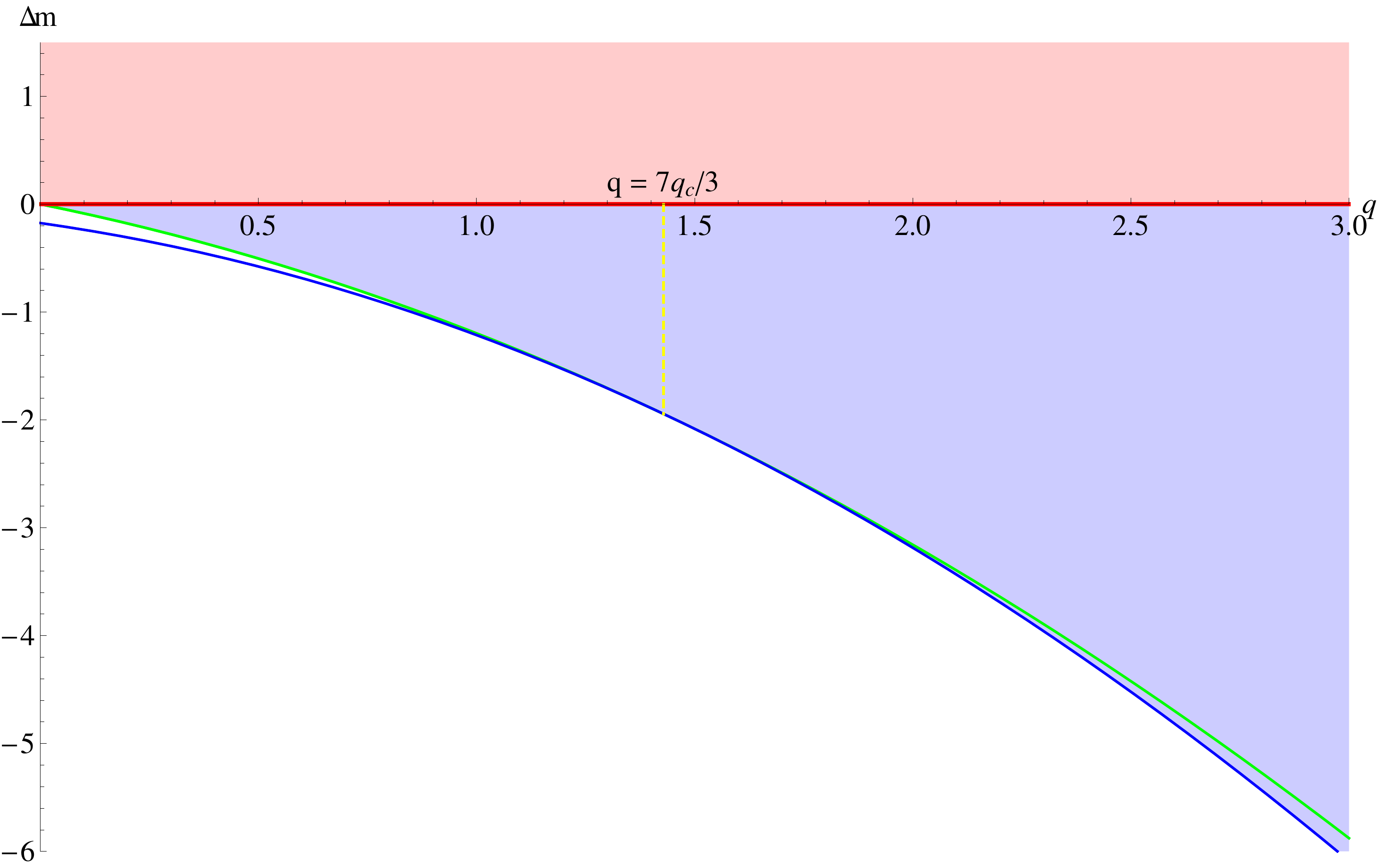} 
\end{center}
\caption{Range of allowed values of $\Delta m$ for hairy
black holes with $e^2 = \frac{32}{3}(1+\theta)$ at leading
order in small $\theta$ (blue shaded). Hairy black holes exist
for all charges $q\geq 0$. The upper limit, $\Delta m=0$ denotes
the onset of superradiant instabilities (red line). This is also
the extremality line for pure RN AdS black holes. The green line
denotes the soliton, $\Delta m =
-\frac{q}{2q_c}-\frac{8q^2}{21}$ and the blue line denotes the
extremal black hole, $\Delta m = -\frac{7}{15}(q+q_c)^2$. Note
how the hairy black hole (at lowest mass) ends in the soliton
for $q\leq \frac{7}{3}q_c$ and in the extremal hairy black hole
for $q>\frac{7}{3}q_c$.}
\label{figure3}
\end{figure}

As before, we consider systems with net charge 
$$\tilde{q} = q\theta\,,$$
and net mass\footnote{In other words, $q$ and $\Delta m$
parametrize our hairy black holes.}
$$m = \frac{q}{q_c}\theta+\left(\frac{2}{3}q^2+\Delta
m\right)\theta^2\,.$$
Eq. \eqref{mueq} along with charge and mass conservations give the
equations
\begin{align}
\alpha &= \frac{1}{a}\left(\frac{3b}{7q_c}-1\right) + {\cal
O}(\theta)\label{alpha1}\,,\\
\label{qeqlarge} q &= \left(2aq_c+b\right)  + {\cal O}(\theta)\,,\\
\label{meqlarge} \Delta m & = \left(-\frac{b}{2q_c}-\frac{a
b}{q_c}-\frac{8b^2}{21}\right) +{\cal O}(\theta)\,.
\end{align}
These equations can be solved for $a$ and $b$
\begin{equation}\label{solforaandb}
\begin{split}
b & = \left(\frac{7}{10}(q+q_c) \pm
\sqrt{\left[\frac{7}{10}(q+q_c)\right]^2+\frac{21}{20}\Delta
m}\right) + {\cal O}(\theta)\,, \\
a & = \frac{q-b}{2q_c} +{\cal O}(\theta)\,.
\end{split}
\end{equation}
Recall that the black hole component of the mix must have
$\alpha\leq2$. From \eqref{alpha1},
this implies that\footnote{The inequality \eqref{bineq1} is
saturated
when the hairy black hole is extremal.\label{commbineq1}}
\begin{equation}\label{bineq1}
b\leq \frac{7}{10}(q+q_c) + {\cal O}(\theta) \,.
\end{equation}
It follows the negative root of \eqref{solforaandb} is the only
acceptable solution.
Positivity of $a$ gives
\begin{equation}\label{bineq2}
 b\leq q +{\cal O}(\theta)\,.
\end{equation}
Inequalities \eqref{bineq1} and \eqref{bineq2} give rise to an
important aspect of
the phase diagram of these hairy black holes which we will now
discuss. Two cases are
possible:
\begin{itemize}
\item[1.] $q \leq \frac{7}{10}(q+q_c) \Rightarrow q \leq
\frac{7q_c}{3}$

In this case, the acceptable inequality for $b$ is given by
$$b\leq q\,.$$
In particular, the above inequality is saturated at $a=0$, i.e.
the hairy
black hole has a smooth soliton ($a=0$) limit. This was
predicted in \cite{Basu:2010uz}.
In this case hairy black holes exist in the mass range (from
\eqref{solforaandb})

\begin{equation}
{\cal O}(\theta)\geq \Delta m\geq
-\frac{q}{2q_c}-\frac{8q^2}{21}+{\cal O}(\theta)\,.
\end{equation}
The upper bound on $\Delta m$ is implied by the positivity of
$b$.

\item [2.] $q > \frac{7}{10}(q+q_c) \Rightarrow q >
\frac{7q_c}{3}$

The acceptable inequality for $b$ is given by 
$$b \leq  \frac{7}{10}(q+q_c). $$
This implies
$$a\geq \frac{3}{10}\left(q-\frac{7q_c}{3}\right) >0. $$
In particular the hairy black hole does  {\it not}  have a smooth
solitonic ($a=0$) limit. Note
that the inequalities above are saturated when the hairy black
hole is extremal
(see footnote \ref{commbineq1} of the previous page). Hairy
black holes exist when
\begin{equation}
{\cal O}(\theta)\geq \Delta m\geq -\frac{7}{15}(q+q_c)^2 +{\cal
O}(\theta).
\end{equation}
At the lower bound the hairy black hole is extremal.
\end{itemize}
The complete phase diagram for this system at leading order is
plotted in
Fig. \ref{figure3}. 

In summary, the non-interacting model for $e^2 =
\frac{32}{3}(1+\theta)$ predicts the following:

\begin{itemize}
\item [1.] Near extremal hairy black holes exist for all charges
$\tilde{q}>0$.
\item [2.] For hairy black hole systems parametrized by $q$ and
$\Delta m$ (defined above)
the soliton and black hole portions of the non-interacting mix
have
\begin{equation}\label{solaandb2}
 \begin{split}
b &=
\left(\frac{7}{10}(q+q_c)-\sqrt{\left[\frac{7}{10}(q+q_c)\right]^2+\frac{21}{20}\Delta
m}\right) + {\cal O}(\theta) \,,\\
a&= \frac{1}{2q_c}\left(\frac{3q}{10}-\frac{7q_c}{10} +
\sqrt{\left[\frac{7}{10}(q+q_c)\right]^2+\frac{21}{20}\Delta
m}\right) + {\cal O}(\theta)\,.
 \end{split}
\end{equation}
\item [3.] Hairy black holes exist for mass ranges
\begin{equation}
 \begin{split}
\label{largemassrange}
{\cal O}(\theta)\geq \Delta m\geq
-\frac{q}{2q_c}-\frac{8q^2}{21}+{\cal O}(\theta), ~~ \hbox{if}\:\: q \leq
\frac{7q_c}{3}\,; \\
\:\:{\cal O}(\theta)\geq \Delta m\geq -\frac{7}{15}(q+q_c)^2 +{\cal
O}(\theta), ~~   \hbox{if}\:\:q>\frac{7q_c}{3}\,.
 \end{split}
\end{equation}
The lower limit of $\Delta m (q)$ undergoes a `phase transition'
at the critical value of charge $q_{c_2} = \frac{7q_c}{3}$. 
Below this critical value of charge, the hairy black hole at its lowest
mass reduces to the soliton. Above this critical value of
charge, the black hole component of the hairy black hole
is extremal at its lowest allowed value of $\Delta m$ at any
$q$. At this value of charge the lowest value of mass,
$\Delta m_c = \Delta m(q_{c_2}) = -\frac{35}{18}$.

\end{itemize}

\subsubsection{Isotherms}\label{subsubsec:isotherm}

As we have explained in the previous section, nature of
the lowest mass
hairy black hole as a function of charge $\tilde{q}$ undergoes a
`phase transition'
at $\tilde{q}_{c_2} = \frac{7}{3}q_{c}\theta$. For smaller
values of $\tilde{q}$
the chemical potential of the solitonic solution is smaller than
$\sqrt{\frac{3}{2}}$
\footnote{$\sqrt{\frac{3}{2}}$ is the chemical 
potential of the zero size extremal black hole}.
In this case the lowest mass hairy black hole is a soliton. Upon
slightly
raising the mass of these solutions at fixed $\tilde{q}$ we
nucleate a vanishingly small
non extremal RN AdS black hole at the core of the soliton. The
temperature
of this black hole diverges in the zero size limit, so the
solitonic
solution is assigned temperature $T=\infty$ in the hairy black
hole
phase diagram.  

At $\tilde{q}=\tilde{q}_{c_2}$ the chemical potential of the
soliton reaches the critical value
$\sqrt{\frac{3}{2}}$. Upon raising the mass of the solution at
fixed charge
we nucleate a small extremal black hole at the center of the
soliton.

When $\tilde{q}>\tilde{q}_{c_2}$ the chemical potential of the
soliton exceeds the critical value
$\sqrt{\frac{3}{2}}$. No zero size RN AdS black hole has a
chemical potential larger
than $\sqrt{\frac{3}{2}}$. As a consequence, at these values of
$\tilde{q}$, the soliton
cannot be obtained as a limit of hairy black hole solutions. The
lowest mass
hairy black hole solutions in this limit are hairy extremal
black holes,
closely analogous to the solutions described in the previous
section. These
solutions are extremal and so have zero temperature.

Note that the temperature of the lowest mass solution at charge
$\tilde{q}$ jumps from $T=\infty$ to $T=0$ as $\tilde{q}$ increases
past $\tilde{q}_{c_2}$. It is
thus interesting to plot lines of constant temperature
(isotherms) in
the phase diagram for hairy black holes, in the neighbourhood 
of the lowest mass black hole at $\tilde{q}=\tilde{q}_{c_2}$. As
$T=0$ and $T=\infty$
isotherms intersect at this point, it is natural to expect
isotherms of every temperature to intersect at the same 
point. This is indeed the case as we now demonstrate.

\begin{figure}
\begin{center}
\includegraphics[height=90mm]{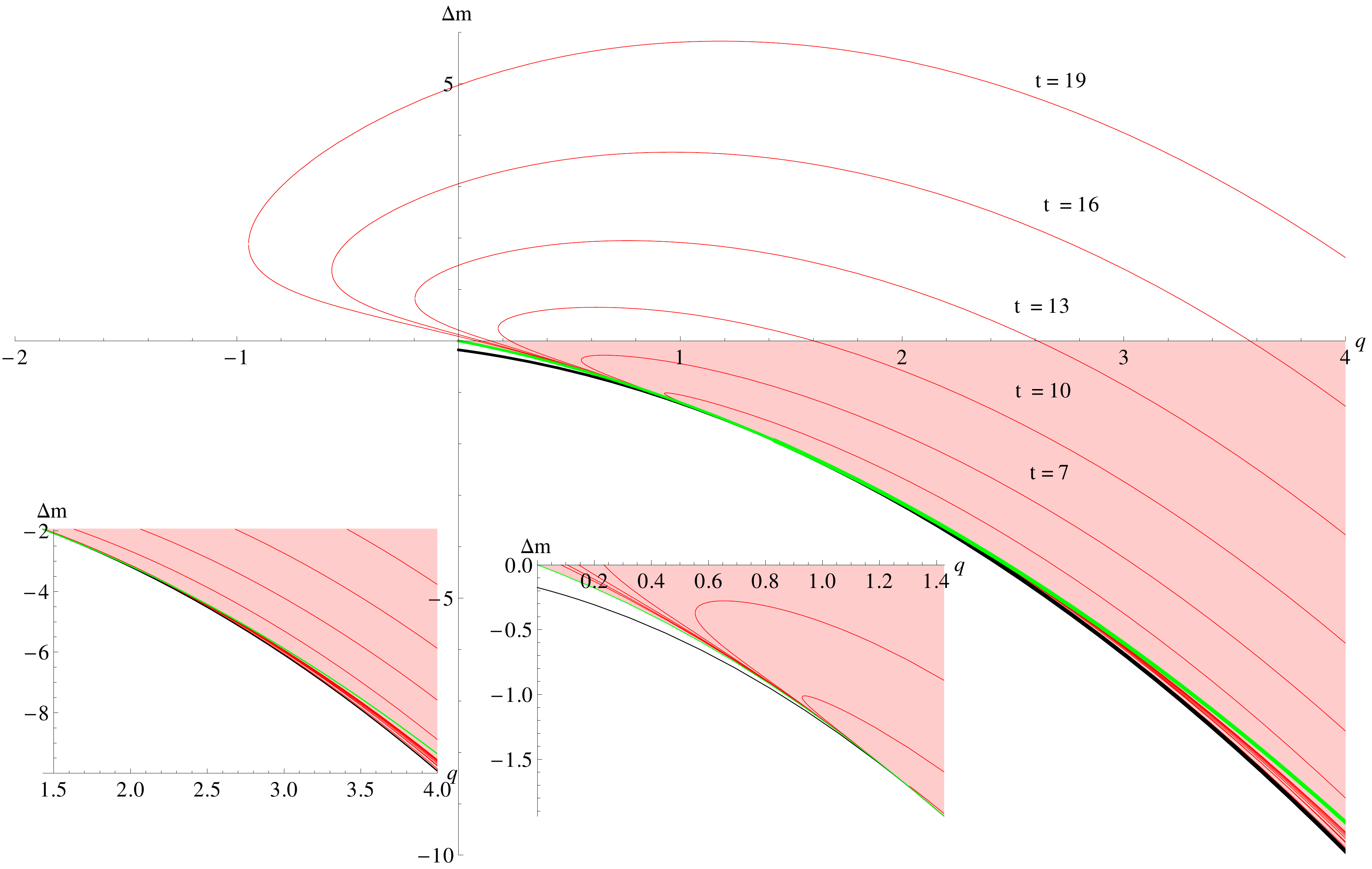} 
\end{center}
\caption{Isotherms for hairy black holes at leading
order. These have been plotted for various values of $t$ as
indicated on the graph. The black curve is the extremal black hole 
$\Delta m =
-\frac{7}{15}(q+q_c)^2$. The green curve is the soliton line
$\Delta m = -\frac{q}{2q_c}-\frac{8q^2}{21}$. Hairy black holes
exist in the pink shaded region. The two insets show zoomed in
regions of the plot. Note that for $q\leq \frac{7}{3}q_c$, the
isotherms lie above the soliton, and approach it as
$t\rightarrow\infty$. For $q>\frac{7}{3}q_c$, the isotherms line
between the soliton ($t=\infty$ limit) and extremal hairy black
hole ($t=0$ limit). In particular, all the isotherms intersect
the soliton and extremal hairy black hole curve at
$q=\frac{7}{3}q_c,~\Delta m = -\frac{35}{18}$.}
  \label{isotherm-plot}
\end{figure}

We define the parameter $t$ as 
$$t = \left(\frac{2\pi T}{\sqrt{\theta}}\right)^2 = a(2-\alpha)^2 = \frac{1}{a}\left(2a-\frac{3b}{7q_c}+1\right)^2$$
and look for curves of constant $t$.
Using \eqref{solaandb2}, we can write
\begin{equation}
t = \frac{20 \left(120 \Delta m+7 \left(4 q \left(2 q+\sqrt{6}\right)+3\right)\right)}{7 \left(6 \sqrt{6} q+\sqrt{21}
   \sqrt{120 \Delta m+7 \left(4 q \left(2 q+\sqrt{6}\right)+3\right)}-21\right)}+\Or(\theta).
\end{equation}
Let us analyze the nature of this function in the neighbourhood
of the critical point $(\frac{7}{3}q_c, -\frac{35}{18})$.
Note that at this point the numerator and denominator
both go to zero and hence $t$ is indeterminate. 
Slightly away from this point however, at 
$q=\frac{7}{3}q_c(1+\epsilon_1),~\Delta m= -\frac{35}{18}(1-\epsilon_2)$, $t$
becomes (keeping upto linear terms in $\epsilon_1$ and $\epsilon_2$)
\begin{equation}
t = \frac{400(7\epsilon_1+5\epsilon_2)}{21(3\epsilon_1+\sqrt{20}\sqrt{7\epsilon_1+5\epsilon_2})} = \frac{400c}{21(3\epsilon_1 + \sqrt{20c})}
\end{equation}
where $c = 7\epsilon_1+5\epsilon_2$. If we choose $\epsilon_1 = -\frac{1}{3}\sqrt{20c}+\gamma c$ and take $c\to0$, $t$ reduces
to 
\begin{equation}
t =  \frac{400}{63\gamma}\,.
\end{equation}
The analysis above clearly demonstrates that it is possible to approach the critical point $(\frac{7}{3}q_c, -\frac{35}{18})$ 
via some path so as to attain any temperature $t$ as is required (by an appropriate choice of $\gamma$). In other words, all
isotherms of this system must pass through this point. This is more clearly seen in Fig. \ref{isotherm-plot}.

\subsection{Perturbative analysis of hairy black holes}
\label{subsec:largeepert}
We have followed the  procedure described in Appendix 
\ref{ap:middlee} to construct the hairy black hole for $e^2 =
\frac{32}{3}(1+\theta)$
in a perturbative expansion in $\theta$. We have listed the
explicit
results of our construction in  Appendix \ref{ap:largee}. 
With these solutions in hand the evaluation of their
thermodynamic charges and
potentials is straightforward. 
Using formulas given in \S\ref{subsec:middleepert}, we find
\begin{equation}\label{largetherm}
\begin{split}
M &= \frac{3\pi}{8}\bigg[ \theta \left(2 a+\frac{8
k}{9}\right)+\theta ^2 \left(a^2+\frac{1024 a
k}{189}-a+\frac{9232 k^2}{59535}\right)\bigg]+{\cal
O}\left(\theta ^3\right) ,\\
Q &= \frac{\pi}{4}\sqrt{\frac{3}{2}}\bigg[\theta \left(2
a+\frac{8 k}{9}\right)+\frac{\theta ^2 \left(322560 a k-59535
a+4192 k^2+26460 k\right)}{59535}\bigg]+{\cal O}\left(\theta
^3\right), \\
\mu & = \sqrt{\frac{3}{2}}+\frac{(8 k-21) \theta }{14
\sqrt{6}}+\frac{\theta ^2 \left(11088 (1642 a+525) k+4584195 (2
a+1)^2+135136k^2\right)}{4074840 \sqrt{6}}+{\cal O}\left(\theta
^3\right), \\
T & = \frac{\sqrt{\theta } (42 a-8 k+21)}{42 \pi  \sqrt{a}} \\
&~~~+\frac{\theta ^{3/2} \left(-5544 (2291 a+945) k-1528065 (3 a
(a+1)+1)+965368
k^2\right)}{3056130 \pi \sqrt{a}}+{\cal O}\left(\theta
^{5/2}\right).
\end{split}
\end{equation}
We have verified that these quantities satisfy the first law of
thermodynamics
$$ dM = T dS + \mu dQ\,.$$
To verify the results of \S\ref{subsec:largeenonint}, 
we work with rescaled and shifted
mass $\Delta m$ and rescaled charge $q$,
\begin{equation}\begin{split}\label{actthermlarge}
q &= \frac{9a+4k}{3\sqrt{6}}+ {\cal O}(\theta), \\
\Delta m &= -\frac{4}{567} k (126 a+16 k+63)+ {\cal O}(\theta).
\end{split}\end{equation}
As before, identifying $b = \frac{k}{3q_c}$ (see \S\ref{subsubsec:therminmeless}), the formulae \eqref{actthermlarge}
reduce to 
\begin{equation}\begin{split}
q &= (2a q_c +b )+ {\cal O}(\theta), \\
\Delta m &= -\left(\frac{b}{2q_c}+\frac{ab}{q_c}+\frac{8b^2}{21}\right)+ {\cal O}(\theta).
\end{split}\end{equation}
These are exactly the formulae \eqref{qeqlarge} and \eqref{meqlarge}. 
The hairy black hole can therefore be considered as a non-interacting
mix of an RN AdS black hole and a soliton.

\paragraph{Breakdown of perturbation theory at extremality}

For $q<q_{c_2}$ our perturbative construction of hairy black
holes applies at
every value of the charge. When $q>q_{c_2}$, however, the value
of the
scalar field blows up at the horizon of the extremal hairy black
hole,
which indicates a breakdown in our perturbative expansion near
extremality,
exactly as in the previous section. Also as in the previous section
our perturbative
construction (and so perturbative predictions for
thermodynamics)
is reliable down to exponentially low temperatures $T\sim
e^{-k/4a\theta}$.

\subsection{Hairy black holes at $e^2=\frac{32}{3}$}
\label{subsec:tz}

As we have explained above, when $e^2=\frac{32}{3}(1-\theta)$ with 
$\theta$ small and positive, \eqref{sysaction} hosts hairy black holes 
of charge greater than or equal to $q_c \theta$. The hairy black hole 
of the lowest mass at any given charge has vanishing temperature. 
It may roughly be thought of as a small extremal RN AdS black hole 
embedded in a soliton. 

On the other hand when $e^2=\frac{32}{3}(1+\theta)$ with 
$\theta$ small and positive, \eqref{sysaction} hosts hairy black holes 
at all values of the charge. The nature of lowest mass hairy black holes, 
at any given value of the charge, however changes discontinuously
at charge equal to $ \frac{7}{3} q_c \theta$. Below this charge, the 
lowest mass hairy black hole is a regular soliton and has infinite 
temperature. Above this charge the lowest mass hairy black hole
is at zero temperature and may roughly be thought of as a small extremal 
RN AdS black hole embedded inside the soliton. 

What is the situation at $e^2=\frac{32}{3}$? This special value of the 
charge may be reached as the $\theta \to 0$ limit of either 
$e^2=\frac{32}{3}(1-\theta)$ or $e^2=\frac{32}{3}(1+\theta)$; either 
limit suggests that we should find hairy black holes at all values of 
the charge, and that the lowest mass hairy black hole at any given charge
should be a zero temperature solution that may roughly be thought of 
as a small extremal RN AdS black hole embedded inside the soliton. 
In Appendix \ref{ap:ecritical} we have verified that this picture is correct, by 
explicitly constructing all hairy black holes at small charge in the
theory with $e^2=\frac{32}{3}$. 

\subsection{$q_{c_2}(e^2)$ at larger values of $e$}
\label{sec:largeeGenExp}

We have used our perturbative expansion above to demonstrate that 
\begin{equation}\label{qcritlarge}
\tilde{q}_{c_2}(e^2) = \frac{7}{3}q_c \left(\frac{3e^2}{32}-1\right) -\frac{9127q_c}{792}\left(\frac{3e^2}{32}-1\right)^2+\Or\left(\left(\frac{3e^2}{32}-1\right)^3\right).
\end{equation}
It is, moreover, an interesting fact that ${\tilde q}_{c_2}(e^2)$ is related 
to a simple property of the solitonic solutions of \eqref{sysaction}, and 
so may be computed quite easily even at larger values of $e^2$ using the 
following considerations.  
When ${\tilde q}<\tilde{q}_{c_2}(e^2)$ the solitonic solution 
is a limit of hairy black hole solutions. In other words, at these charges 
 there exist hairy black holes that may be thought of as an infinitesimal 
RN AdS black hole sitting at the center of the soliton. As the RN AdS black 
hole is infinitesimal, the resultant solution may accurately be thought 
of as a mix of the soliton and the RN AdS black hole at {\it every} value 
of the charge ${\tilde q}$ (not just at small ${\tilde q}$). As we have explained in earlier
sections, the conditions for stationarity of such a mix are that the 
chemical potential of the small RN AdS black hole matches the chemical 
potential of the soliton.  However the chemical potential of infinitesimal 
RN AdS black holes is bounded from above by $\sqrt{\frac{3}{2}}$. 
It follows that if ${\tilde q}< {\tilde q}_{c_2}$ then the chemical potential of the soliton 
at that charge is less than $\sqrt{\frac{3}{2}}$. The charge ${\tilde q}_{c_2}$ is 
determined by the requirement that the chemical potential of the soliton 
at charge ${\tilde q}_{c_2}$ precisely equals $\sqrt{\frac{3}{2}}$.

\subsection{Numerical results}
\label{subsec:largeenum}

The set-up for constructing hairy black holes numerically was already detailed in subsection~\ref{subsec:middleenum}. Here, we present the results for the range $e^2>\frac{32}{3}$. Recall that, for this range, the soliton family of solutions exists for all charges.

\begin{figure}[t]
\begin{center}
\includegraphics[scale=0.45]{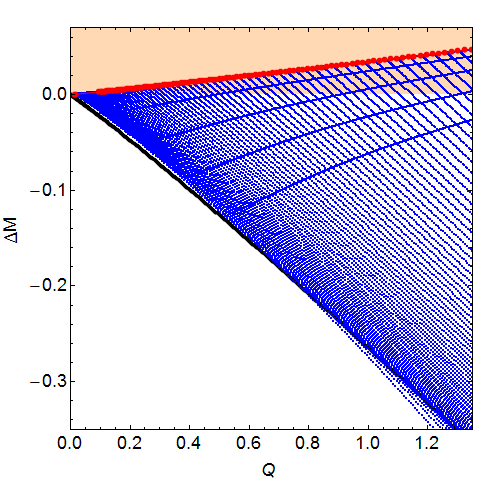}
\includegraphics[scale=0.45]{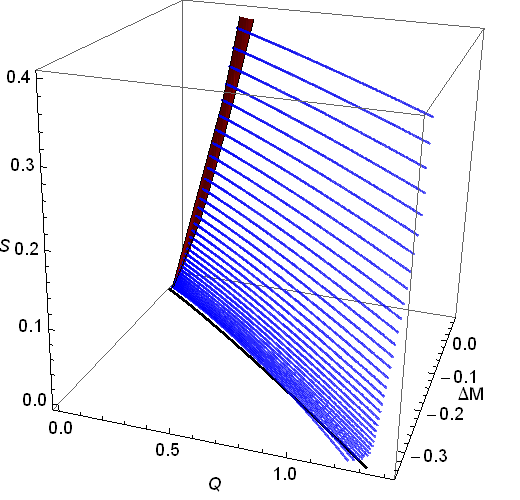}
\end{center}
\caption{\small{{\it Left:} $\Delta M$ vs. $Q$ for $e=4$. The black curve corresponds to the soliton and the blue region corresponds to hairy black holes. The red curve is the line of marginal modes of the linear problem  discussed in section \ref{subsec:linear}, and the shaded region is the area occupied by RN AdS black holes. The black curve now lies entirely below the region occupied by RN AdS black holes and it continues for arbitrarily large values of $Q$. The blue curves end at the black curve for small charges ($Q\lesssim 0.75$), indicating that the soliton branch is the endpoint of the hairy black holes family. However, for large charges ($Q\gtrsim 0.75$), the lower mass bound of the hairy black holes is below the soliton. These numerical results reproduce the behaviour predicted in Fig.~\ref{figure3}, now for $e=4$. {\it Right:} same data with the entropy as third axis. The red surface is the strip of RN AdS black holes between the marginal line of stability and extremality.}}
\label{fig:e4_dM_Q}
\end{figure}

\begin{figure}[t]
\begin{center}
\includegraphics[scale=0.6]{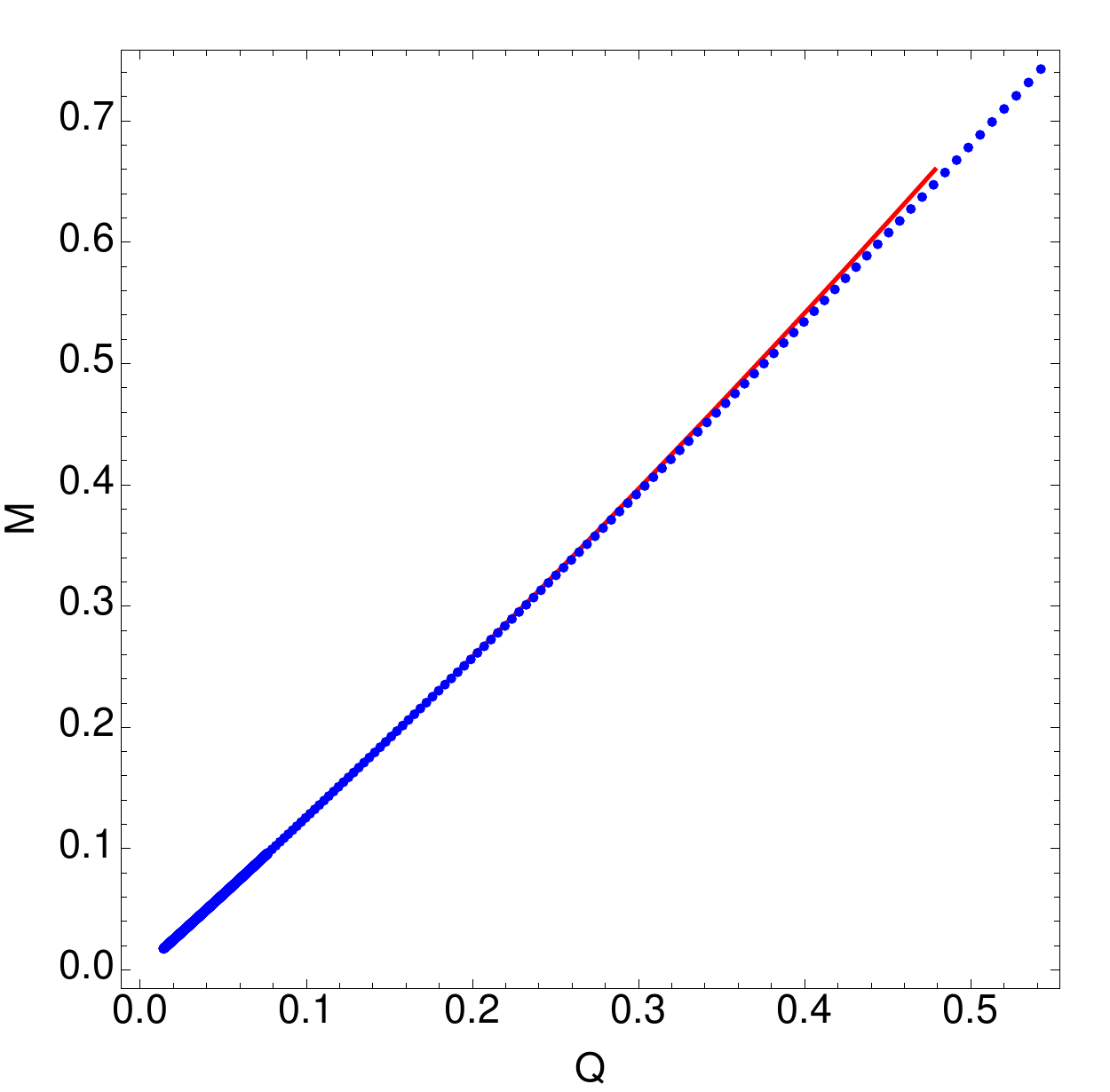}
\hspace{0.5cm}
\includegraphics[scale=0.6]{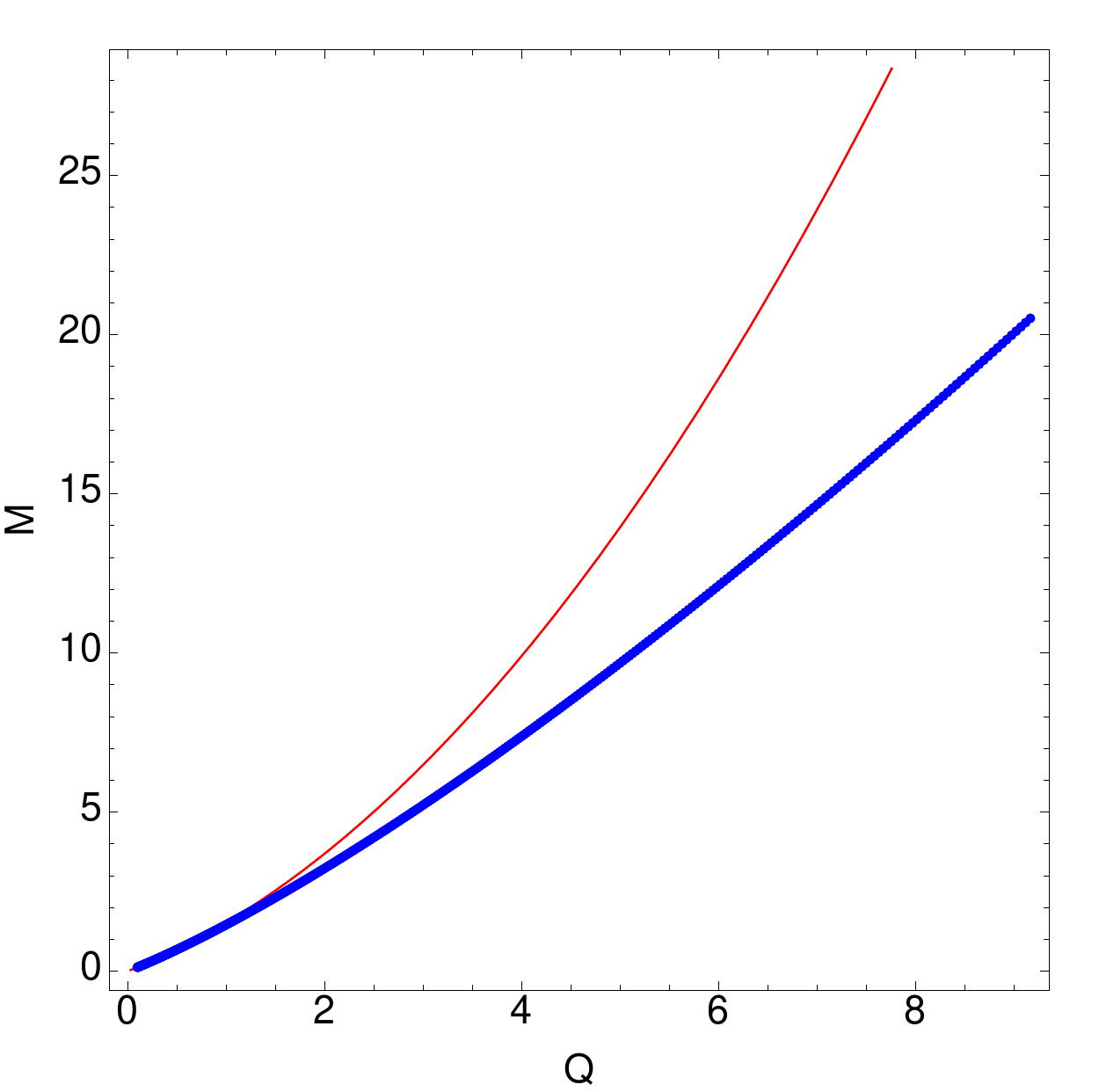}
\end{center}
\caption{\small{These plots $M$ vs. $Q$ represent the agreement of the numerical data (blue lines, made of data points) with the perturbative analysis (red lines) for small charge. \textit{Left:} Merger between hairy black holes and RN AdS black holes along the line of marginal stability for $\theta=0.01$. \textit{Right:} Hairy black holes along a line of constant boundary condensate $\epsilon=0.1$ for $e=3.33$.}}
\label{fig:nonlinearcomparasionthetap}
\end{figure}

The most remarkable feature of the phase diagram, which we represent in Fig.~\ref{fig:e4_dM_Q}, is the lower mass bound of the hairy black holes. The Fig. confirms the phase transition predicted from the perturbative analysis. It should be compared to Fig.~\ref{figure3}; there, $e^2$ close to $\frac{32}{3}$ was used, while we now have $e=4$. For charges smaller than a critical value, the hairy black holes have a zero size limit which is the soliton branch. For higher charges, however, the lower mass bound consists of extremal but singular solutions, as happened already for $3<e^2<\frac{32}{3}$. This behaviour, which should occur for arbitrarily large charges, was described by Ref.~\cite{Horowitz:2009ij} in the planar case. It was found there too that the extremal solution is singular. In order to make a quantitative comparison with the perturbative prediction of \eqref{largetherm}, we represent in Fig.~\ref{fig:nonlinearcomparasionthetap} hairy black holes along the merger line (upper mass bound) and along a line of constant boundary condensate $\epsilon=0.1$, for values of $e^2$ close to $\frac{32}{3}$. The agreement for small charge is clear.

\begin{figure}[t]
\begin{center}
\includegraphics[scale=0.5]{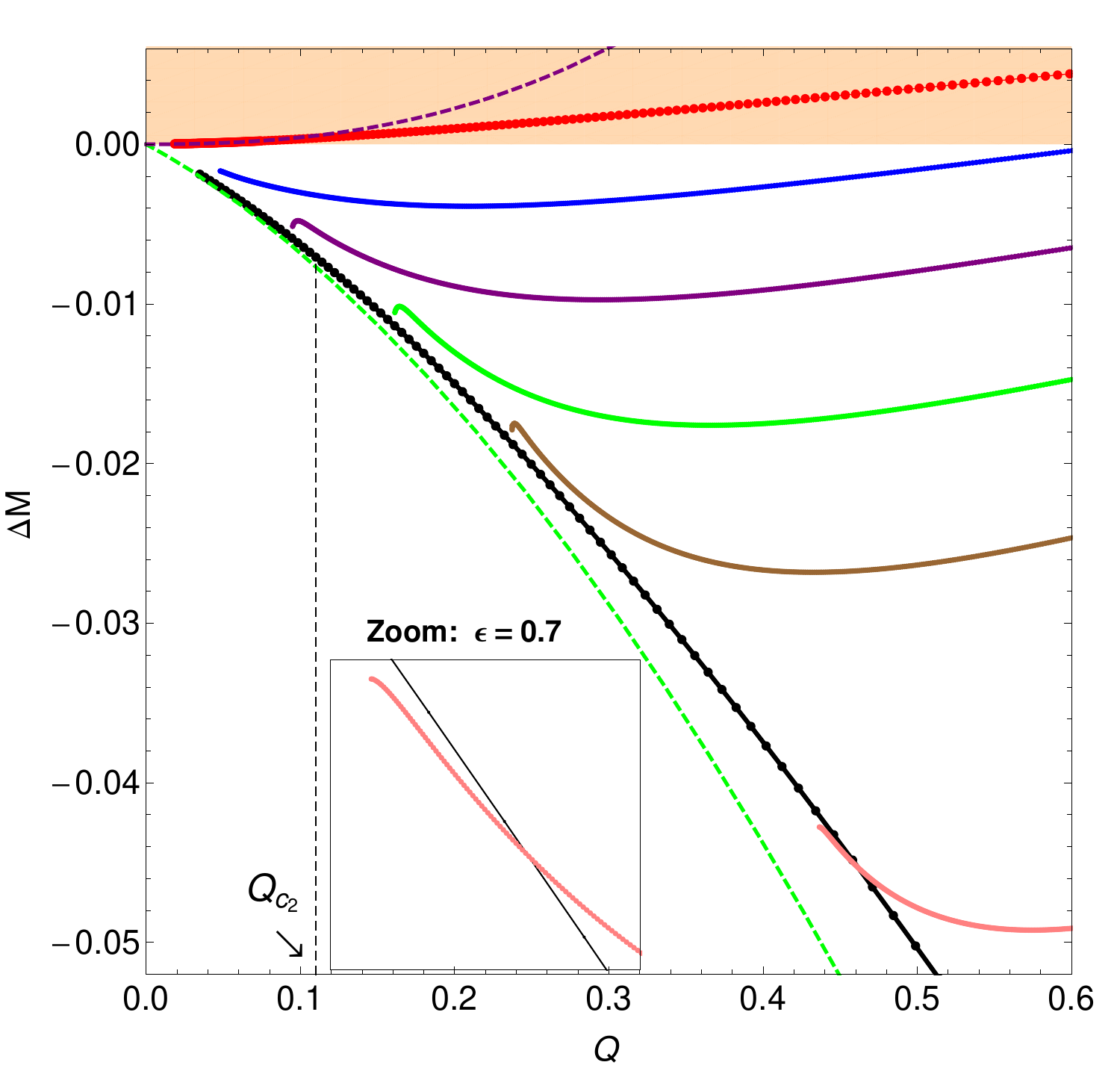}
\includegraphics[scale=0.5]{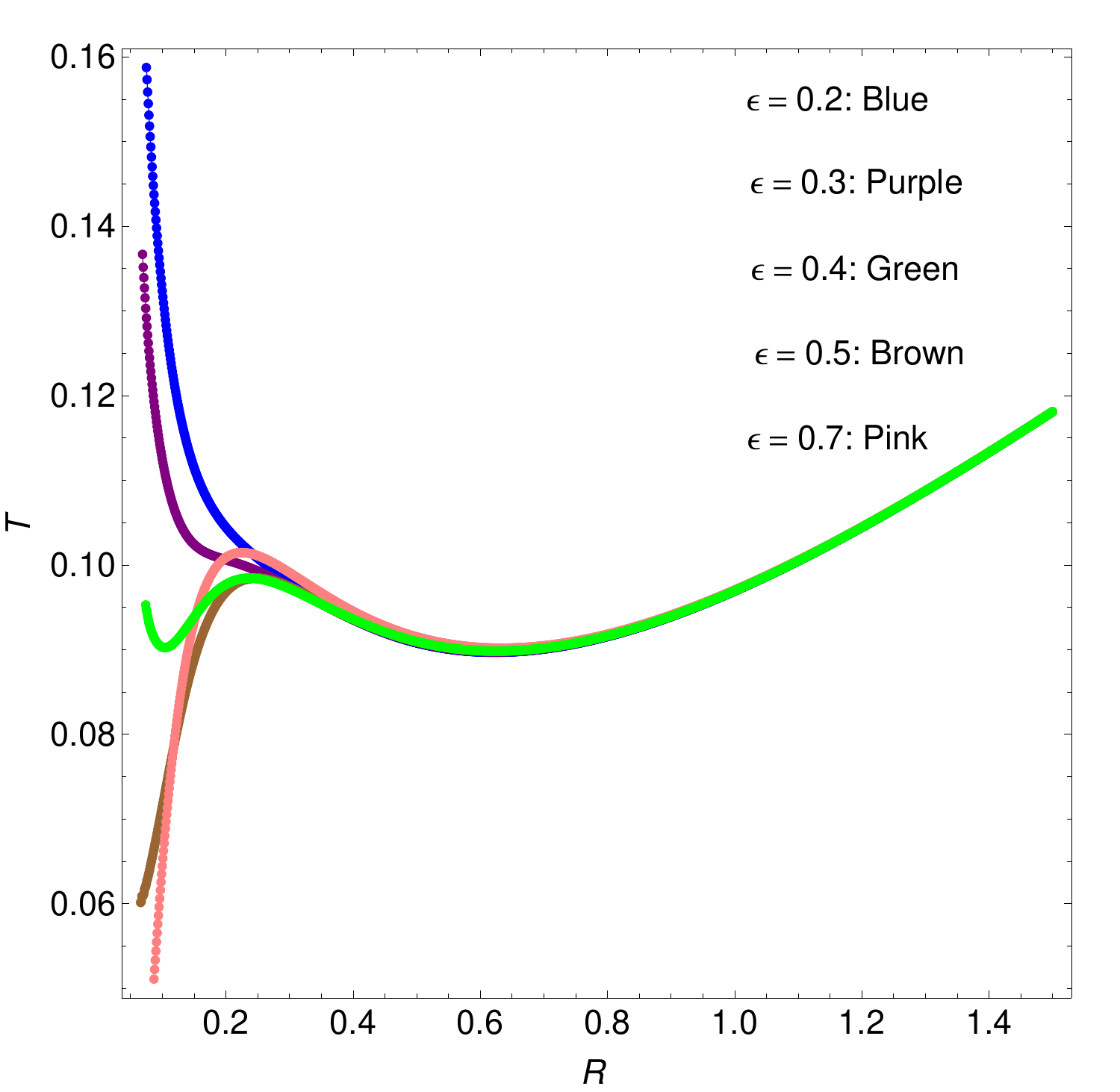}
\end{center}
\caption{\small{{\it Left:} $\Delta M$ vs. $Q$ for $e=3.4$. The black curve corresponds to the soliton, which agrees for small charge with the perturbative prediction represented by the green dashed line. The red curve corresponds to the line of marginal modes of the linear problem  discussed in section \ref{subsec:linear}, which agrees for small charge with the perturbative prediction represented by the magenta dashed line. The other lines correspond to hairy black holes at different fixed values of the boundary condensate $\epsilon$ (see legend on the right plot). $Q_{c_2}$ represents the perturbative prediction for the change in the lower mass bound of hairy black holes. It is clear that, for $\epsilon=0.7$, the hairy black holes extend to the the left of the soliton curve; the same should happen for $\epsilon=0.4$ and $\epsilon=0.5$, but it was not possible to reach a sufficiently small value of the charge with our numerical approach. {\it Right:} $T$ vs. $R$ for $e=3.4$. The curves correspond to hairy black holes at the same fixed values of $\epsilon$. The temperature should diverge for $\epsilon=0.2$ and $\epsilon=0.3$ as $R$ vanishes, and it should vanish in the other cases at a small but non-zero value of $R$. The numerical results are inconclusive because there is an intricate behaviour around the transition, so that we cannot reach sufficiently small values of $R$ with our technique. Nevertheless, that does seem to be the case at least for $\epsilon=0.5$ and $\epsilon=0.7$; for $\epsilon=0.3$ the curve might oscillate but end up turning down at smaller $R$.}}
\label{fig:e4phaseT}
\end{figure}

We perform another check of the numerical results in Fig.~\ref{fig:e4phaseT}, choosing $e=3.4$, close to $\frac{32}{3}$. On the left, we verify that the soliton and the merger (between RN AdS and hairy black holes) curves agree for small charge with the perturbative predictions of \eqref{largemassrange} and \eqref{largetherm}. We also want to check that curves of hairy black holes whose limit is a singular extremal black hole rather than the soliton, i.e. terminating for $Q>Q_{c_2}$ predicted in \eqref{qcritlarge}, cross the soliton line and have a zero-temperature endpoint; see right plot. This seems consistent for $\epsilon=0.7$, and the opposite situation (endpoint is the soliton and temperature diverges) seems consistent for $\epsilon=0.2$ and $\epsilon=0.3$. Unfortunately, for the curves closer to the transition, $\epsilon=0.4$ and $\epsilon=0.5$, the results are inconclusive. The intricate behaviour near the transition presented a technical obstacle in our effort to approach numerically the endpoints of the curves.

\begin{figure}[t]
\begin{center}
\includegraphics[scale=0.55]{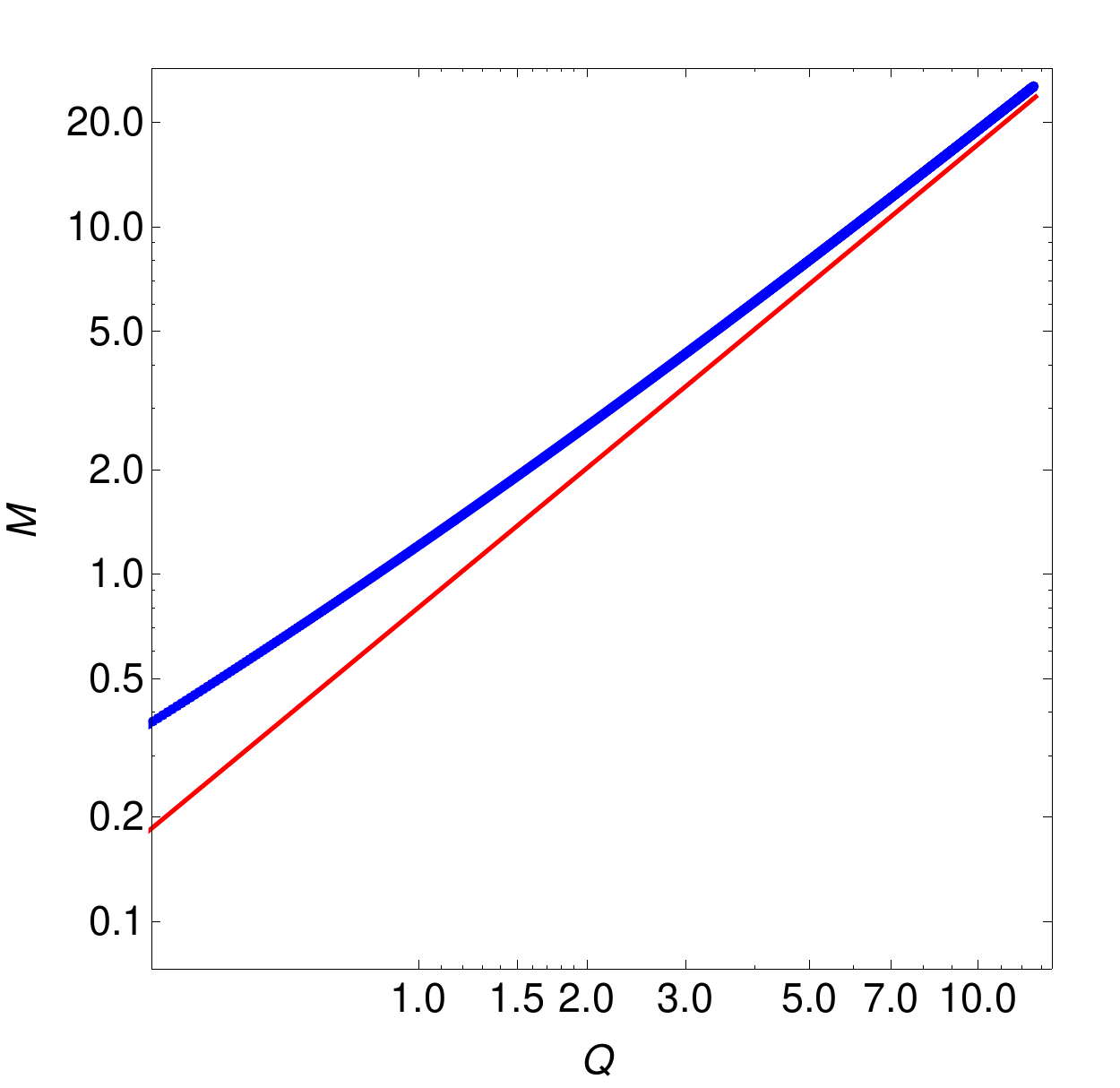}
\hspace{0.5cm}
\includegraphics[scale=0.55]{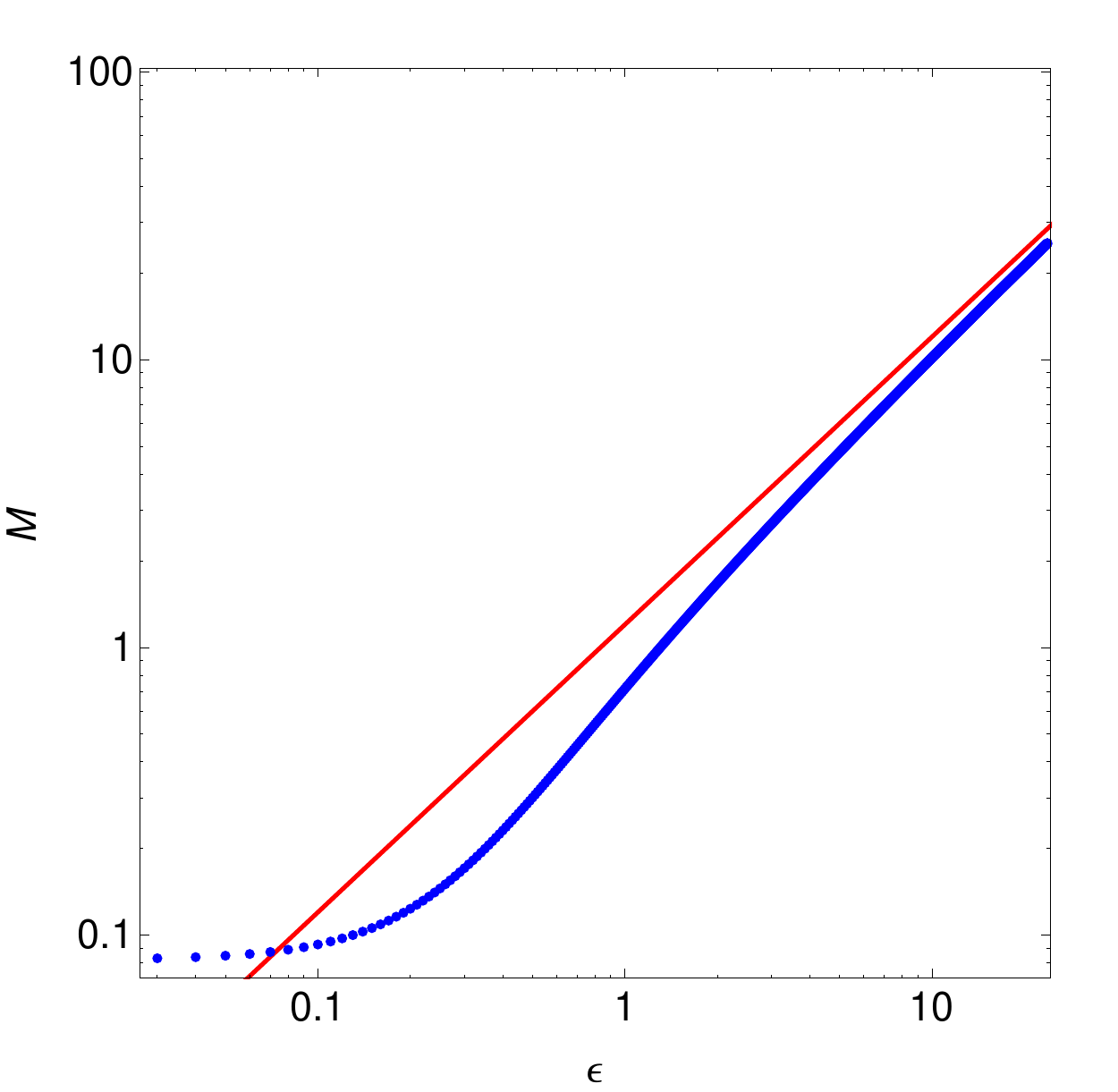}
\end{center}
\caption{\small{These plots $M$ vs. $Q$ \textit{(left)} and $M$ vs. $\epsilon$ \textit{(right)} for $e=4$, represent our best estimate for where the extremal singular curve lies in the phase diagram, for large charges. In order to estimate the shape of this singular curve in the phase diagram, we take the points of smallest temperature for each line of constant $\epsilon$. The blue dots correspond to our numerical data and the red curves correspond to the best fit of our data to functions of the form $M = A \,Q^{4/3}$ and $M = B\,\epsilon^{C}$, respectively. We find that $A \simeq 0.8$, $B \simeq C \simeq 1.0$. These results agree with the scaling found in Fig.~\ref{fig:scalingsolitonq4} for the soliton.}}
\label{fig:e4scalingextre}
\end{figure}

As in the previous section, we expect that hairy black holes at $e^2 \geq 
\frac{32}{3}$ admit the planar scaling limit \eqref{sfp}. We provide numerical evidence that such a limit exists for $e=4$ in Fig.~\ref{fig:e4scalingextre}. In particular, we confirm that $M$ scales as $Q^{\frac{4}{3}}$ and, equivalently, as $\epsilon$ for large charges. These results agree with the soliton scaling, according to Fig.~\ref{fig:scalingsolitonq4}.

\begin{figure}[t]
\begin{center}
\includegraphics[scale=0.55]{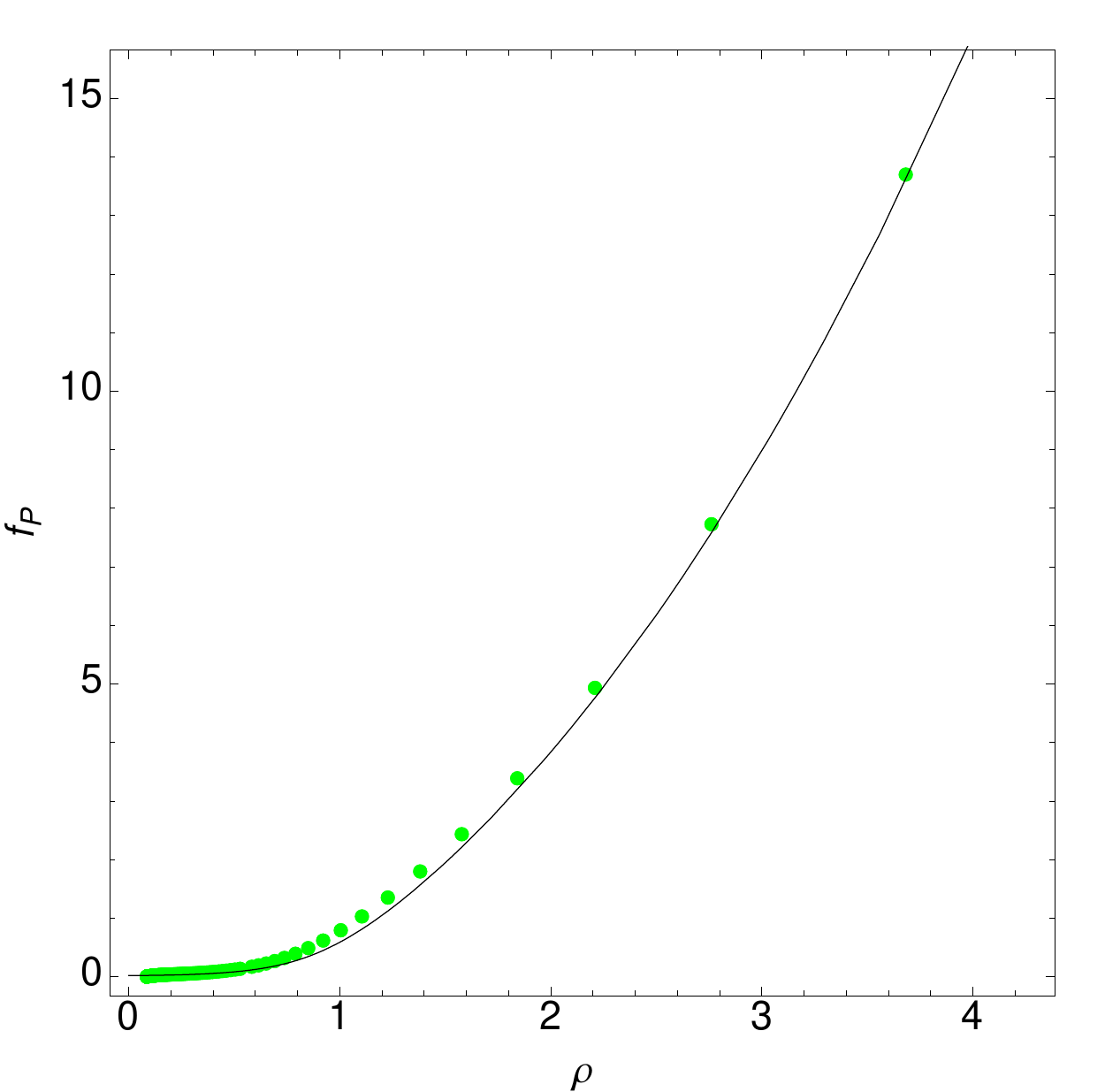}
\end{center}
\caption{\small{Metric function $f_P$ of a planar solution \eqref{scsop}, constructed numerically from the scaling \eqref{sfp}, for $e=4$, {\it i.e.} $\zeta=0.8$. The black line represents the scaling solution of the solitons, and the green dots represent the scaling solution of the extremal hairy black holes.}}
\label{fig:planarf}
\end{figure}

We can do better and contruct approximately the one parameter set of planar solutions \eqref{scsop} through an analysis for large charges. Our results are presented in Fig.~\ref{fig:planarf} for $e=4$. We show the behaviour of the metric function $f_P$ for fixed $\zeta$ (here $\zeta \simeq 0.8$, which corresponds to $e=4$, according to the scaling in Fig.~\ref{fig:e4scalingextre}). We present the results for both the solitons and the extremal hairy black holes, which we approximate by the lowest mass black holes we can find numerically for a given charge. The two limits appear to coincide, within our accuracy. This provides evidence that the solitons and the extremal hairy black holes have the same large charge limit, which we expect to be the zero temperature limit of the black branes studied by Hartnoll \emph{et al.} \cite{Hartnoll:2008vx,Hartnoll:2008kx}. We expect also that our non-extremal hairy black holes have as a planar limit the black branes at finite temperature.

\section*{Acknowledgements}

We would like to acknowledge useful discussions with J. Bhattacharya, 
S. Bhattacharyya, M. Rangamani and S. Trivedi. We would like to thank 
M. Rangamani for intimating us of his results prior to publication. 
OJCD acknowledges financial support provided by the European
Community through the Intra-European Marie Curie contract
PIEF-GA-2008-220197; DAMTP, Cambridge where part of this work was done; and the organizers and participants of the workshops 
``Numerical Relativity and High Energy Physics", Madeira (Portugal), and 
``Recent Advances in Gravity", Durham (UK) for discussions. 
PF is supported by an EPSRC postdoctoral
fellowship [EP/H027106/1]. SM is supported by a Swarnajayanti Fellowship. 
SM would also like to thank the organizers of the 
the meeting `Numerical Approaches to AdS/CFT, large $N$ and Gravity' 
for hospitality. PM would like to thank BITS-Pilani
for giving him the opportunity to work at TIFR, and TIFR for hospitality
while the bulk of this work was completed.  RM is supported by the 
Danish Council for Independent Research - Natural Sciences (FNU). 
SM and PM would also like to express their gratitude to the people of 
India for their steady support to research in the basic sciences.

\appendix

\section{Non-existence of regular planar solitons}
\label{sec:AppPlanarSol}

In this appendix we will provide an argument for the non-existence of regular solitons in the planar limit. 

In the planar case, the ansatz for the metric, gauge field and scalar field is
\begin{equation}
ds^2=-f(r)\,dt^2+\frac{dr^2}{g(r)}+r^2\,(dx^i)^2\,,\quad
A_a=A(r)\,(dt)_a\,\qquad \phi=\phi(r)\in\mathbb R
\end{equation}  

Following the same procedure as in the main text we can derive 2nd order (quasi-linear) equations of motion for $\phi(r)$, $A(r)$ and $\phi(r)$. The general structure of the equations is as follows:
\begin{equation}
Q_i^{''}+M_{(1)\,i}^{\phantom{(1)\,i}kl}(Q)\,r\,Q_k'(Q_l')^2+M_{(2)\,i}^{\phantom{(1)\,i}kl}(Q)\,Q_k'\,Q_l'+M_{(3)\,i}^{\phantom{(1)\,i}j}(Q)\,\frac{1}{r}\,Q_j'+\frac{1}{r^2}\,M_{(4)\,i}(Q)=0\,,
\label{eqn:eomsA}
\end{equation}
where the objects $M_{(\ell)}$, $\ell=1,\ldots, 4,$ are non-linear functions of the variables $Q_i(=f,A,\phi)$ and $M_{(1)}$ and $M_{(2)}$ need not have any special symmetry in the pair of indices $(kl)$.

The idea is to solve the equations near the origin $r=0$, \textit{assuming} that the functions $Q_i$ have an expansion in non negative integer powers of $r$:\footnote{Non-integer powers of $r$ would lead to singularities in the derivatives of the metric, and therefore in the curvature tensor (or derivatives of it).}
\begin{equation}
Q_i=\sum_{\ell=0}^{N}q_{i,\ell}\,r^\ell+O(r^{N+1})\,
\end{equation}
for $r\to 0$ and where the $q_{i,\ell}$ are constants that are determined from the expansion, \textit{i.e.,} the values of the functions and their first derivatives at $r=0$. By an induction argument then we will show that in fact the expansion above is in even powers of $r$ only. 

Indeed, solving \eqref{eqn:eomsA} near $r=0$ we find that the first few terms in the expansions contain only even powers of $r$:
\begin{equation}
\phi(r)=\frac{2}{e}+O(r^2)\,,\quad A(r)=O(r^2)\,,\quad f(r)=O(r^2)\,.
\end{equation}
As discussed in \cite{Horowitz:2009ij}, one can use the scaling symmetries of theory and the various fields in the \textit{planar} case to set the leading terms in $A$ and $f$ to 1. Therefore, we have:
\begin{equation}
\phi(r)=\frac{2}{e}-\frac{e}{6}\,r^2+O(r^3)\,,\quad A(r)=r^2+O(r^3)\,,\quad f(r)=r^2+O(r^3)\,.
\end{equation}
Now we assume that up to $O(r^{2n})$ the various functions have an expansion in \textit{even} powers of $r$, but at the next order there is an odd power of $r$:
\begin{equation}
Q_i=\sum_{\ell=0}^nq_{i,2\ell}\,r^{2\ell}+q_{i,2n+1}\,r^{2n+1}+O(r^{2n+2})\,,
\end{equation}
Next we have to show that the $O(r^{2n+1})$ term above vanishes because of the equations of motion \eqref{eqn:eomsA}.  

From \eqref{eqn:eomsA} and the fact that, by assumption, we have determined the solution up to $O(r^{2n})$, we see that the leading term appears at $O(r^{2n-1})$.  Clearly the terms in \eqref{eqn:eomsA} with $M_{(1)}$ and $M_{(2)}$ cannot contribute at this order: either they give terms which are of higher order or terms which are even in $r$. Therefore, the only contribution at this order can come from $Q_i''$ and the last two terms. Note that the combination $\frac{1}{r}\,Q_j'$ gives terms with positive even powers of $r$ up to $O(r^{2n-2}$).  Depending on the form of $M_{(3)\,i}^{\phantom{(3)\,i}j}$ we may also get a contribution from these other terms. Using the explicit expressions, one finds that the $O(r^{2n-1})$ contribution of these terms is
\begin{equation}
\begin{aligned}
&i=\phi:\,\qquad 3(2n+1)\,q_{\phi,2n+1}+\frac{e}{2}\big[4\,q_{A,2n+1}+(2n-3)q_{f,2n+1}\big]\,,\\
&i=A:\,\qquad \big[3(2n+1)+4\big]q_{A,2n+1}+2(2n-1)q_{f,2n+1}\,,\\
&i=f:\,\qquad (2n+1)q_{f,2n+1}\,.
\end{aligned}
\end{equation}
To close the argument we need the explicit form of $M_{(4)}$. The components are:
\begin{equation}
M_{(4)\,\phi}=\frac{3\, e^2 A^2 \phi}{6 f+e^2 A^2 \phi^2}\,,\quad M_{(4)\,A}=-\frac{6 e^2 A\, f \phi^2}{6 f+e^2 A^2 \phi^2}\,,\quad M_{(4)\,f}=-\frac{12 f\left(2 f+e^2 A^2 \phi^2\right)}{ 6 f+e^2 A^2 \phi^2}\,.
\end{equation}
One can see that the $O(r^{2n-1})$ contribution that these terms give rise to is:
\begin{equation}
M_{(4)\,\phi}^{{(2n-1)}}=e(2\,q_{A,2n+1}-q_{f,2n+1})\,,\quad M_{(4)\,A}^{(2n-1)}=-4\,q_{A,2n+1}\,,\quad M_{(4)\,f}^{(2n-1)}=-4\,q_{f,2n+1}\,.
\end{equation}

Therefore, at $O(r^{2n-1})$ we are led to a homogeneous system of linear equations for the $q_{i,2n+1}$'s:
\begin{equation}
\begin{aligned}
& (2n+1)2n\,q_{\phi,2n+1}+3(2n+1)\,q_{\phi,2n+1}+\frac{e}{2}\big[4\,q_{A,2n+1}+(2n-3)q_{f,2n+1}\big]+e(2\,q_{A,2n+1}-q_{f,2n+1})=0\,,\\
&(2n+1)2n\,q_{A,2n+1}+3(2n+1)\,q_{\phi,2n+1}+\big[3(2n+1)+4\big]q_{A,2n+1}+2(2n-1)q_{f,2n+1}-4\,q_{A,2n+1}=0\,,\\
&(2n+1)2n\,q_{f,2n+1}+(2n+1)q_{f,2n+1}-4\,q_{f,2n+1}=0\,.
\end{aligned}
\end{equation}
But clearly this system is non-degenerate and the only solution is the trivial one: $q_{\phi,2n+1}=q_{A,2n+1}=q_{f,2n+1}=0$. Therefore, by induction, we have shown that the near $r=0$ expansion can only contain even powers of $r$ if we demand regularity. 

Now we can integrate out the equations of motion and we find that all coefficients in this expansion are be determined in terms of $e$ only (we have constructed such an expansion up to $O(r^{12})$. Therefore we see that this near horizon expansion has no free parameters. Ultimately this is a consequence of the regularity requirements that we have imposed. If there was a solution in the whole spacetime, then this near ``horizon'' solution should be matched to a far region solution with the correct asymptotic behavior. However, one can easily check (numerically) that this is only possible at isolated values of $e$. This is precisely the result that \cite{Horowitz:2009ij} found: only at one particular value of the scalar charge there is regular soliton. Therefore, we conclude that for generic values of the scalar charge $e$ there is no regular planar soliton.

\section{Perturbative construction of small hairy black holes
for $e^2= \frac{32}{3}(1-\theta)$}
\label{appert}
In this section, we present a detailed description of
the perturbative procedure used to construct small 
charge hairy black holes in an expansion
in $\theta$. We also present a detailed listing of our 
findings.
\subsection{Setting up the perturbation theory}\label{ap:setup}
As discussed in \S\ref{subsubsec:natureofpert}, RN AdS black
holes form the starting
point of our perturbative expansion. In the following paragraph,
we briefly review
RN AdS black holes.
\paragraph{RN AdS Black Holes}
\label{ap:blackholes}
The AdS-Reissner-Nordstr\"om black holes constitute a very well
known
two parameter set of solutions to the equations of \S
\ref{subsubsec:mp}.
These solutions are given by 
\begin{equation}\label{bhsol}
\begin{split}
ds^2&= -V(r) dt^2+ \frac{dr^2}{V(r)}+ r^2 d \Omega_3^2 \\
V(r)&\equiv
1+r^2-\frac{R^2}{r^2}\left[1+R^2+\frac{2}{3}\mu^2\right] +
\frac{2}{3}\mu^2\frac{R^4}{r^4} \\
&=
\left[1-\frac{R^2}{r^2}\right]\left[1+r^2+R^2-\frac{2}{3}\frac{\mu^2
R^2}{r^2}\right]\\
A(r)&= \mu \left[1-\frac{R^2}{r^2}\right]\\
\phi(r)&=0 \\
\end{split}
\end{equation}
where $\mu$ is the chemical potential of the RN AdS black hole.
The function $V(r)$ in \eqref{bhsol} vanishes at $r=R$ and 
consequently this solution has a horizon at $r=R$. We will
require that $R$ is the outer horizon of our solution; this imposes the
restriction (see \cite{Basu:2010uz}) 
\begin{equation}\label{mucond}
 \mu^2 \leq \frac{3}{2}(1+2R^2)
\end{equation}
In this paper we are interested in small, near extremal RN AdS
black holes.
Following
\S\ref{subsubsec:middleeleadord}, we set
\begin{equation}\label{Randmueq}
 \begin{split}
  R^2 &=a\theta \\
\mu^2 & = \frac{3}{2}\left(1+\alpha a \theta\right)
 \end{split}
\end{equation}
The RN AdS black hole becomes
\begin{equation}\label{bhsol1}
 \begin{split}
  ds^2&= -V(r) dt^2+ \frac{dr^2}{V(r)}+ r^2 d \Omega_3^2 \\
V(r) &=\left(1-\frac{a \theta }{r^2}\right) \left(-\frac{a
\theta (a \alpha \theta +1)}{r^2}+a \theta +r^2+1\right) \\
A(r) &=\sqrt{\frac{3}{2}\left(1+\alpha a
\theta\right)}\left(1-\frac{a\theta}{r^2}\right)
 \end{split}
\end{equation}
The condition \eqref{mucond} reduces to
\begin{equation}
 \alpha \leq 2
\end{equation}
The thermodynamics of these RN AdS black holes was 
reviewed in \S\ref{subsubsec:middleeleadord}. 

As we have explained in the main text (see \S\ref{subsubsec:natureofpert}), 
our perturbative construction of small
hairy black holes uses \eqref{bhsol1} as the starting point. 
Our construction proceeds by systematically correcting
the background solution \eqref{bhsol1} in a power series expansion
in $\sqrt{\theta}$ in three different regions (near field,
intermediate field
and far field) and finally patching the full solution together
by matching
in regions of overlap. We now describe our procedure in more
detail in
each of the three regions.

\subsubsection{The far field region $ r \gg \sqrt{\theta}$} 

In this region the starting RN AdS solution \eqref{bhsol1} is a
small
perturbation about global AdS$_5$ space. 
The scalar condensate - of order $\sqrt{\theta}$ 
- is also a small perturbation about global AdS$_5$ space.
Consequently
we seek solutions of the form 
\begin{equation}\label{ffexp}
 \begin{split}
  f^{out}(r) = &\sum_{n=0}^{\infty} \theta^n f^{out}_{n}(r) \\
g^{out}(r) = &\sum_{n=0}^{\infty} \theta^n g^{out}_{n}(r)  \\
A^{out}(r) = &\sum_{n=0}^{\infty} \theta^n A^{out}_{n}(r) \\
\phi^{out}(r) = & \sqrt{\theta} \sum_{n=0}^{\infty} \theta^n
\phi^{out}_{(2n+1)/2}(r) \\
 \end{split}
\end{equation}
As the starting point of our perturbation theory is the 
RN AdS black hole, we have
\begin{equation}\label{leadordout}
 \begin{split}
f^{out}_0(r) = &\left(1-\frac{a \theta }{r^2}\right)
\left(-\frac{a \theta (a \alpha \theta +1)}{r^2}+a \theta
+r^2+1\right) \\
g^{out}_0(r) = &\frac{1}{f_0(r)} \\
A^{out}_0(r) = &\sqrt{\frac{3}{2}\left(1+\alpha a
\theta\right)}\left(1-\frac{a\theta}{r^2}\right)
 \end{split}
\end{equation}
The superscript $out$ emphasises that this expansion is good at
large $r$. Notice that $f^{out}_0,~g^{out}_0$ and $A^{out}_0$ depend on two parameters
$\alpha$ and $a$. While $a$ is one of the parameters of 
the hairy black hole solution that we
wish to find, $\alpha$ is not. Our perturbative procedure will
determine  $\alpha$
as a function of $a$ and $k$ (see \eqref{kdef}) in a power
series expansion in $\theta$.\footnote{Our 
perturbative expansion has an ambiguity, corresponding to a 
$\theta$ dependent redefinition of $\alpha$ in the RN AdS
solution used
as the starting point of the perturbative expansion. This
ambiguity
of course has no physical significance: it corresponds to
different
partitionings of the same final solution into the `zero order'
part
and the `perturbation'. In this paper 
we fix this ambiguity by requiring that the correction to the
gauge field at
infinity vanishes at every nonzero order in perturbation theory.
This
convention removes the ambiguity in perturbation theory, and
gives
physical significance to $\alpha$. Concretely, the chemical
potential of
our  solution is given  in terms of $\alpha$ via
\eqref{Randmueq} \label{gaugefix}.}
\begin{equation} \label{cp} 
 \alpha = \sum\limits_{n=0}\alpha_n(a, k) \theta^n
\end{equation}
We pause to explain a slightly confusing aspect of the notation of 
\eqref{leadordout}. 
As is apparent from \eqref{leadordout}, the starting point of
our perturbative
expansion,  $f^{out}_0$, $g^{out}_0$ and $A^{out}_0$ are each 
{\it given} functions of $\theta$, each of which starts out at
$\Or(1)$ and admits a regular expansion in small $\theta$. The
higher order corrections, however, $f^{out}_n$, $g^{out}_n$
and $A^{out}_n$ ($n>0$) in the expansion
\eqref{ffexp} are each taken to be independent of $\theta$.
We adopt this convention to east matching with the intermediate field 
region.\footnote{This convention differs from the 
usual one (that all
$f^{out}_n$, $g^{out}_n$ and $A^{out}_n$ are independent of $\theta$)
by a simply shift of all higher order terms in the $\theta$ expansion.}

The far field perturbation expansion is expected to break down
when
the base (zero order) solution deviates substantially from the
metric
of AdS$_5$ space. A cursory inspection of \eqref{leadordout}
reveals that this
happens at $r \sim \sqrt{a \theta}$.

\subsubsection{The intermediate field region $r \ll 1$ and 
$r-R \gg R-R_{in} \sim \zeta R^3$} 
The intermediate region is defined by  
$\zeta \sqrt{a \theta}^3\ll r-\sqrt{a \theta} \ll 1$ (see \eqref{Randmueq})
. Over these length scales the small black hole is far from a 
small perturbation about AdS$_5$ space. Instead the
simplification
in this region stems from the fact that we focus on 
distances of order ${\cal O}(\sqrt{\theta})$. Over these small
length scales  the background gauge field, which is of order unity, is 
negligible compared to the derivatives which are of 
order $\frac{1}{\sqrt{\theta}}$. A second simplification
results from
the fact that we insist that $r-R \gg R-R_{in}$,
i.e. we do not let our length scales become too small. At these
distances
 the black hole that we perturb around are effectively 
extremal at leading order. Moreover the black hole may 
also be thought of (at leading order) as a small black 
hole in flat rather than global AdS space.

In this 
region it is convenient to work in a rescaled radial 
coordinate $y=\frac{r}{\sqrt{a\theta}}$ and a rescaled time
 coordinate $\tau = \frac{t}{\sqrt{a\theta}}$. Note that the
 intermediate field region consists of space time points with $y$ of
 order ${\cal O}(1)$. Points with $y$ of order 
$\frac{1}{\sqrt{\theta}}$ (or larger) and $y-1$ of 
${\cal O}(\theta)$ (or smaller) are excluded from 
the considerations of this subsection.

The metric and the gauge field of the background 
black hole take the form
\begin{equation}\label{ifbg}
\begin{split}
ds^2 &= a\theta \left[-V(y)d\tau^2 + \frac{dy^2}{V(y)} + y^2
d\Omega_3^2\right] \\
V(y) &= \left(1-\frac{1}{y^2}\right)
\left(1-\frac{1}{y^2}+a\theta(1+y^2-\frac{\alpha}{y^2})\right)
\\
{A_\tau}(y) &= \sqrt{\frac{3}{2}a\theta(1+\alpha a \theta)
}\left(1-\frac{1}{y^2}\right) \\
\alpha &= \sum\limits_{n=0}\alpha_n\theta^n
\end{split}
\end{equation}

Note the metric and gauge field in \eqref{ifbg} are not
independent of
$\theta$ even at leading order in small $\theta$. However the
leading
order dependence of the metric and gauge field on $\theta$ is
very simple;
$g_{\mu \nu} \sim {\cal O}(\theta)$ and $A_\mu \sim {\cal
O}(\sqrt{\theta})$.
Now the action for our system 
\begin{equation}
S = \frac{1}{8\pi G_5}\int d^5x \sqrt{g}\left[\frac{1}{2}{\cal
R}[g]-\frac{1}{4}F_{\mu\nu}F^{\mu\nu}+\left|D_\mu\phi\right|^2 +
6 \right]
\end{equation}
can be rewritten in terms of the rescaled variables 
\begin{equation}\label{rescalings1}\begin{split}
g_{\mu\nu}&= \theta g_{\mu\nu}^{mid}\\
A_\mu & =\sqrt{\theta} A_\mu^{mid}\\
\phi &= \phi^{mid}\\
\end{split}
\end{equation}
as
\begin{equation} \begin{split}
 S &=\frac{\theta^{3/2}}{8\pi G_5}
\int d^5x \sqrt{-det(g_{\mu\nu}^{mid})}
\left[\frac{1}{2}{\cal
R}^{mid}[g]-\frac{1}{4}F^{mid}_{\mu\nu}(F^{mid})^{\mu\nu}
+|D^{mid}_\mu\phi^{mid}|^2 + 6 \theta \right]\\
D_\mu^{mid}&=\partial_\mu -i e \sqrt{\theta} A^{mid}_\mu
\end{split}
\end{equation}
The net effect of the rescaling is three fold. First we multiply
the action
of the system by a constant factor, which does not affect
classical equations
in any way. Second the cosmological constant term is multiplied
by a
factor of $\theta$, implying that the cosmological 
constant is a small perturbation to the dynamics in the
intermediate field
region. Third the coupling of the Maxwell term in the covariant
derivative is
multiplied by a factor of $\sqrt{\theta}$, implying that the
scalar and
gauge field are almost decoupled in this region.

Now $g_{\mu\nu}^{mid}$ and $A_\mu^{mid}$ have a good $\theta \to
0$ limit
on the background RN AdS solution. This suggests that
$g_{\mu\nu}^{mid}$ and
$A_\mu^{mid}$ and $\phi$
should admit a standard expansion in the amplitude of the scalar
field
in the intermediate field region, when expressed as functions of
the
rescaled variable $y$. In other words if we define 
\begin{equation}
 \begin{split}
ds^2 &= a \theta \left[ -f^{mid}(y) d\tau^2+g^{mid}(y)dy^2+
y^2d\Omega_3^2 \right] \\
A^{mid}_\tau &= \sqrt{a \theta}
A^{mid}(y),~A^{mid}_y=A^{mid}_i=0,~\phi=\phi^{mid}(y)
 \end{split}
\end{equation}
then we can expand
\begin{equation}
 \begin{split}\label{ifexp}
  f^{mid}(y) = &\sum_{n=0}^{\infty} \theta^n f^{mid}_{n}(y) \\
g^{mid}(y) = &\sum_{n=0}^{\infty} \theta^n g^{mid}_{n}(y)  \\
A^{mid}(y) = &\sum_{n=0}^{\infty} \theta^n A^{mid}_{n}(y)  \\
\phi^{mid}(y) = & \sqrt{\theta} \sum_{n=0}^{\infty} \theta^n
\phi^{mid}_{(2n+1)/2}(y) \\
 \end{split}
\end{equation}
where 
\begin{equation}
 \begin{split}
f_0^{mid}(y) = &\left(1-\frac{1}{y^2}\right)
\left(1-\frac{1}{y^2}+a\theta(1+y^2-\frac{\alpha}{y^2})\right)
\\
g_0^{mid}(y) = &\frac{1}{f_0(y)} \\
A_0^{mid}(y) = & \sqrt{\frac{3}{2}(1+\alpha a \theta)
}\left(1-\frac{1}{y^2}\right) \\
\phi_0^{mid}(y)=&0
\end{split}
\end{equation} 
\footnote{As in the previous subsection $f^{mid}_0$, $g^{mid}_0$ and $A^{mid}_0$
have a specified dependence on $\theta$, whereas all other higher 
order terms in the perturbation theory $f^{mid}_n$, $g^{mid}_n$ 
and $A^{mid}_n$ ($n>0$) do not depend on $\theta$.
As before, this convention eases matching the intermediate
field regions to the far-field and near-field regions.}
The perturbative procedure that determines all the higher order
terms in
\eqref{ifexp} is now just a standard expansion in the amplitude
of
the scalar field $\phi$, very much like in the far field region.
After solving for the hairy black hole in the far and
intermediate field
regions we will need to match the two solutions. In order to
facilitate
this matching it is useful to rewrite the intermediate field
metric and
gauge field in terms of the far field coordinates. We find 
\begin{equation}
 \begin{split}
ds^2 &=  -f^{mid}(r/R) dt^2+g^{mid}(r/R)dr^2+ r^2 d\Omega_3^2 \\
A_t &= A^{mid}(r/R),~A_r=A_i=0,~\phi(r)=\phi^{mid}(r/R)
 \end{split}
\end{equation}
Comparing with \eqref{eqn:ansatz} we see that we must directly
match the functions
$f^{out}(r)$ with $f^{mid}\left(r/R\right)$, $g^{out}(r)$ with
$g^{mid}\left(r/R\right)$, $A^{out}(r)$ with
$A^{mid}\left(r/R\right)$ and
$\phi^{out}(r)$ with $\phi^{mid}\left(r/R\right)$. 

Finally let us examine when we expect the intermediate field
expansion to
break down. We expect a break down when $y^2$ is order $
\frac{1}{a \theta}$
or greater, as the last term in the expression for $V(y)$ in
\eqref{ifbg}
cannot be treated as a small perturbation in this region.
Moreover the
two roots of the equation $V(y)=0$ are given by $y=1$ and 
$y=1-\frac{1}{2}a\zeta \theta $ for 
$y-1 \gg\frac{1}{2} a \zeta\theta $ 
the two roots appear coincident, and dynamics effectively 
takes place in the background of an extremal black hole. At 
$y -1 \sim \frac{1}{2}a\zeta\theta$, however, the equations see the
deviation from
extremality in an important manner, and the intermediate field
expansion is expected to break down. 

\subsubsection{The near field region $r-R \ll R$}
 \label{nfexplanation}

In this subsection we now turn to the near field 
region $r-\sqrt{a \theta} \ll \sqrt{a \theta}$ (see \eqref{Randmueq}. 
We will work in terms of a further rescaled radial coordinate
$ z = \frac{y-1}{a \zeta\theta }$.
Note that the black hole horizon occurs at $z=0$ and that 
points with $z$ of order unity are located at 
$r-R\sim R-R_{in} \sim \zeta R^3 \sim$ or $y-1 \sim \zeta R^2\sim a\zeta \theta$. 
It is also convenient to work with the new time coordinate 
$T=\sqrt{a\theta} \zeta t = a\zeta\theta \tau$. As in the previous
subsection, the background gauge field makes a small direct contribution to
dynamics in this region. However deviation of the black hole metric
from extremality (and the difference between an AdS and flat
space black hole metric) are all important in this region, and have to
be  dealt with exactly rather than perturbatively. 
Let us now see all of this in more detail.

In the new coordinates, the RN AdS metric and gauge field take
the form
\begin{equation}\label{nfbg}
 \begin{split}
{ds^2} &= {a\theta}\left[-V(z) dT^2 + \frac{dz^2}{V(z)} + (1 +
a\zeta \theta z)^2 d\Omega_3^2\right]\\
A_T(y) &= \frac{1}{\zeta }\sqrt{\frac{3}{2}\frac{(1+\alpha
a\theta)}{a\theta}}\left( 1 - \frac{1}{\left(1 + a\zeta \theta
z\right)^2}\right)
= \sqrt{6a\theta}z +{\cal O}(\theta^{3/2}) \\
V(z) &= \frac{1}{a^2\zeta ^2 \theta ^2}\left(1-\frac{1}{(a z \zeta \theta
+1)^2}\right)
\left(1+a\theta+a\theta(1+az\zeta \theta)^2-\frac{1+\alpha a
\theta}{(1+az\zeta \theta)^2}\right)
\\
& =2z(1+2z) + {\cal O}(\theta) \\
\alpha &= \sum\limits_{i=0}^\infty \alpha_i\theta^i
\end{split}
\end{equation}\label{ineq}
As in the previous subsubsection the metric and gauge field do
not directly
have a good $\theta \to 0$ limit. However the rescaled gauge
field and
metric 
\begin{equation}\label{rescalings2}\begin{split}
g_{\mu\nu}&= \theta g_{\mu\nu}^{near}\\
A_\mu & =\sqrt{\theta} A_\mu^{near}\\
\phi &= \phi^{near}\\
\end{split}
\end{equation}
are well defined in the small $\theta$ limit; moreover exactly
as in the
previous section, the rescaled variables obey the same equations
of motion
as in rescaled quantities apart from a rescaling of the
cosmological constant
(which sets it to ${\cal O}(\theta))$. It follows that the
functions defined
by the expansions  
\begin{equation}\label{metnear}
 \begin{split}
ds^2 &= a \theta \left[ -f^{near}(z) dT^2+g^{near}(z)dz^2+ (1 +
a\zeta\theta z)^2d\Omega_3^2 \right] \\
A_T &= \sqrt{a \theta}
A^{near}(z),~A_z=A_i=0,~\phi=\phi^{near}(z)
 \end{split}
\end{equation}
should admit a power series expansion in $\sqrt{\theta}$, 
\begin{equation}\label{en}
 \begin{split}
f^{near}(z) = &\sum_{n=0}^{\infty} \theta^n f^{near}_{n}(z) \\
g^{near}(z) = &\sum_{n=0}^{\infty} \theta^n g^{near}_{n}(z) \\
A^{near}(z) = &\sum_{n=0}^{\infty} \theta^n A^{near}_{n}(z) \\
\phi^{near}(z) = & \sqrt{\theta} \sum_{n=0}^{\infty} \theta^n
\phi_{(2n+1)/2}^{near}(z) \\
 \end{split}
\end{equation}
where 
\begin{equation}
 \begin{split}
f_0^{near}(z) = &\frac{1}{a^2\zeta^2 \theta ^2}\left(1-\frac{1}{(a z\zeta 
\theta +1)^2}\right)
\left(1+a\theta+a\theta(1+az\zeta \theta)^2-\frac{1+\alpha a
\theta}{(1+az\zeta \theta)^2}\right)\\
g_0^{near}(z) = &\frac{1}{f_0^{near}(z)} \\
A_0^{near}(z) = &\frac{1}{\zeta } \sqrt{\frac{3}{2}\frac{(1+\alpha
a\theta)}{a^2\theta^2}}\left( 1 - \frac{1}{\left(1 + a\zeta\theta
z \right)^2}\right) \\
\phi_0^{near}(z) = &0
\end{split}
\end{equation} 
The perturbative procedure that recursively determines higher
order terms in \eqref{en} is now simply a standard expansion in the amplitude
of the scalar field $\phi$, just as in the intermediate and far field
regions.

In order to match with the functions $f^{mid}$, 
$g^{mid}$, $A^{mid}$ and $\phi^{mid}$,
it is useful to rewrite the configuration
\eqref{metnear} in the original coordinates $r$ and $t$. We find
\begin{equation}
\begin{split}
ds^2 &= -(a\zeta\theta)^2 f^{near}\left(\frac{r-R}{\zeta R^3}\right) d
t^2+ \frac{1}{a^2\zeta^2 \theta^2} g^{near}\left(\frac{r-R}{\zeta R^3}\right)
dr^2 + r^2 d \Omega_3^2  \\
A_t &= (a \zeta \theta)
A^{near}\left(\frac{r-R}{\zeta R^3}\right),~A_r=A_i=0,~\phi=
\phi^{near}\left(\frac{r-R}{\zeta R^3}\right)
\end{split}
\end{equation}

We conclude that, for matching purposes
\begin{equation}
 \begin{split}
f^{out} \sim& f^{mid} \sim (a\zeta  \theta)^2 f^{near} \\
g^{out} \sim& g^{mid} \sim \frac{g^{near}}{a^2\zeta ^2\theta^2} \\
A^{out} \sim& A^{mid} \sim a \zeta \theta A^{near} \\
\phi^{out} \sim& \phi^{mid} \sim\phi^{near} 
 \end{split}
\end{equation}

In order to facilitate this matching we simply define further
rescaled
functions 
\begin{equation} \label{fr}
\begin{split}
f^{in}(z)=  & (a\zeta  \theta)^2 f^{near}(z) \\
g^{in}(z) = & \frac{g^{near}(z)}{a^2\zeta ^2\theta^2} \\
A^{in}(z) = & a\zeta  \theta A^{near}(z) \\
\phi^{in}(z) =& \phi^{near}(z)
\end{split}
\end{equation}
 so that (for matching purposes) 
$$ f^{out} \sim f^{mid} \sim f^{in},~g^{out} \sim g^{mid} \sim
g^{in},~A^{out} \sim A^{mid} \sim A^{in}~\text{and}~\phi^{out}
\sim\phi^{mid} \sim \phi^{in} $$
Below we will find it convenient to use the functions $f^{in}$,
in actually
implementing our perturbative expansion. 
It follows from \eqref{en} and \eqref{fr} that $f^{in}$,
$g^{in}$ and $A^{in}$
admit the expansions
\footnote{The appropriate factor of $\zeta$ has been absorbed into the functions
$f^{in}_n(z)$, $g^{in}_n(z)$ and $A^{in}_n(z)$}
\begin{equation}\label{nfexp}
 \begin{split}
f^{in}(z) &= (a\theta)^2\sum\limits_{n=0}\theta^{n}f^{in}_n(z)
\\
g^{in}(z) &=
\frac{1}{a^2\theta^2}\sum\limits_{n=0}\theta^{n}g^{in}_n(z) \\
A^{in}(z) &= a\theta\sum\limits_{n=0}\theta^{n}A^{in}_n(z) \\
\phi^{in}(z) &=
\sqrt{\theta}\sum\limits_{n=0}\theta^{n}\phi^{in}_n(z)
 \end{split}
\end{equation}

where
\begin{equation}
 \begin{split}
f_0^{in}(z) = &\frac{1}{a^2 \theta ^2}\left(1-\frac{1}{(a z\zeta 
\theta +1)^2}\right)
\left(1+a\theta+a\theta(1+az\zeta \theta)^2-\frac{1+\alpha a
\theta}{(1+az\zeta \theta)^2}\right)\\
g_0^{in}(z) = &\frac{1}{f_0^{in}(z)} \\
A_0^{in}(z) = & \sqrt{\frac{3}{2}\frac{(1+\alpha
a\theta)}{a^2\theta^2}}\left( 1 - \frac{1}{\left(1 + a\zeta \theta
z\right)^2}\right) \\
\phi_0^{in}(z) = &0
 \end{split}
\end{equation}

\subsubsection{Subtleties in the near field region}

There are two related further subtleties
 in working out the perturbative expansion 
in the near region. First recall that the radial coordinate $r$
employed in this paper 
has geometrical significance; it parametrizes the volume of the
surrounding $S^3$ at that point. For this reason re-parametrizations of $r$ do
not, in general form a symmetry of the equations in this paper. At leading
order in the near field region, however, the metric metric takes the
form
\begin{equation}\label{lidmet}
\begin{split}
ds_0^2 = a \theta\left[-2 z(1 + 2z) dT^2 + \frac{
dz^2}{2 z(1 + 2z)} + d\Omega_3^2\right]
\end{split}
\end{equation}
Note in particular that the size of three sphere (at leading
order) is a
constant independent of $z$. Therefore, at this order, 
coordinate re-parametrizations of $z$ are a symmetry of the
problem.

Now consider working out the equations of motion for the near
field metric
and gauge field perturbatively in $\theta$. At leading order in
this
expansion, it follows from our discussion above that the 
equations in the near field region will admit a whole functions
worth (instead
of 4 numbers worth) of solutions, parametrized by any 
${\cal O}(\theta)$ redefinition of $z$ coordinate. In order for
this
to work out in practice, it must clearly be the case that the 
differential equations of \S\ref{subsubsec:middleeleadord} are
not all independent at leading order
in $\theta$. Indeed the relation that we expect between the
equations
is easy to deduce. A coordinate redefinition of $z$ 
$$z = z'+\delta(z')$$ induces the following variable changes
\begin{equation}\label{varchange}
 \begin{split}
\delta g_{zz} &= -\delta(z)\partial_{z} g_{zz}(z) +
2\delta'(z)g_{zz}(z) \\
\delta g_{TT} &= -\delta(z)\partial_zg_{TT} \\
\delta A_{T} &= -\delta(z)\partial_zA_{T} 
 \end{split}
\end{equation}
The change in the action induced by these variable changes must
vanish (at
leading order in $\theta$) because it is effectively a
coordinate
transformation. Independently, however, the change in the action
due to these
field variations is given by 
$$\delta S \propto E^{TT}\delta g_{TT} + E^{zz} \delta g_{zz} +
M^T \delta A_T = 0$$
where $E^{TT}$, $E^{zz}$ and $M^T$ are the $TT$, $zz$ of the
Einstein
Equation and the $T$ component of the Maxwell Equation. 
Using \eqref{varchange}, this gives
\begin{equation}\label{nearcond1}
2\partial_z \left(V_0^{in}E_{zz}\right)
-\frac{2}{V^{in}_0}\left(\partial_zA^{in}_0\right) M_T =
\frac{\partial_z V_0^{in}}{{(V_0^{in})}^2}\left(E_{TT} +
{(V_0^{in})}^2E_{zz}\right)
\end{equation}
where $V_0^{in}(z) = 2 z(1 + 2z)$. The net upshot of
this discussion
is that we expect \eqref{nearcond1} to be an identity at leading
order
in $\theta$, and it is not difficult to directly verify (see
below)
that that is indeed the case. 

The equations admit a second subtlety of similar nature. It
turns out that
to leading order in $\theta$, the action is a function only of 
the square root of the determinant of the metric in the $z$ and
$T$
directions, rather than separately a function of $g_{TT}$ and
$g_{zz}$.\footnote{In order to see this, we note that the metric
at leading order
is simply $\text{AdS}_2 \times S_3$. 
Therefore, the term ${\cal R}[g]$ in the metric 
simply topological. Further, we will later see that 
the ${\cal O}(\sqrt{\theta})$ calculations at near field set
$\phi^{in}_{1/2}(z)=1$. Consequently, the only non-trivial 
part of the Lagrangian at leading order ${\cal O}(\theta)$ is 
$\sqrt{g}F_{\mu\nu}F^{\mu\nu} = \sqrt{g}g^{TT}g^{zz}(F_{Tz})^2$,
which is a
function only of $\sqrt{g}$}
For this reason metric variations of the form 
$$g^{TT} \rightarrow g^{TT}(1-\epsilon),~ g^{zz} \rightarrow
g^{zz}(1+\epsilon)$$
leave the action unchanged (at leading order). Consequently it
must be that
\begin{equation}\label{nearcond2}
 E_{TT} +{(V_0^{in})}^2E_{zz} = 0
\end{equation}
is an identity at leading order, and this is easy to directly
verify.

Thus we have deduced, without even doing any calculations
that two linear combinations of the three functions in the
metric and
the gauge field should be undetermined at leading order in the
perturbative
expansion. 

How do we deal with these subtleties. It is very important that
the subtleties
referred to above are purely leading order phenomena; at higher
orders
in $\theta$ coordinate redefinitions in $z$ are not symmetries
of our
system because the size of the sphere is not constant at
subleading orders in $\theta$;
furthermore at higher orders in $\theta$
the action depends separately on $g_{TT}$ and
$g_{zz}$ instead of only on their product.

In order to see the import of all of this, consider perturbation
theory at ${\cal O}(\theta^2)$. As the 
homogeneous part of the equations are same at every order,
the same linear combinations \eqref{nearcond1} and
\eqref{nearcond2} of the ${\cal O}(\theta^2)$ fluctuations 
disappears from (i.e. is undetermined by) the second order
equations. However, the two functions that were left undetermined
in the $\Or(\theta)$ calculations now appear in the 
$\Or(\theta^2)$ calculations.

More precisely, at every order
 other than the leading, we actually do have as many equations
 as variables. The variables, however, consist of one unknown 
function at that order coupled with two unknown functions
of the previous order.\footnote{These are the unknown coordinate
transformations and unknown ratio $g_{TT}/g_{zz}$}
It turns out that the perturbative near field equations at 
order ${\cal O}(\theta^m)$ take the form
\begin{equation}\label{rec}
 \begin{split}
  \frac{d^2}{dz^2}A_{m-1}(z) &= T^A_{m-1}(z) \\
\sqrt{\frac{2}{3}}\frac{d^2}{dz^2}\left[V^{in}_0(z)\frac{d}{dz}A_{m-1}(z)\right]
+ \frac{d^2}{dz^2}f_{m-1}(z) &= T^f_{m-1}(z) \\
g_{m}(z) +
\frac{1}{V^{in}_0(z)^2}f_{m}(z)+\sqrt{\frac{2}{3}}\frac{1}{\zeta  V_0^{in}(z)}\frac{d}{dz}A_{m}(z)
&=T^g_{m}(z)
 \end{split}
\end{equation}
where
$$V^{in}_0(z) = 2\zeta^2 z (1+2z)$$
Terms on the RHS of the equation above are referred to as source
terms.
The source terms $T^A_{m-1}(z)$, $T^f_{m-1}(z)$ are 
determined completely by $A_{n-1}$, $f_{n-1}$ and $g_n$ for $n <
m$.
On the other hand he source term $T^g_{m}(z)$ 
depends on $A_{n}$, $f_{n}$ and $g_n$ for $n<m$. 
In order to implement perturbation theory, we solve the first
two equations
in \eqref{rec} at $m=2$ along with the last equation in
\eqref{rec} at $m=1$
(this determines $A_1$, $f_1$ and $g_1$). At next order we solve
the first
two equations in \eqref{rec} at $m=3$ along with the last
equation in
\eqref{rec} at $m=2$. This procedure continues recursively. 

Finally, let us examine when we expect the near field
expansion to
break down. We expect a break down when $z$ is order
$\frac{1}{\theta}$
or greater, i.e. $r-\sqrt{a\theta} \sim \sqrt{a\theta}$, as the
last two
terms in $V(z)$ in \eqref{nfbg} cannot be treated a small
perturbation
in this region.

\subsubsection{Summary}
According to the discussion above, in order to evaluate the
hairy black hole solution we must implement the perturbative procedure 
discussed above separately in each of the three regions (near,
intermediate and far), implement boundary conditions of normalisability at
infinity \eqref{gCondns} in the far field region
and regularity at horizon in the near field region, and 
then match all the solutions into a smooth whole. 
We will now implement this 
procedure in detail upto order $\theta$.

\subsection{Perturbation theory at ${\cal O}(\sqrt{\theta})$}
\label{ap:leadordpertmethod}

In this subsection we work to order ${\cal O}(\sqrt{\theta})$. 
The procedure described 
in this subsection applies with minor modifications 
to the perturbative construction at ${\cal
O}(\sqrt{\theta}^{2m+1})$.
The only non-trivial equation at ${\cal O}(\sqrt{\theta})$ is 
$$D^2\phi=0$$ where $D_\mu = \nabla_\mu-ieA_\mu$ is 
the linearised gauge covariantised Laplace equation 
about the background \eqref{bhsol}. We will now solve 
this equation subject to the constraints of normalisability
 of infinity, regularity at the horizon, and the requirement 
that $\phi(r)\sim\frac{\sqrt{k\theta}}{r^4}+{\cal O}(1/r^6)$ at
large $r$.

\subsubsection{Far field region $ r \gg \sqrt{\theta}$}
Let us first focus on the region $r \gg \sqrt{\theta}$. 
In this region the background \eqref{bhsol1} is a small 
perturbation about global AdS space 
with $A_t = \sqrt{\frac{3}{2}}$. A stationary linearised
 fluctuation about this background is gauge equivalent 
to a linearised fluctuation with time dependence
\footnote{The time dependence is $e^{-i \mu e t } = e^{-i \sqrt{3/2}\sqrt{32/3}t} = e^{-4it}$
where we have used $e^2 = \frac{32}{3}(1-\theta)$ and worked to leading order in $\theta$}
$e^{-4it}$ about global AdS space with $A_t=0$.
The required solution is simply the ground state excitation of a massless
 minimally coupled scalar field about global AdS.
The far-field scalar equation at $\Or(\sqrt{\theta})$ is 
\begin{equation}\label{scalareqpert-1}
\frac{d}{dr}\left(\frac{r^3}{(1+r^2)^3}\frac{d}{dr}\left[(1+r^2)^2\phi_{1/2}^{out}(r)\right]\right)
= 0
\end{equation}
The normalizable solution is 
\begin{equation}\label{phi1}
 \phi_{1/2}^{out}(r) = \frac{\sqrt{k}}{(1+r^2)^2}
\end{equation}
the overall normalisation of the mode is set by the requirement
\eqref{kdef}
$$\phi^{out}(r) = \frac{\sqrt{k\theta}}{r^4}+{\cal
O}\left(1/r^6\right)$$
This completes our perturbation analysis at 
$\Or(\sqrt{\theta})$. 

The far-field scalar equation at 
${\cal O}\left(\sqrt{\theta}^{2m+1}\right)$
generalises to 
\begin{equation}\label{scalareqpert}
\frac{d}{dr}\left(\frac{r^3}{(1+r^2)^3}\frac{d}{dr}\left[(1+r^2)^2\phi_{(2m+1)/2}^{out}(r)\right]\right)
= P_{(2m+1)/2}^{out}(r)
\end{equation}
The source term $P_{(2m+1)/2}^{out}(r)$ is determined by the
lower order solutions. We could therefore iterate this process
 to generate $\phi_{(2m+1)/2}^{out}(r)$ till any desired $m$.
 
It turns out that the expressions $\phi_{(2m+1)/2}^{out}(r)$ are
increasingly singular as $r\rightarrow0$ and the most singular
piece
scales like $\frac{1}{r^{2(m-1)}}$ ($m>0$). In other words, the
expansion
of $\phi^{out}$ is really an expansion in $\frac{\theta}{r^2}$
 (upto log corrections and an overall factor).
 The expansion therefore breaks down at $r\sim\sqrt{\theta}$ 
as expected

\subsubsection{Intermediate field region $r\ll1$ and 
$r-R\gg \zeta R^3$}

The equation $(D^{mid})^2\phi^{mid} =0$ about the background 
\eqref{ifbg} is
\begin{equation}\label{ifscaleq-1}
\frac{d}{dy}\left[y^3\left(1-\frac{1}{y^2}\right)^2\frac{d}{dy}\phi^{mid}_{1/2}(y)\right]
= 0
\end{equation}
The solutions must match the far-field solution of the previous
subsection, and the near-field solution of the next subsection,
but are subject to no intrinsic boundary regularity
requirements.

The solution to this equation is 
\begin{equation}
 \phi^{mid}_{1/2}(y) = c_1 + \frac{c_2}{y^2-1}
\end{equation}
where $c_1$ and $c_2$ are constants. It is easy to check 
that the matching of $\phi^{mid}_{1/2}(y)$ with 
$\phi_{1/2}^{out}(r)$ sets $c_1=\sqrt{k}$. 
It follows on general grounds that matching with the (as yet undetermined)
 near field solution forces $c_2$ to vanish. This is
 because, were $c_2$ to be non-zero, it would match onto 
a near field solution of order ${\cal
O}\left(\frac{1}{\theta}\right)$
 (see the next subsection for details), 
violating the requirement that the solution has a smooth
$\theta\rightarrow0$
limit. Thus
\begin{equation}
 \phi_{1/2}^{mid}(y)=\sqrt{k}
\end{equation}
This completes our intermediate field calculations at $\Or(\sqrt{\theta})$.

At higher orders in $\theta$, i.e. at $\Or(\sqrt{\theta}^{2m+1})$, the 
equation \eqref{ifscaleq-1} generalises to
\begin{equation}\label{ifscaleq}
\frac{d}{dy}\left[y^3\left(1-\frac{1}{y^2}\right)^2\frac{d}{dy}\phi^{mid}_{(2m+1)/2}(y)\right]
= P^{mid}_{(2m+1)/2}(y)
\end{equation}
where $P^{mid}_{(2m+1)/2}(y)$ is a source term determined by the
perturbative procedure. 

It turns out that the expressions $\phi^{mid}_{(2m+1)/2}(y)$ are
increasingly singular as $y\rightarrow\infty$ and
$y\rightarrow1$.
At large $y$, $\phi^{mid}_{(2m+1)/2}(y)$ scales like $y^{2m}$. 
In other words the expansion of $\phi^{mid}$ is really an
expansion
in $\theta y^2$. Further at $y\sim1$, 
$\phi^{mid}_{(2m+1)/2}(y)$ scales like $\frac{1}{(y-1)^{m-1}}$
($m>0$),
 i.e. the expansion for $\phi^{mid}$ is really an expansion in 
$\frac{\theta}{y-1}$. The perturbation theory thus breaks down 
when $y-1\sim\theta$ and $y\sim\frac{1}{\theta}$.

\subsubsection{Near field region $r - R \ll R$ or $y -1\ll 1$}
We plug in \eqref{nfexp} to the linearised equation
$(D^{in})^{2}\phi^{in} = 0$ calculated in background 
\eqref{nfbg}. The equation reduces to
 \begin{equation}\label{inequ-1}
\frac{d}{dz}\left[2\zeta^2 z(1+2z)\frac{d}{dz}\phi^{in}_{1/2}(z)\right]
= 0
\end{equation}
The near field solutions are subject to the requirement of regularity at the
horizon $z=0$. The leading order solution (after
imposing regularity at the horizon) to this equation for
$\phi^{in}_{1/2}(z)$
is a constant. Matching determines 
the value of the constant to be $\sqrt{k}$.
\begin{equation}
\phi^{in}_{1/2}(z)=\sqrt{k}
\end{equation}

At $\Or(\sqrt{\theta}^{2m+1})$, the near-field
scalar equation takes the form
 \begin{equation}\label{inequ}
\frac{d}{dz}\left[2\zeta^2 z(1+2z)\frac{d}{dz}\phi^{in}_{(2m+1)/2}(z)\right]
= P^{in}_{(2m+1)/2}(z)
\end{equation}
As usual $P_{(2m+1)/2}^{in}(z)$ is a source term whose 
form is determined from the results of perturbation 
theory at lower orders. 
\subsubsection{Summary}

We started this section with a small RN AdS black hole
characterized by two parameters, $a$ and $\alpha$. In this subsection we have
constructed an ${\cal O}(\sqrt{\theta})$ static correction to the starting
vacuum solution. The correction lies entirely in the scalar field, and
its amplitude is characterized by a new parameter $k$. 

The reader may be puzzled that we appear to have constructed a 3
rather than two parameter set (as expected on general grounds) of black
hole solutions. The resolution to this puzzle is that $\alpha$ will
turn out not to be a free parameter, but will be determined as a
function
of $a$ and $k$ (this determination will, infact, justify the 
non interacting thermodynamic model of the previous section).
However
the constraint that determines $\alpha$ as a function of $a$ and
$k$ shows up only at order $\theta^{3/2}$, as we will see below.
\subsection{Perturbation theory at ${\cal O}(\theta)$}
\label{ap:ordonepertmethod}
We now briefly outline the procedure used to evaluate 
the solution at ${\cal O}(\theta)$. We proceed in close 
imitation to the previous subsection. The main difference 
is that at this order (and all integer orders) in the $\theta$ 
expansion, perturbation theory serves to determine the 
corrections to the functions $f$, $g$, and $A$ rather 
than $\phi$. The procedure described here applies with 
minor modifications, to the perturbative construction 
at ${\cal O}(\theta^m)$ for all $m$. The non-trivial equations
at this order are the Einstein Equations \eqref{einsteineq} and
the Maxwell equation \eqref{maxwelleq}.

\subsubsection{Far field region $r\gg \sqrt{\theta}$}
Plugging in \eqref{ffexp} in \eqref{maxwelleq} and
\eqref{einsteineq}, we
get a set of three equations which we solve subject to boundary
conditions
\eqref{gCondns}. The equations take are
\begin{equation}\label{bahir-1}
 \begin{split}
\frac{d}{dr}\bigg(r^2(1 + r^2)^2 g^{out}_{1}(r)\bigg) &= \frac{32kr^3}{3(1+r^2)^4}\\
\frac{d}{dr}\left(\frac{f^{out}_{1}(r)}{1+r^2}\right) &=
\frac{2(1 + 2 r^2)}{r}g^{out}_{1}(r) +\frac{32kr^3}{3(1+r^2)^5}\\
\frac{d}{dr}\left(r^3\frac{dA^{out}_{1}(r)}{dr}\right) &=
\frac{32\sqrt{\frac{2}{3}}kr^3}{(1+r^2)^5}\\
 \end{split}
\end{equation}
One of the integration 
constants in the first equation is fixed by the requirement that
$f^{out}_{m}(r)$ is normalizable (see \eqref{gCondns}). Another
constant
from the last equation is fixed by requiring that the gauge
field
$A^{out}_m(r)$ at infinity 
(see footnote \ref{gaugefix}).
The remaining two integration constants (one in the first
equation and one
in the last) will be fixed by matching with the intermediate
field
solution below. With these conditions the solutions at ${\cal
O}(\theta)$ are
\begin{equation}
 \begin{split}
A_{1}^{out}(r) &= k\left(\frac{4 \sqrt{6}-9 c_1
\left(r^2+1\right)^3}{18 r^2 \left(r^2+1\right)^3}\right) \\
f_{1}^{out}(r) &= k\left(\frac{8-9 c_2 \left(r^2+1\right)^3}{9
r^2 \left(r^2+1\right)^3}\right) \\
g_{1}^{out}(r) &= k\left(\frac{9 c_2 \left(r^2+1\right)^3-8
\left(3 r^2+1\right)}{9 r^2 \left(r^2+1\right)^5}\right) \\
\end{split}
\end{equation}
where $c_1$ and $c_2$ are constants to be obtained by matching
with the
intermediate field solutions. To facilitate this matching, 
we note here the small $r$ expansions of $A^{out}$, $f^{out}$
and $g^{out}$
\begin{equation}\begin{split}
A^{out} &= \sqrt{\frac{3}{2}}+\theta \left(\frac{-9 \sqrt{6}
a-9c_1 k+4 \sqrt{6} k}{18 r^2}+\frac{3 a \alpha_0 -8
k}{2 \sqrt{6}}+{\cal O}\left(r\right)\right)+{\cal
O}\left(\theta ^2\right) \\
f^{out} &= \left(1+{\cal O}\left(r\right)\right)+\theta
\left(\frac{-2 a-c_2 k+\frac{8 k}{9}}{r^2}-\frac{8
k}{3}+{\cal O}\left(r\right)\right)+{\cal O}\left(\theta
^2\right) \\
(g^{out})^{-1} &=\left(1+{\cal O}\left(r\right)\right)+\theta
\left(\frac{-2 a-c_2k+\frac{8
k}{9}}{r^2}+{\cal O}\left(r\right)\right)+{\cal
O}\left(\theta^2\right)
\end{split}
\end{equation}
This completes the far-field calculations at $\Or(\theta)$.

At each higher integer order in $\theta$, the far-field 
equations take the form (generalisations of \eqref{bahir-1}
to $\Or(\theta^m)$)
\begin{equation}\label{bahir}
 \begin{split}
\frac{d}{dr}\bigg(r^2(1 + r^2)^2 g^{out}_{m}(r)\bigg) &=
P^g_{m}(r)\\
\frac{d}{dr}\left(\frac{f^{out}_{m}(r)}{1+r^2}\right) &=
\frac{2(1 + 2 r^2)}{r}g^{out}_{m}(r) + P^f_{m}(r)\\
\frac{d}{dr}\left(r^3\frac{dA^{out}_{m}(r)}{dr}\right) &=
P^A_{m}(r)\\
 \end{split}
\end{equation}
where $P^f_{m}$, $P^g_{m}$ and $P^A_{m}$ are source 
terms determined by the solutions at all previous orders in 
perturbation theory.
\subsubsection{Intermediate field region $r\ll1$ and
$r-R\gg \zeta R^3$}
We plug in \eqref{ifexp} to \eqref{maxwelleq} and
\eqref{einsteineq} to
obtain the intermediate field equations. It turns out that the
equations are slightly simpler when rewritten in terms of a 
new function 
$$K_{m}(y) = V_0(y) g^{mid}_{m}(y) +
\frac{f^{mid}_{m}(y)}{V_0(y)}$$ where
$$V_0(y) = \left(1-\frac{1}{y^2}\right)^2$$
In terms of this function the final set of equations are
\begin{equation}\label{samikaran}
 \begin{split}
\frac{dK_{1}(y)}{dy} &= 0\\
\sqrt{\frac{2}{3}}\frac{d}{dy}\left(y^3\frac{dA^{mid}_{1}(y)}{dy}\right)
&= \frac{d K_{1}(y)}{dy}\\
\frac{d}{dy}\left(y^2f^{mid}_{1}(y)\right) &= 2 y
K_{1}(y) -2\sqrt{\frac{2}{3}}\left(
\frac{dA^{mid}_{1}(y)}{dy}\right)\\
\end{split}
\end{equation}
These equations are all easily solved by integration, upto four
undetermined
integration constants (one each from the first and third
equation, and
two for the second). It will turn out that two of these
constants are
determined by matching with the far field solution while the
other two
are determined by matching with the near field solution. The
solution
at leading order is
\begin{equation}
 \begin{split}
A_{1}^{mid} &=k\left(b_2-\frac{b_1}{2 y^2}\right) \\
f_{1}^{mid} &= k\left(\frac{\sqrt{\frac{2}{3}}
b_1}{y^4}+b_3+\frac{b_4}{y^2}\right)\\
g_{1}^{mid} &= -\frac{ky^4 \left(\sqrt{6} b_1+3 y^2 (2
b_3+b_4)-3 b_3\right)}{3
   \left(y^2-1\right)^4}
\end{split}
\end{equation}
Matching can be done by substituting
$y=\frac{r}{\sqrt{a\theta}}$
and expanding around large $r$. 
These expansions are
\begin{equation}
 \begin{split}
A^{mid} &= \sqrt{\frac{3}{2}}+\theta \left[\left(b_2
k+\frac{1}{2} \sqrt{\frac{3}{2}} a \alpha_0
\right)-\frac{\sqrt{\frac{3}{2}} a}{r^2}+{\cal
O}\left(1/r^3\right)\right]+{\cal O}\left(\theta
   ^2\right) \\
f^{mid} &= \left[r^2+1+{\cal O}\left(1/r^3\right)\right]+\theta
\left[b_3 k-\frac{2
a}{r^2}+{\cal O}\left(1/r^3\right)\right]+{\cal O}\left(\theta
^2\right) \\
(g^{mid})^{-1} &= \left[r^2+1+{\cal
O}\left(1/r^3\right)\right]+\theta \left[-\frac{2
a}{r^2}+{\cal O}\left(1/r^3\right)\right]+{\cal O}\left(\theta
^2\right)
\end{split}
\end{equation}
Matching determines
\begin{equation}
 \begin{split}
  c_1 &= \frac{4}{3}\sqrt{\frac{2}{3}} \\
  c_2 &= \frac{8}{9} \\
  b_2 &= -2\sqrt{\frac{2}{3}} \\
  b_3 &= -\frac{8}{9}
 \end{split}
\end{equation}
To facilitate matching with the near field region in the 
next subsection, we substitute $y=1+a\theta z$ and 
expand $A^{mid},~f^{mid}$ and $(g^{mid})^{-1}$ around $z=0$.
\begin{equation}\label{nearexpmid}
 \begin{split}
A^{mid} &= \theta \left(-\frac{k}{2}\left(4\sqrt{\frac{2}{3}}
+b_1\right)+\sqrt{6} a
z+{\cal O}\left(z^3\right)\right)+{\cal O}\left(\theta ^2\right)
\\
f^{mid} &= \frac{1}{3} k \theta \left(\sqrt{6} b_1+3b_4-8\right)
\\
&~~~+\theta ^2 \left(-\frac{2}{3}a
\left(3 \alpha _0 a-6 a+2 \sqrt{6} b_1k+3 b_4 k\right) z+4 a^2z^2+{\cal O}\left(z^3\right)\right)+{\cal O}\left(\theta
^3\right) \\
(g^{mid})^{-1} &= {\cal O}\left(\theta ^2\right)
 \end{split}
\end{equation}
This completes our intermediate field calculations at 
$\Or(\theta)$.

At all higher orders in $\theta$, i.e. at $\Or(\theta^m)$,
the equations \eqref{samikaran} generalise to
\begin{equation}\label{samikaran-1}
 \begin{split}
\frac{dK_{m}(y)}{dy} &= S^{K}_{m}(y)\\
\sqrt{\frac{2}{3}}\frac{d}{dy}\left(y^3\frac{dA^{mid}_{m}(y)}{dy}\right)
&= S^{A}_{m}(y)+ \frac{d K_{m}(y)}{dy}\\
\frac{d}{dy}\left(y^2f^{mid}_{m}(y)\right) &= S^{f}_{m}(y) + 2 y
K_{m}(y) -2\sqrt{\frac{2}{3}}\left(
\frac{dA^{mid}_{m}(y)}{dy}\right)\\
\end{split}
\end{equation}
where $S^{K}_{m}(y)$, $S^{A}_{m}(y)$ and $S^{f}_{m}(y)$ are 
source terms determined by the perturbative procedure

\subsubsection{Near field region $r - R \ll R$ or $y-1\ll 1$}\label{nearfieldcalc}
To obtain the equations in the near field region, we plugin
\eqref{nfexp} in
\eqref{maxwelleq} and \eqref{einsteineq}. However, as explained
in
\S\ref{nfexplanation}, two linear combinations of these three
equations
\eqref{nearcond1} and \eqref{nearcond2} are trivially zero. 
At leading order ${\cal O}(\theta)$, we can therefore 
obtain only the relation for $g^{in}_1(z)$ in terms of
$A^{in}_1(z)$
and $f^{in}_1(z)$. This is 
\begin{equation}\label{gcond1}
 g^{in}_{1}(z) = -\frac{f^{in}_{1}(z)}{V^{in}_0(z)^2}
-\sqrt{\frac{2}{3}}\frac{1}{\zeta V_0^{in}(z)}\frac{d}{dz}A^{in}_{1}(z)
\end{equation}
The functions $f^{in}_1(z)$ and $A^{in}_1(z)$ are obtained at 
${\cal O}(\theta^2)$, 
which in turn determines $g^{in}_1(z)$. For the sake of
completeness however,
we list here the solutions for $f^{in}_1(z)$, $A^{in}_1(z)$ and
$g^{in}_1(z)$
that are determined at ${\cal O}(\theta^2)$.
\begin{equation}\label{nfsol1}
 \begin{split}
A_1^{in}(z) &= \frac{4 \sqrt{\frac{2}{3}} k ((-42 a+8 k+21) z-42 a \log (2 z+1))}{21 a} \\
f_1^{in}(z) &= \\
&-\frac{16 k (-42 a+8 k+21) (2 z (856 k+315 (-42 a+8 k+21) z)-6615a (4 z +1) \log (2 z+1))}{416745 a^2} \\
g_1^{in}(z) &=\frac{28 a^2 k ((-39690 a+808 k+6615) z+6615a (4 z+1) \log (2 z+1))}{15 (42 a-8 k-21)^3 z^2 (2 z+1)^2}
 \end{split}
\end{equation}
It can be verified that the \eqref{nfsol1} indeed satisfies
\eqref{gcond1}.
Matching can be done expanding by \eqref{nfsol1} at large $z$
and comparing
with \eqref{nearexpmid}. Expanding around large $z$, we get
\begin{equation}
\begin{split}
A^{in} = &~\theta \left(\sqrt{6} a z+{\cal
O}\left(1/z\right)\right)+ \theta ^2 \left(-3
\sqrt{\frac{3}{2}} a^2z^2+\frac{1}{7} \sqrt{\frac{3}{2}} a (7-16
k) z\right. \\
  &\left.~~~~+8 \sqrt{\frac{2}{3}} a k
\log \left(\frac{42 a-8 k-21}{42 a z}\right)+{\cal
O}\left(1/z\right)\right)+{\cal O}\left(\theta
   ^3\right) \\
f^{in} = &~\theta ^2 \left(4 a^2 z^2+2a^2(2-\alpha_0)z+{\cal
O}\left(1/z\right)\right)+ \theta
   ^3 \left(-12 a^3 z^3+a^2 (-2 a-8 k+7) z^2+ \right. \\
&\left.\left(\frac{64}{3} k \log \left(\frac{42 a-8 k-21}{42 a
z}\right)
a^2+\frac{1930736 k^2 a}{1528065}-2 (3 (a-1) a+1) a \right.
\right. \\
&\left.\left.+\frac{16 (210-821 a) k a}{2205}\right)
z+\frac{16}{63} a
(42 a-8 k-21) k \left(\log \left(\frac{42 a-8
k-21}{42az}\right)-2\right) \right. \\
&\left.+{\cal O}\left(1/z\right)\right)+{\cal O}\left(\theta
^4\right) \\
(g^{in})^{-1} = &\theta ^2 \left(4 a^2 z^2+\frac{2}{21} a (42
a-8 k-21)
z+{\cal O}\left(1/z\right)\right)+{\cal O}\left(\theta ^3\right)\end{split}
\end{equation}
Matching clearly determines
\begin{equation}
 \begin{split}
b_1 & = -4\sqrt{\frac{2}{3}} \\
b_4 & = \frac{16}{3}  
\end{split}
\end{equation}
The generalised near-field equations at every higher
integer order in $\theta$ were listed in \eqref{rec}.

In the previous subsections, we have completely determined the
first order corrections to the starting RN AdS black hole. However, 
in order to completely characterize the hairy black hole
solution, we must also determine the leading order term in $\alpha$ as 
a function of $a$ and $k$.

\subsubsection{Determination of $\alpha_0$ from
the scalar equation at ${\cal O}(\theta^{3/2})$}
\label{ap:alphadetermine}
In this section, we briefly present the far-field calculation at
${\cal O}(\theta^{3/2})$ which determines the leading order term
in $\alpha$ as a function of $a$ and $k$. We plug \eqref{ffexp},
\eqref{leadordout} and \eqref{phi1} into the scalar equation 
$D^2\phi^{out} =0$ and obtain an equation of the form
\eqref{scalareqpert}.
The source term in \eqref{scalareqpert} is determined completely by the far-field
solutions at ${\cal O}(\sqrt{\theta})$ and ${\cal O}(\theta)$, i.e. by
known functions $\phi^{out}_{1/2}(r)$, $f^{out}_1(r)$, $g^{out}_1(r)$
and $A^{out}_1(r)$ and is explicitly
\begin{equation}
\begin{split}
P^{out}_{3/2}(r) &= \frac{16 \sqrt{k} r}{9
   \left(r^2+1\right)^9} \left[r^2 \left(9 r^8+4 (4 k+9) r^6+(44 k+54) r^4+12 (k+3) r^2+9\right) \right.\\
   &\left.-9 a   \left(r^2+1\right)^3 \left(\alpha_0 r^4+(\alpha_0-4) r^2+1\right)\right]
\end{split}
\end{equation}

The solution to \eqref{scalareqpert}
is then given by
\begin{equation}\label{phi3}
\begin{split}
\phi_{3/2}^{out}(r) = &\frac{\sqrt{k}}{189(1+r^2)^2}\left[9 \log
\left(\frac{1}{r^2}+1\right) (-21 a (\alpha_0-4)+8 k+21)+63
\left(\frac{1-a \alpha_0}{r^2}-\frac{12 a}{r^2+1}\right) \right.
\\
&\left.+\frac{2 k \left(-36 r^8-42 r^6+68 r^4+65
r^2+12\right)}{r^2
   \left(r^2+1\right)^4}\right]
\end{split}
\end{equation}
The small $r$ expansion of $\phi^{out}(r)$ at ${\cal
O}(\theta^{3/2})$ is
\begin{equation}\label{phi3smallr}
\sim \theta^{3/2}\phi_{3/2}^{out}(r) =
\frac{\sqrt{k}\theta^{3/2} (-21 a\alpha_0+8 k+21)}{63
r^2}+O\left(r^0\right)
\end{equation}
Note that term of form \eqref{phi3smallr} matches to a term of
the
form $\frac{\sqrt{\theta}}{y^2}$ in the intermediate field
region.
However, we know that the intermediate field solution at 
${\cal O}(\sqrt{\theta})$ is simply $\phi^{mid}_{1/2}(y) =
\sqrt{k}$
and in particular, possesses no term of the form $\frac{1}{y^2}$
at large $y$.
It follows that the corresponding term in \eqref{phi3smallr}
must vanish,
giving 
\begin{equation}
 \alpha_0 = \frac{1}{a}+\frac{8k}{21a}
\end{equation}
This determines the leading order term in $\alpha$ as a function
of $a$
and $k$ and hence, completely determines our hairy black hole
solution
 to ${\cal O}(\theta)$

\subsubsection{Summary} 
To leading order the metric and gauge field of the hairy black
hole is
simply that of an RN AdS black hole parameterized by $a$ and
$\alpha$.
In this subsection we have completely constructed the leading
order
(${\cal O}(\theta)$) correction to the metric and gauge field of
the
hairy black hole solution. Importantly, we have also used one
equation
at ${\cal O}(\theta^{3/2})$ to determine $\alpha$ as a function
of $a$ and $k$ (recall $k$ parametrizes the scalar expectation
value). In particular the results of this section establish that hairy
black holes appear in a 2 parameter class, and justify the thermodynamical
model of the previous section.
We have carried out the perturbative construction of our
solution to higher orders. In Appendix \ref{ap:middlee}
 we note the full solution up to ${\cal O}(\theta^2)$, and 
the far field solution for the metric and gauge field at ${\cal
O}(\theta^3)$.


\subsection{Listing of the final perturbative results}
\label{ap:middlee}
\subsubsection{Far field solution}
\textit{Scalar Field}
\begin{equation}
\begin{split}
\phi_{1/2}^{out}(r) &= \frac{\sqrt{k}}{\left(r^2+1\right)^2} \\
\phi_{3/2}^{out}(r) &=\frac{\sqrt{k}\left(756 a
\left(r^2+1\right)^4 \log \left(\frac{1}{r^2}+1\right)-2
\left(378 a \left(r^2+1\right)^3+k
\left(48 r^6+90 r^4+4
r^2-17\right)\right)\right)}{189\left(r^2+1\right)^6} \\
\phi_{5/2}^{out}(r) &=\frac{8\sqrt{k}a (21 a+4
k)}{21\left(r^2+1\right)^2}\log^2\left(1+\frac{1}{r^2}\right)-\frac{128
ak^{3/2}}{21
\left(r^2+1\right)^2}\left(\log^2(r)+\frac{1}{2}\text{Li}_2(-r^2)\right)
\\
&+\frac{2a\sqrt{k}}{189(1+r^2)^6}\left[-378 a \left(5
r^2+9\right) \left(r^2+1\right)^3 \right.\\
&\left.+4 k \left(2 \left(r^2+2\right) \left(6 r^8+48 r^6+112
r^4+56
r^2+71\right) r^2+107\right)+567 \left(r^2+1\right)^4\right] +
\\
&-\frac{2\sqrt{k}}{68762925 r^2
\left(r^2+1\right)^{10}}\left[-137525850 a^2 \left(4 r^4+8
r^2+1\right) \left(r^2+1\right)^6 \right.\\
&\left.+3465 a r^2 \left(r^2+1\right)^3 \left(2 k
\left(2520 r^{12}+26460 r^{10}+104124 r^8+156336 r^6+91024
r^4\right.\right. \right.\\
&\left.\left.\left.+29891 r^2 +2520 \pi ^2
\left(r^2+1\right)^5+1789\right)+19845
\left(r^2+1\right)^4\right)+k r^2 \left(k \left(2676720
   r^{14} \right.\right.\right.\\
&\left.\left.\left.+10296810 r^{12}+21833780 r^{10}+47080525
r^8+54295556 r^6+10445188 r^4 \right.\right.\right. \\
&\left.\left.\left.-8980072r^2-5013164\right)-5821200
\left(r^2+1\right)^5 \left(9 r^4+21 r^2+13\right)\right)\right]
\end{split}
\end{equation}

\textit{Metric and Gauge Field}
\begin{equation}
 \begin{split}
A_1^{out}(r) &=-\frac{2k \sqrt{\frac{2}{3}} \left(r^4+3
r^2+3\right)}{3 \left(r^2+1\right)^3} \\
A_2^{out}(r) &=-\frac{\sqrt{\frac{2}{3}}k}{19845 r^2
\left(r^2+1\right)^7}\left[105840 a r^2 \left(r^4+3 r^2+3\right)
\left(r^2+1\right)^4 \log \left(\frac{1}{r^2}+1\right) \right.
\\
&~~\left.+4410 a \left(17
r^8+44 r^6+18 r^4-36 r^2-9\right) \left(r^2+1\right)^3+r^2
\left(8 k \left(131 r^{12}+917 r^{10}+2751
   r^8\right.\right.\right. \\
&~~\left.\left.\left.+4270 r^6+4333 r^4+2541 r^2-747\right)-6615
\left(r^2+1\right)^4 \left(r^4+3 r^2+3\right)\right)\right]
\end{split}
\end{equation}
\begin{equation}
 \begin{split}
A_3^{out}(r) & = \frac{8k \sqrt{\frac{2}{3}} a}{19845 r^2
\left(r^2+1\right)^7}\left[2205 a \left(27 r^{10}+113 r^8+230
r^6+306 r^4+225 r^2+27\right) \left(r^2+1\right)^3 \right. \\
&\left.+r^2 \left(8 k \left(4911
r^{14}+30825 r^{12}+79527 r^{10}+106533 r^8+75845 r^6+22715 r^4
\right. \right. \right. \\
&\left.\left.\left.-1995 r^2+735\right)+6615 \left(3 r^6+7
r^4+3r^2-3\right)
\left(r^2+1\right)^4\right)\right]\log\left(1+\frac{1}{r^2}\right)
\\
&-\frac{64 \sqrt{\frac{2}{3}} ak \left(21 a r^2 \left(r^4+3
r^2+3\right)-2 k\right)}{63 r^2
   \left(r^2+1\right)^3}\log^2\left(1+\frac{1}{r^2}\right)\\
&-\frac{512 k^2\sqrt{\frac{2}{3}} a}{63 r^2
\left(r^2+1\right)^3}\left(\frac{1}{2}\text{Li}_2(-r^2)+\log^2r\right)
+
\\
&\frac{k}{31768471350 \sqrt{6}r^2
\left(r^2+1\right)^{11}}\left[508295541600 a^2
\left(r^2+1\right)^{11} \log \left(a\theta^2\left(2 a-\frac{8
k}{21}-1\right)\right) \right. \\
&\left.-2353220100 a^2 \left(221 r^{10}+901 r^8+1922 r^6+3105
r^4+2835 r^2+540\right)
   \left(r^2+1\right)^6 \right.\\
&\left.+91476 a \left(r^2+1\right)^3 \left(8 k \left(r^2
\left(-22355646 r^2+58800 \pi ^2
\left(r^4+3 r^2+3\right)
\left(r^2+1\right)^5-\right.\right.\right.\right. \\
&\left.\left.\left.\left.r^4 \left(2238406 r^{10}+16225148
r^8+50527008 r^6+87899161
r^4+93015307 r^2+60271386\right)\right.\right. \right.\right.\\&\left.\left.\left.\left.-4384217\right) -26145\right)-77175
\left(r^2+1\right)^5 \left(142 r^6+342
   r^4+222 r^2+9\right)\right)\right. \\
&\left.+r^2 \left(-16 k^2 \left(\left(\left(\left(\left(3
\left(\left(\left(55861681
\left(r^4+11 r^2+55\right) r^2+9207422235\right)
r^2\right.\right. \right.\right.\right.\right.\right.
\right.\right. \\
&\left.\left.\left.\left.\left.\left.\left.\left.\left.+18541209786\right)
r^2+26393430910\right)
r^2+81319155082\right) r^2+61776138545\right)
r^2\right.\right.\right. \right.\right.\\
&\left.\left.\left.\left.\left.+31707662899\right)
r^2+6259111287\right)r^2+748943493\right)
\right.\right. \\ 
&\left.\left.-268939440 k \left(\left(\left(\left(361
\left(r^4+7 r^2+21\right) r^2+12740\right)
r^2+12677\right) r^2+7049\right) r^2 \right.\right.\right. \\
&\left.\left.\left.+1757\right) \left(r^2+1\right)^4+5294745225
r^2 \left(r^{18}+11r^{16}+55 r^{14}+164 r^{12}+322 r^{10}+434
r^8 \right.\right.\right. \\
&\left.\left.\left.+406 r^6+260 r^4+109
r^2+27\right)\right)+15884235675 r^2\right]
\end{split}
\end{equation}

\begin{equation}
\begin{split}
f_1^{out}(r) &= -\frac{8k \left(r^4+3 r^2+3\right)}{9
\left(r^2+1\right)^3} \\
f_2^{out}(r) &=\frac{64 ak \left(r^4+3 r^2+3\right) \log
\left(\frac{r^2}{r^2+1}\right)}
{9 \left(r^2+1\right)^3}-\frac{8 ak
\left(17 r^8+41 r^6+6 r^4-54 r^2-18\right)}{27 r^2
\left(r^2+1\right)^4} \\
&~~~~-\frac{16 k^2\left(577 r^{12}+3304
r^{10}+6972 r^8+3248 r^6-7966 r^4-12966 r^2-8787\right)}{59535
\left(r^2+1\right)^7}
   \end{split}
   \end{equation}
   
 \begin{equation}
 \begin{split}
f_3^{out}(r) &=\frac{32 ak}{59535 r^2
\left(r^2+1\right)^7}\left[2205 a \left(36 r^{12}+180 r^{10}+365
r^8+434 r^6+402 r^4+270 r^2+ \right. \right. \\
&\left.\left.45\right) \left(r^2+1\right)^3+r^2 \left(8 k
\left(\left(\left(r^2 \left(3 \left(\left(\left(501
\left(r^2+8\right) r^2+12949\right) r^2+21168\right)
   r^2   \right.\right.\right.\right.\right.\right.\right. \\
&\left.\left.\left.\left.\left.\left.\left.+17171\right)
r^2+14420\right)-2800\right) r^2 +2730\right) r^2+6405\right)
\right.\right. \\
   &\left.\left.+19845 \left(r^2+2\right)
\left(r^3+r-1\right) \left(r^3+r+1\right)
\left(r^2+1\right)^4\right)\right]\log\left(1+\frac{1}{r^2}\right)
\\
&-\frac{256 a k\left(21 a r^2 \left(r^4+3 r^2+3\right)-2
k\right)}{189 r^2 \left(r^2+1\right)^3}\log^2
\left(1+\frac{1}{r^2}\right)
\\&-\frac{2048 ak^2}{189 r^2
\left(r^2+1\right)^3}\left(\frac{1}{2}\text{Li}_2(-r^2)+\log^2r\right)
\\
&+\frac{8k}{47652707025 r^4
\left(r^2+1\right)^{11}}\left[63536942700 a^2 r^2
\left(r^2+1\right)^{11} \log \left(a\theta^2\left(2 a-\frac{8
k}{21}-1\right)\right)\right. \\
&\left.-588305025 a^2 \left(432 r^{14}+2176 r^{12}+4337
r^{10}+4531 r^8+3162 r^6+1818 r^4\right.\right. \\
&\left.\left.+405
r^2+27\right) \left(r^2+1\right)^6-22869 a r^2
\left(r^2+1\right)^3 \left(4 k \left(\left(15536824
r^2\right.\right.\right.\right. \\
   &\left.\left.\left.\left.-58800
\pi ^2 \left(r^4+3 r^2+3\right) \left(r^2+1\right)^5+r^4
\left(420840 r^{12}+4514806 r^{10}+19830323
   r^8 \right.\right.\right.\right.\right. \\
&\left.\left.\left.\left.\left.+47657218 r^6+69544482
r^4+64721769 r^2+39221561 \right)+4359566\right)
r^2+262605\right)\right.\right. \\
&\left.\left.+231525 r^2
\left(12 r^8+67 r^6+134 r^4+113 r^2+31\right)
\left(r^2+1\right)^4\right) \right. \\
&\left.+2 k r^4 \left(-k
\left(\left(\left(\left(\left(3 \left(67989951 r^{10}+716821501
r^8+3397699745 r^6+9447771883
r^4\right.\right.\right.\right.\right.\right.\right.\right. \\&\left.\left.\left.\left.\left.\left.\left.\left.
+17302667096 r^2 +22075104832\right) r^2+59792970787\right)
r^2+39889757930\right) r^2\right.\right.\right.\right.\right.
\\&\left.\left.\left.\left.\left.+19763824564\right)
   r^2+5824785222\right) r^2 +2930284173\right)
- \right.\right. \\
&\left.\left.800415 \left(6767 r^{12}+48104 r^{10}+147252
r^8+256816
   r^6\right.\right. \right.\\
&\left.\left.\left.+271348 r^4+163008 r^2+42426\right)
\left(r^2+1\right)^4\right)\right]
\end{split}
\end{equation}
\begin{equation}
\begin{split}
g_1^{out}(r) &= \frac{8 kr^2 \left(r^2+3\right)}{9
\left(r^2+1\right)^5} \\
g_2^{out}(r) &=\frac{8k}{59535 \left(r^2+1\right)^9}\left[52920
a r^2 \left(r^2+3\right) \left(r^2+1\right)^4 \log
\left(\frac{1}{r^2}+1\right) +2205 a \left(17 r^6+41
   r^4\right.\right.\\
&\left.\left.+18 r^2 +54\right) \left(r^2+1\right)^3+2 k r^2
\left(577 r^{10}+3304 r^8+9912 r^6+29960 r^4+41930
   r^2+7350\right)\right]
\end{split}
\end{equation}
\begin{equation}
\begin{split}
g_3^{out}(r) &= \frac{32 ak}{59535 r^2
\left(r^2+1\right)^9}\left[-2205 a \left(r^2+1\right)^3 \left(5
r^8+74 r^6+108 r^4-72 r^2+9\right)+ \right. \\
&\left.8 k r^2 \left(3237 r^{12}+18144
r^{10}+43932 r^8+70840 r^6+68950 r^4+16590
r^2+1260\right)\right.\\
&\left.+19845 \left(r^2+3\right)
\left(r^3+r\right)^4\right]\log\left(1+\frac{1}{r^2}\right) \\
&+ \frac{256 a k \left(21 a r^4 \left(r^2+3\right)-2 k \left(3
r^2+1\right)\right)}{189 r^2
   \left(r^2+1\right)^5}\log^2\left(1+\frac{1}{r^2}\right) \\
&+\frac{2048k^2 \left(3 a r^2+a\right)}{189 r^2
\left(r^2+1\right)^5}\left(\frac{1}{2}\text{Li}_2(-r^2)+\log^2r\right)
\\
&+\frac{8k}{47652707025 r^2
\left(r^2+1\right)^{13}}\left[-63536942700 a^2
\left(r^2+1\right)^{11}
\log \left(a\theta^2\left(2 a-\frac{8 k}{21}-1\right)\right)
\right. \\
&\left.-588305025 a^2 \left(200 r^{10}+991 r^8+1403 r^6+936
r^4+1269 r^2-135\right)
   \left(r^2+1\right)^6 \right. \\
&\left.-22869 a \left(r^2+1\right)^3 \left(4 k \left(\left(58800
\pi ^2-937666\right)
r^{16}+\left(470400 \pi ^2-6433583\right)
r^{14}+\right.\right.\right. \\
&\left.\left.\left.14 \left(105000 \pi ^2-1435547\right)
r^{12}+672 \left(3500
\pi ^2-56243\right) r^{10}+105 \left(19600 \pi ^2-400069\right)
r^8\right.\right.\right. \\
&\left.\left.\left.+175 \left(5376 \pi ^2-150641\right)
r^6+1960 \left(90 \pi ^2-6533\right) r^4-1558830
r^2-21210\right)\right.\right. \\
&\left.\left.-231525 \left(r^2+1\right)^4 \left(13
   r^8+40 r^6+38 r^4+24 r^2+3\right)\right)+\right. \\
&\left.2 k r^4 \left(k \left(\left(\left(\left(3 \left(67989951
r^{10}+716821501 r^8+3671382385 r^6+12206966475 r^4 +\right.
\right.\right.\right.\right.\right.\right.\\
&\left.\left.\left.\left.\left.\left.\left.
   27226931600 r^2+45360308224\right)
r^2+192955373842\right) r^2+174685548422\right)
r^2\right.\right.\right.\right. \\
&\left.\left.\left.\left.+52823111110\right)
r^2+6712351338\right)
+800415\left(6767 r^{10}+48104 r^8+147252
r^6\right.\right.\right. \\
&\left.\left.\left.+229600 r^4+180460 r^2+58800\right)
\left(r^2+1\right)^4\right)\right]
 \end{split}
\end{equation}

\subsubsection{Intermediate field solution}
\textit{Scalar Field}
\begin{equation}
 \begin{split}
  \phi_{1/2}^{mid}(y) &= \sqrt{k}\\
\phi_{3/2}^{mid}(y) &= -\frac{2\sqrt{k}}{189}\left(378 a \log
\left(a \left(y^2-1\right) \theta \right)+189 a
\left(y^2+2\right)-17 k\right) \\
\phi_{5/2}^{mid}(y) &=\frac{\sqrt{k}}{68762925
\left(y^2-1\right)}\left[68762925 a^2 \left(3 y^6+13 y^4+8
y^2-40\right) \right. \\
&\left.+727650 a \left(\left(y^2-1\right) (4158 a-428 k-567)
\log (a
\theta )+\log \left(y^2-1\right) \left(y^2 (4158 a-428 k
\right.\right.\right. \\
&\left.\left.\left.-567)+1512 a \left(y^2-1\right) \log \left(a\left(y^2-1\right) \theta \right)-4914 a+572 k+945\right)
\right.\right. \\
&\left.\left.+18 (42 a-8 k-21) \log \left(\theta  \left(2
a-\frac{8 k}{21}-1\right)\right)+756 a \left(y^2-1\right) \log
^2(a \theta )\right.\right.\\
&\left.\left.+ 756 a \left(y^2-1\right)^2 \log
\left(a \left(y^2-1\right) \theta \right)-756 a
\left(y^2-1\right) \log ^2\left(y^2-1\right)\right)\right. \\
&\left.-6930 a \left(2 k \left(5565 y^4-3776 y^2+2520 \pi ^2
\left(y^2-1\right)-16909\right)+19845
   \left(y^2-5\right)\right) \right.\\
&\left.+104 k (96407 k+1455300) \left(y^2-1\right)\right]
 \end{split}
\end{equation}
\textit{Metric and Gauge Field}
\begin{equation}
 \begin{split}
A_1^{mid}(y) &= -\frac{2k \sqrt{\frac{2}{3}}
\left(y^2-1\right)}{y^2} \\
A_2^{mid}(y) &= \frac{\sqrt{\frac{2}{3}}k}{2205 y^2}\left[8820 a
\left(2 \log \left(2 a-\frac{8 k}{21}-1\right)+4 y^2 \log
\left(a \left(y^2-1\right) \theta \right)-4
   \log (a)+y^4 \right.\right. \\
&~~\left.\left.-6 \log \left(y^2-1\right)-2 \log (\theta
)-1\right)+(664 k+2205) \left(y^2-1\right)\right] \\
f_1^{mid}(y) &=-\frac{8 k\left(y^2-1\right)^2}{3 y^4} \\
f_2^{mid}(y) &=\frac{16k
\left(y^2-1\right)}{19845y^4}\left[-13230 a \left(-\log \left(2
a-\frac{8 k}{21}-1\right)-2 y^2 \log \left(a \left(y^2-1\right)
\theta \right)+2
   \log (a)+\right.\right. \\
&~~\left.\left.3 \log \left(y^2-1\right)+\log (\theta
)\right)+6615 a \left(y^4-1\right)+k \left(2929
   y^2-2073\right)\right] \\
g_1^{mid}(y) &= 0 \\
g_2^{mid}(y) &=\frac{8 ky^4 \left(6615(
a \left(y^2-2\right) \left(y^2+5\right)+1)+808 k-26460 a \log
\left(\left(y^2-1\right) \theta \left(2 a-\frac{8
k}{21}-1\right)\right)\right)}{19845 \left(y^2-1\right)^3}
 \end{split}
\end{equation}

\subsubsection{Near field solution}
\textit{Scalar Field}
\begin{equation}\label{scalarin3}
 \begin{split}
  \phi_{1/2}^{in}(z) &= \sqrt{k} \\
\phi_{3/2}^{in}(z) &= -\frac{2}{189} \sqrt{k} \left(189 a \left(2 \log \left(\frac{1}{21} a (2 z+1) \theta ^2 (42 a-8 k-21)\right)+\log ^2(2 z+1)\right) \right.\\
&~~\left.+378 a \text{Li}_2(-2 z)+63 \left(9+\pi^2\right) a-17 k\right)
 \end{split}
\end{equation}
\textit{Metric and Gauge Field}
\begin{equation}
 \begin{split}
A_1^{in}(z) &=\frac{4 \sqrt{\frac{2}{3}} k (z (-42 a+8 k+21)-42 a \log (2 z+1))}{21 a} \\
f_1^{in}(z) &= -\frac{16 k (-42 a+8 k+21)}{416745 a^2} \left[2 z (315 z (-42 a+8 k+21)+856 k) \right.\\
&\left.-6615 (4 a z+a) \log (2 z+1)\right]\\
g_1^{in}(z) &=\frac{28 a^2 k (z (-39690 a+808 k+6615)+6615 (4 a z+a) \log (2 z+1))}{15 z^2 (2 z+1)^2 (42 a-8 k-21)^3}
 \end{split}
\end{equation}
Matching also sets the parameter $\alpha$ as
\begin{equation}\label{alphaeq}
 \begin{split}
\alpha =
\left(\frac{1}{a}+\frac{8k}{21a}\right)+\left(\frac{89224
k^2}{1528065 a}-\frac{16 k}{21 a}+3 a+\frac{1}{a}+\frac{6568
k}{2205}-3\right)\theta +{\cal O}(\theta^2)
 \end{split}
\end{equation}

\paragraph{Near-horizon expansions of solution}
In order to see the breakdown of perturbation theory at 
extremality, we list the near horizon expansions of 
the near field solutions (in terms of the AdS radial coordinate $r$)
\begin{equation}\label{nfexpr}
 \begin{split}
\phi^{in}_{3/2}(r) =& \frac{2}{189} \sqrt{k} \left(-378 a \log \left(\frac{1}{21} a \theta ^2 (42 a-8 k-21)\right)-63 \left(9+\pi ^2\right) a+17 k\right) \\
&-\frac{1764 \sqrt{k}
\left(r-\sqrt{a \theta }\right)^2}{\theta ^3 (42 a-8 k+21)^2}+\frac{82320 k \left(r-\sqrt{a \theta }\right)^3}{\theta ^4 (42 a-8 k+21)^3 \sqrt{a \theta
    k}}+\Or\left(\left(r-\sqrt{a \theta }\right)^4\right)  \\
A^{in}_1(r) =&\frac{4 \sqrt{\frac{2}{3}} k (-126 a+8 k+21) \left(r-\sqrt{a \theta }\right)}{(a \theta )^{3/2} (42 a-8 k+21)}+\frac{2352 \sqrt{6} k \left(r-\sqrt{a
   \theta }\right)^2}{a \theta ^3 (42 a-8 k+21)^2} \\
&~~~~-\frac{65856 \left(\sqrt{6} k\right) \left(r-\sqrt{a \theta }\right)^3}{a^{3/2} \theta ^{9/2} (42 a-8
   k+21)^3}+\Or\left(\left(r-\sqrt{a \theta }\right)^4\right) \\
f^{in}_1(r) =& \frac{32 k (856 k-6615 a) \left(r-\sqrt{a \theta }\right)}{19845 a^{5/2} \theta ^{3/2}}+\frac{32 k (-105 a+8
   k+21) \left(r-\sqrt{a \theta }\right)^2}{3 a^3 (42 a-8 k-21) \theta ^3} \\
   &~~~~+\frac{12544 k \left(r-\sqrt{a
   \theta }\right)^3}{a^{5/2} (-42 a+8 k+21)^2 \theta ^{9/2}}+\Or\left(\left(r-\sqrt{a \theta }\right)^4\right)\\
g^{in}_1(r) =&\frac{4 a k (-26460 a+808 k+6615) (a \theta )^{3/2}}{45 (-42 a+8 k+21)^2 \left(r-\sqrt{a \theta
   }\right)}+\frac{56 a^2 (6615 (11 a-2)-1616 k) k}{15 (42 a-8 k-21)^3} \\
   &~~~~-\frac{2352 \left((735 (58 a-9)-808 k)
   k (a \theta )^{3/2}\right) \left(r-\sqrt{a \theta }\right)}{5 \left((-42 a+8 k+21)^4 \theta
   ^3\right)}+\Or\left(\left(r-\sqrt{a \theta }\right)^2\right)
 \end{split}
\end{equation}

\section{Perturbative expansion for small charge black holes
when $e^2 =\frac{32}{3}$}
\label{ap:ecritical}
\subsection{Far field solution}
\textit{Scalar Field}
\begin{equation}
\begin{split}
\phi_{1/2}^{out}(r) &= \frac{\sqrt{k}}{(1+r^2)^2} \\
\phi_{3/2}^{out}(r) &= -\frac{2 \sqrt{k} \left(378 a
\left(r^2+1\right)^3+378 a \left(r^2+1\right)^4 \log
\left(\frac{r^2}{r^2+1}\right)+k \left(48 r^6+90 r^4+4
r^2-17\right)\right)}{189
   \left(r^2+1\right)^6} \\
\phi_{5/2}^{out}(r) &=
\frac{-2\sqrt{k}}{68762925r^2(1+r^2)^{10}}\bigg[-137525850 a^2
\left(4 r^4+8 r^2+1\right) \left(r^2+1\right)^6 \\
&+104781600 a k r^2 \left(r^2+1\right)^8
\text{Li}_2\left(-r^2\right)+6930 a k r^2 \left(2520 \pi ^2
   \left(r^2+1\right)^5\right. \\
&\left. +r^2 \left(4 \left(630 r^8+6615 r^6+26031 r^4+39084
r^2+22756\right) r^2+29891\right)+1789\right)
\left(r^2+1\right)^3 \\
&-727650 a r^2 \left(r^2+1\right)^4
\left(\log \left(\frac{1}{r^2}+1\right) \left(18
\left(r^2+1\right)^4 (21 a+4 k) \log
\left(\frac{1}{r^2}+1\right) \right. \right. \\
&\left.\left.-189 a \left(5 r^2+9\right) \left(r^2+1\right)^3+4
k r^2
\left(r^2+2\right) \left(6 r^8+48 r^6+112 r^4+56
r^2+71\right)+214 k\right)\right. \\
&\left.-288 k \left(r^2+1\right)^4 \log ^2(r)\right)+k^2 r^2
\left(r^2 \left(r^2 \left(\left(5
\left(535344 r^6+2059362 r^4+4366756 r^2
\right.\right.\right.\right.\right. \\
&\left.\left.\left.\left.\left.+9416105\right)
r^2+54295556\right)
r^2+10445188\right)-8980072\right)-5013164\right)\bigg]
\end{split}
\end{equation}
\textit{Metric and Gauge Field}
\begin{equation}
\begin{split}
A_1^{out}(r) &=-\frac{2 \sqrt{\frac{2}{3}} k \left(r^4+3
r^2+3\right)}{3 \left(r^2+1\right)^3} \\
A_2^{out}(r) &=
\frac{2k}{19845r^2(1+r^2)^7}\sqrt{\frac{2}{3}}\bigg[52920 a r^2
\left(r^4+3 r^2+3\right) \left(r^2+1\right)^4 \log
\left(\frac{r^2}{r^2+1}\right) \\
&-2205 a \left(17 r^8+44 r^6+18 r^4-36 r^2-9\right)
\left(r^2+1\right)^3 \\
&-4 k r^2\left(r^2 \left(\left(\left(131 \left(r^4+7
r^2+21\right) r^2+4270\right) r^2+4333\right)
r^2+2541\right)-747\right)\bigg]
\end{split}
\end{equation}
\begin{equation}
\begin{split}
f_1^{out}(r) &=-\frac{8 k \left(r^4+3 r^2+3\right)}{9
\left(r^2+1\right)^3} \\
f_2^{out}(r) &= \frac{8k}{59525(1+r^2)^7}\bigg[52920 a
\left(r^4+3 r^2+3\right) \left(r^2+1\right)^4 \log
\left(\frac{r^2}{r^2+1}\right) \\
&-\frac{2205 a \left(17 r^8+41 r^6+6 r^4-54 r^2-18\right)
\left(r^2+1\right)^3}{r^2} \\
&-2k \left(577 r^{12}+3304 r^{10}+6972 r^8+3248 r^6-7966
r^4-12966 r^2-8787\right)\bigg]
\end{split}
\end{equation}
\begin{equation}
\begin{split}
g_1^{out}(r) &=\frac{8 k r^2 \left(r^2+3\right)}{9
\left(r^2+1\right)^5} \\
g_2^{out}(r) &= \frac{8k}{59525(1+r^2)^9}\bigg[-52920 a r^2
\left(r^2+3\right) \left(r^2+1\right)^4 \log
   \left(\frac{r^2}{r^2+1}\right) \\
&+2205 a \left(17 r^6+41 r^4+18
   r^2+54\right) \left(r^2+1\right)^3\\
&+2 k r^2 \left(577 r^{10}+3304 r^8+9912
   r^6+29960 r^4+41930 r^2+7350\right)\bigg]
\end{split}
\end{equation}
\subsection{Intermediate field solution}
\textit{Scalar Field}
\begin{equation}
\begin{split}
\phi_{1/2}^{mid}(y) &= \sqrt{k} \\
\phi_{3/2}^{mid}(y) & = \frac{2}{189} \sqrt{k} \left(-378 a \log
\left(a \left(y^2-1\right) \theta
   \right)-189 a \left(y^2+2\right)+17 k\right)
\end{split}
\end{equation}
\textit{Metric and Gauge Field}
\begin{equation}
\begin{split}
A_1^{mid}(y) &=-\frac{2 \sqrt{\frac{2}{3}} k
\left(y^2-1\right)}{y^2}\\
A_2^{mid}(y) &=\frac{4\sqrt{\frac{2}{3}}k}{2205y^2}\bigg[2205 a
\left(2 \log \left(2a-\frac{8 k}{21}\right)+4 y^2 \log \left(a
\left(y^2-1\right) \theta \right)-2\log (a^2\theta)+y^4 \right.
\\
&~~~~\left. -6 \log \left(y^2-1\right)-1\right)+166 k
\left(y^2-1\right)\bigg]
\end{split}
\end{equation}
\begin{equation}
\begin{split}
f_1^{mid}(y) &= -\frac{8 k \left(y^2-1\right)^2}{3 y^4} \\
f_2^{mid}(y) &= \frac{16k}{19845y^4}\bigg[13230 a
\left(\left(y^2-1\right) \log \left(2a-\frac{8
k}{21}\right)+y^2\left(2 y^2 \log (a \theta )\right. \right. \\
&~~~\left.\left.+\left(2 y^2-5\right) \log
\left(y^2-1\right)- 3 \log \theta \right)+\left(2-4 y^2\right)
\log a+3
   \log \left(y^2-1\right)+\log (\theta )\right) \\
   &+\left(y^2-1\right)
\left(6615 a \left(y^4-1\right) +k \left(2929
y^2-2073\right)\right)\bigg] \\
g_1^{mid}(y) &=0 \\
g_2^{mid}(y) &= \frac{8 k y^4 \left(-26460 a \log \left(\frac{42
a \theta -8 k \theta }{21(y^2-1)}\right)+6615 a
\left(y^2-2\right) \left(y^2+5\right)+808
   k\right)}{19845 \left(y^2-1\right)^3}
\end{split}
\end{equation}

\subsection{Near field solution}
\begin{figure}[ht]
 \begin{center}
\includegraphics[height=75mm]{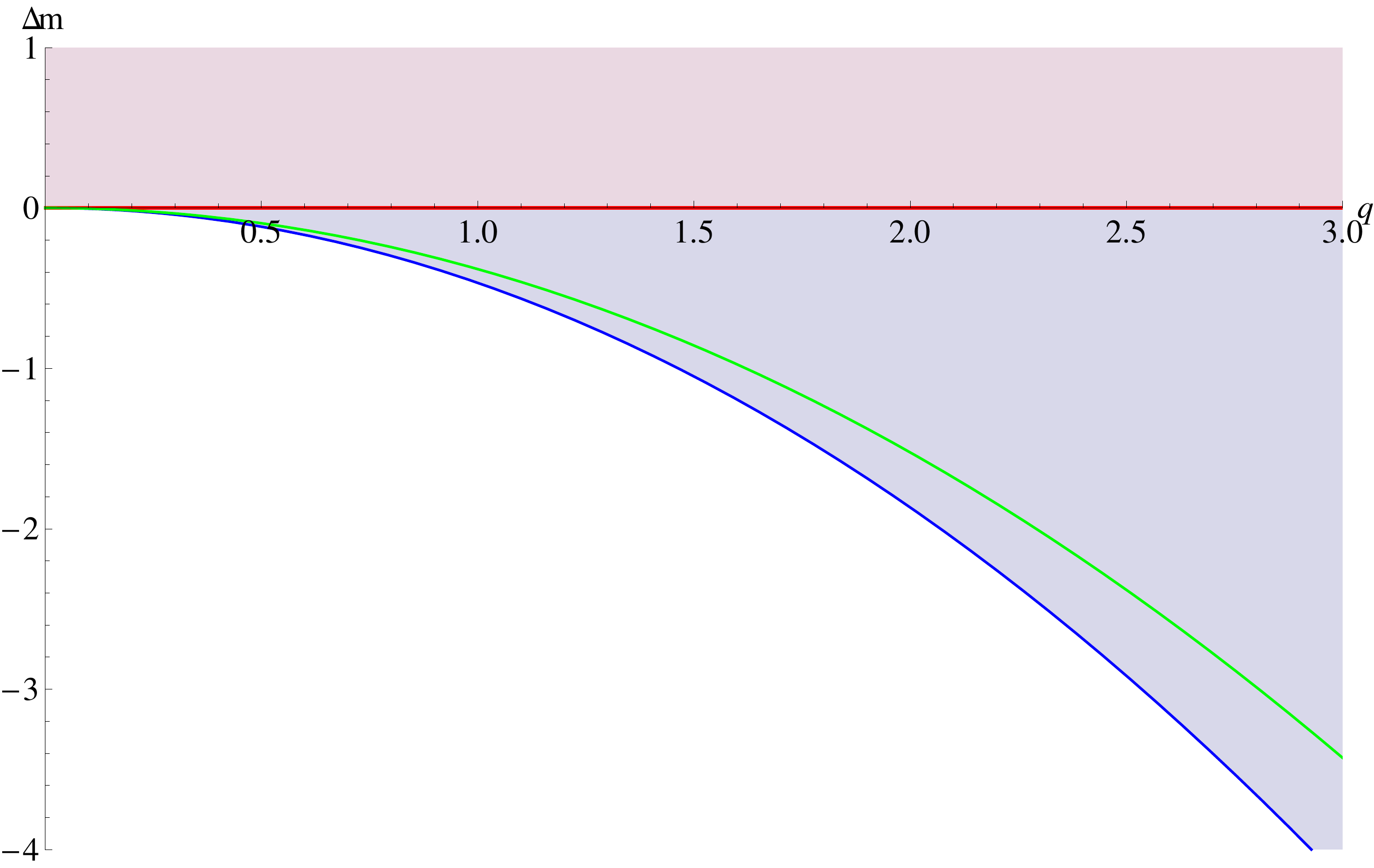} 
\end{center}
\caption{\textbf{Range of allowed values of $\Delta m$ for hairy black holes with $e^2 = \frac{32}{3}$} at leading order in small $\theta$ (blue shaded). Hairy black holes exist for all charges $q\geq 0$. The upper limit, $\Delta m=0$ denotes the onset of superradiant instabilities (red line). This is also the extremality line for pure RN AdS black holes. The lower limit $\Delta m = -\frac{7}{15}q^2$ corresponds to the extremal hairy black hole with $\alpha =2$ (blue line). The green line denotes the soliton, $\Delta m = -\frac{8q^2}{21}$. This phase diagram is quite similar to Fig. \ref{figure1}, the only difference being that the hairy black hole now exist at all charges $q\geq0$.}
\label{figure4}
\end{figure}

\textit{Scalar Field}
\begin{equation}
\begin{split}
\phi_{1/2}^{in}(z) &= \sqrt{k} \\
\phi_{3/2}^{in}(z) &= -\frac{2}{189} \sqrt{k} \left(378 a \text{Li}_2(-2 z)
+189 a (4 \log (a \theta )+\log (2 z+1) (\log (2 z+1)+2)) \right.\\
&\left.+63 \left(9+\pi ^2\right) a-17 k\right)
\end{split}
\end{equation}
\textit{Metric and Gauge Field}
\begin{equation}
\begin{split}
A_1^{in}(z) &=-\frac{8 \sqrt{\frac{2}{3}} k (21 a z+21 a \log (2 z+1)-4 k z)}{21 a} \\
f_1^{in}(z) &= -\frac{32 k (21 a-4 k)}{416745 a^2} \left[6615 a (4 (z-1) z-1)+6615 (4 a z+a) \log (2 z+1) \right. \\
&\left.~~+4 k (4 (208-315 z) z+315)\right]\\
g_1^{in}(z) &= \frac{7 a^2 k (-6615 a (10 z+1)+6615 (4 a z+a) \log (2 z+1)+4 k (1462 z+315))}{30 z^2 (2 z+1)^2 (21 a-4 k)^3}
\end{split}
\end{equation}
Matching also sets the parameter $\alpha$ as
\begin{equation}
\alpha = \frac{8 k}{21 a}+\theta \left(\frac{89224 k^2}{1528065
a}+3 a+\frac{6568 k}{2205}\right)
\end{equation}
\paragraph{Leading order thermodynamics in perturbation theory}
Once we have the solutions from the previous subsections,
the evaluation of their thermodynamic charges and potentials
is straightforward. Using formulas \eqref{thermform}, 
we find
\begin{equation}
\begin{split}
M &= \frac{3\pi}{8}\bigg[\theta  \left(2 a+\frac{8 k}{9}\right)+\theta ^2 \left(a^2+\frac{1024 a k}{189}+\frac{9232 k^2}{59535}\right) \bigg]+{\cal O}\left(\theta ^3\right)\\
Q & = \frac{\pi}{4}\sqrt{\frac{3}{2}}\bigg[\theta  \left(2 a+\frac{8 k}{9}\right)+\theta ^2 \left(\frac{1024 a k}{189}+\frac{4192 k^2}{59535}\right) \bigg]+{\cal O}\left(\theta ^3\right) \\
\mu &= \sqrt{\frac{3}{2}}+\frac{2}{7} \sqrt{\frac{2}{3}} k \theta +\frac{\theta ^2 \left(4584195 a^2+4551624 a k+33784 k^2\right)}{1018710 \sqrt{6}}+{\cal O}\left(\theta ^3\right)\\
T &= \frac{\sqrt{\theta } (21 a-4 k)}{21 \pi  \sqrt{a}}+\frac{\theta ^{3/2} \left(-4584195 a^2-12701304 a k+965368 k^2\right)}{3056130 \pi  \sqrt{a}}+{\cal O}\left(\theta ^{5/2}\right)\\
S &= \frac{\pi^2(a\theta)^{3/2}}{2}
\end{split}
\end{equation}
We have verified that these quantities obey the first law of thermodynamics 
$$dM = TdS + \mu dQ$$
It is quite easy to show, that in terms of the shifted and rescaled
mass $\Delta m$ and rescaled charge $q$, these formulae reduce to
\begin{equation}
 \begin{split}
  q &= \frac{9a+4k}{3\sqrt{6}}+{\cal O}(\theta) \\
\Delta m &= -\frac{8}{567}k\left(63a+8k\right)+{\cal O}(\theta) \\
 \end{split}
\end{equation}
Positivity and reality of $k$ constrains $\Delta m$ in the ranges
\begin{equation}
{\cal O}(\theta) \geq \Delta m \geq \frac{7q^2}{15}+{\cal O}(\theta) 
\end{equation}
This range is shown in the Fig. \ref{figure4}

\section{Perturbative expansion for $e^2=\frac{32}{3}(1 +
\theta)$}
\label{ap:largee}
\subsection{Far field solution}
\textit{Scalar Field}
\begin{equation}
\begin{split}
\phi^{out}_{1/2}(r) &= \frac{\sqrt{k}}{\left(r^2+1\right)^2} \\
\phi^{out}_{3/2}(r) &= -\frac{2 \sqrt{k} \left(378 a
\left(r^2+1\right)^3-378 a
   \left(r^2+1\right)^4 \log \left(\frac{1}{r^2}+1\right)+k
   \left(48 r^6+90 r^4+4 r^2-17\right)\right)}{189
   \left(r^2+1\right)^6} \\
\phi^{out}_{5/2}(r) &= -\frac{2a
\sqrt{k}}{189(1+r^2)^6}\left(378 a \left(5 r^2+9\right)
\left(r^2+1\right)^3+567 \left(r^2+1\right)^4\right. \\
&\left.-4 k \left(2 \left(r^2+2\right) \left(6 r^8+48 r^6+112
r^4+56 r^2+71\right)
   r^2+107\right)\right)\log \left(1+\frac{1}{r^2}\right) \\
&+\frac{8 a \sqrt{k} (21 a+4 k)}{21 \left(r^2+1\right)^2}\log^2
\left(1+\frac{1}{r^2}\right)
-\frac{64ak^{3/2}}{21(1+r^2)^2}\left(\text{Li}_2(-r^2)+2\log^2
r\right) \\
&-\frac{2\sqrt{k}}{68762925r^2(1+r^2)^{10}}\bigg[-137525850 a^2
\left(4 r^4+8 r^2+1\right)
   \left(r^2+1\right)^6 \\
&+3465 a r^2 \left(r^2+1\right)^3\left(2 k \left(2520 \pi ^2
\left(r^2+1\right)^5+r^2
\left(4 \left(630 r^8+6615 r^6+26031 r^4 \right.\right.
\right.\right.\\
&\left.\left.\left.\left.+39084r^2 +22756\right)
r^2+29891\right)+1789\right)-19845
   \left(r^2+1\right)^4\right) \\
&+k r^2 \left(k \left(r^2\left(r^2 \left(\left(5 \left(535344
r^6+2059362
   r^4+4366756 r^2+9416105\right) r^2+54295556\right)
   r^2 \right.\right.\right.\right. \\
&\left.\left.\left.\left.+10445188\right)-8980072\right)-5013164\right)+5821200
\left(9 r^4+21 r^2+13\right) \left(r^2+1\right)^5\right)\bigg]\end{split}
\end{equation}
\textit{Metric and Gauge Field}
\begin{equation}
\begin{split}
A_1^{out}(r) &= -\frac{2 \sqrt{\frac{2}{3}} k \left(r^4+3
r^2+3\right)}{3
   \left(r^2+1\right)^3} \\
A_2^{out}(r) &=
\frac{\sqrt{\frac{2}{3}}k}{19845r^2(1+r^2)^7}\bigg[-105840 a r^2
\left(r^4+3 r^2+3\right) \left(r^2+1\right)^4
   \log \left(\frac{1}{r^2}+1\right) \\
&-4410 a \left(17 r^8+44
   r^6+18 r^4-36 r^2-9\right) \left(r^2+1\right)^3 \\
&-r^2\left(8 k \left(r^2 \left(\left(\left(131 \left(r^4+7
   r^2+21\right) r^2+4270\right) r^2+4333\right)
   r^2+2541\right)-747\right)\right. \\
&\left.+6615 \left(r^4+3 r^2+3\right)
   \left(r^2+1\right)^4\right)\bigg]
\end{split}
\end{equation}

\begin{equation}
\begin{split}
f_1^{out}(r) &=-\frac{8 k \left(r^4+3 r^2+3\right)}{9
\left(r^2+1\right)^3} \\
f_2^{out}(r) &= \frac{8k}{59535r^2(1+r^2)^7}\bigg[52920 a r^2
\left(r^4+3 r^2+3\right) \left(r^2+1\right)^4 \log
\left(\frac{r^2}{r^2+1}\right) \\
&-2205 a \left(17 r^8+41 r^6+6 r^4-54 r^2-18\right)
\left(r^2+1\right)^3 \\
&-2 k r^2 \left(577 r^{12}+3304 r^{10}+6972 r^8+3248 r^6-7966
   r^4-12966 r^2-8787\right)\bigg] \\
g_1^{out}(r) &= \frac{8 k r^2 \left(r^2+3\right)}{9
\left(r^2+1\right)^5} \\
g_2^{out}(r) &= \frac{8k}{59535r^2(1+r^2)^9}\bigg[52920 a r^2
\left(r^2+3\right) \left(r^2+1\right)^4 \log
   \left(\frac{1}{r^2}+1\right) \\
&+2205 a \left(17 r^6+41
   r^4+18 r^2+54\right) \left(r^2+1\right)^3 \\
&+2 k r^2
   \left(577 r^{10}+3304 r^8+9912 r^6+29960 r^4+41930
   r^2+7350\right)\bigg]
\end{split}
\end{equation}

\subsection{Intermediate field solution}
\textit{Scalar Field}
\begin{equation}
\begin{split}
\phi_{1/2}^{mid} &= \sqrt{k}\\
\phi_{3/2}^{mid} &= \frac{2}{189} \sqrt{k} \left(-378 a \log
\left(a
   \left(y^2-1\right) \theta \right)-189 a
   \left(y^2+2\right)+17 k\right)\\
\end{split}
\end{equation}
\textit{Metric and Gauge Field}
\begin{equation}
\begin{split}
A_1^{mid} (y) &= -\frac{2 \sqrt{\frac{2}{3}} k
\left(y^2-1\right)}{y^2} \\
A_2^{mid} (y) &=
\frac{\sqrt{\frac{2}{3}}k}{2205y^2}\bigg[\left(y^2-1\right)
\left(8820 a \left(y^2+1\right)+664 k+2205\right) \\
&-17640 a \left(-\log \left(2 a-\frac{8
   k}{21}+1\right)-2 y^2 \log \left(a \left(y^2-1\right)
\theta \right)+\log (a^2\theta)+3 \log
\left(y^2-1\right)\right)\bigg] \\
f_1^{mid}(y)  &= -\frac{8 k \left(y^2-1\right)^2}{3 y^4} \\
f_2^{mid}(y) &= \frac{8k(y^2-1)}{19845y^4}\bigg[13230 a
\left(y^4-1\right)+k \left(5858 y^2-4146\right)-6615 \\
&-26460 a \left(3 \log \left(y^2-1\right)-\log \left(2 a-\frac{8
k}{21}+1\right)-2 y^2\log \left(a \left(y^2-1\right) \theta
\right)+2 \log
   (a^2\theta)\right)\bigg] \\
g_1^{mid}(y) &= 0 \\
g_2^{mid}(y) &= \frac{8 k y^4 \left(-26460 a\log
\left[\frac{\theta\left(42
a-8k+21\right)}{21(y^2-1)}\right]+6615 a \left(y^2-2\right)
\left(y^2+5\right)+808
   k\right)}{19845 \left(y^2-1\right)^3}
\end{split}
\end{equation}
\subsection{Near field solution}
\textit{Scalar Field}
\begin{equation}
\begin{split}
\phi_{1/2}^{in}(z) &= \sqrt{k} \\
\phi_{3/2}^{in}(z) &= -\frac{2}{189} \sqrt{k} \left(189 a \left(2 \log \left(\frac{1}{21} a \theta ^2 (2 z+1) (42 a-8 k+21)\right)+\log ^2(2 z+1)\right)\right. \\
&\left.+378 a   \text{Li}_2(-2 z)+63 \left(9+\pi ^2\right) a-17 k\right)
\end{split}
\end{equation}
\textit{Metric and Gauge Field}
\begin{equation}
\begin{split}
A_1^{in}(z) &= -\frac{4 \sqrt{\frac{2}{3}} k (z (42 a-8 k+21)+42 a \log (2 z+1))}{21 a} \\
f_1^{in}(z) &= -\frac{8 k (42 a-8 k+21)}{416745 a^2} \left[1260 z^2 (42 a-8 k+21)+315 (42 a-8 k+21)\right. \\
&\left.+13230 (4 a z+a) \log (2 z+1)+2 (6615-1712 k) z\right]\\
g_1^{in}(z) &= \frac{14 a^2 k (-13230 a (6 z-1)+13230 (4 a z+a) \log (2 z+1)+
8 k (202 z-315)+6615)}{15 z^2 (2 z+1)^2 (42 a-8 k+21)^3}
\end{split}
\end{equation}
Matching also sets the parameter $\alpha$ as 
\begin{equation}
\alpha =\frac{8
k-21}{21 a}+ \theta \left(\frac{\frac{89224
k^2}{1528065}+\frac{16
   k}{21}+1}{a}+3 a+\frac{6568 k}{2205}+3\right)
\end{equation}


\bibliographystyle{utphys}
\bibliography{bibliography}


\end{document}